\documentclass[
        prf ,
         amsmath,amssymb,
        longbibliography]{revtex4-1}
        
        \usepackage{graphicx}
        \usepackage{dcolumn}
        \usepackage{subfigure}
        \usepackage{bm}
        \usepackage{textcomp}
        \usepackage{siunitx}
        \usepackage{hyperref}
        \hypersetup{
            colorlinks=true, 
            citecolor=blue,
            linktoc=all,     
            linkcolor=blue,  
        }
        \usepackage{comment}
        \usepackage{lipsum}
        \usepackage{mathtools}
        \usepackage{float}
        
        \usepackage[normalem]{ulem}
        \usepackage{color}
        \usepackage{appendix}
        \usepackage{enumerate}
    \usepackage[utf8]{inputenc}
        
        \newcommand{\kvsc}[1]{\textcolor{black}{#1}}
        
        \graphicspath{ {images/} }
        \begin{document}
        
        \preprint{AIP/PRE}
        
        
        \title{Wall curvature driven dynamics of a microswimmer}
        
        \author{Chaithanya K. V. S.}
        \author{Sumesh P. Thampi}
        \affiliation{Department of Chemical Engineering, Indian Institute of Technology Madras, Chennai, 600036, India.}

        
        \begin{abstract}
        Microorganisms navigate through fluid, often confined by complex environments, to survive and sustain life. Inspired by this fact, we consider a model system and seek to understand the wall curvature driven dynamics of a squirmer, a mathematical model for a microswimmer, using (i) lattice Boltzmann simulations and (ii) analytical theory by \citet{dario_gareth}.  The instantaneous dynamics of the system is presented in terms of fluid velocity fields, and the translational and angular velocities of the microswimmer, whereas the long time dynamics is presented by plotting the squirmer trajectories near curved boundaries \kvsc{in physical and dynamical space}, as well as characterising them in terms of \kvsc{fixed points and experimentally relevant measures, namely, }(i) proximity parameter, (ii) retention time, (iii) swimmer orientation and (iv) tangential velocity near the boundary, and (v) scattering angle during the collision. Our detailed analysis shows that irrespective of the type and strength, microswimmers exhibit a greater affinity towards a concave boundary due to hydrodynamic interactions compared to a convex boundary. In the presence of additional repulsive interactions with the boundary,  we find that pullers (propel by forward thrust)  have a slightly greater affinity towards the convex--curved walls compared to pushers (propel by backward thrust). Our study provides a comprehensive understanding of the consequence of hydrodynamic interactions in a unified framework that encompasses the dynamics of pullers, pushers, and neutral swimmers in the neighbourhood of flat, concave, and convex walls. In addition, the combined effect of oppositely curved surfaces is studied by confining the squirmer in an annulus. The results presented  in a unified framework and insights obtained are expected to be useful to design geometrical confinements to control and guide the motion of microswimmers in microfluidic applications.
        \end{abstract}
        
        \pacs{Valid PACS appear here}
        \maketitle
        
        
        \section{\label{sec:level1}Introduction}
        
        The natural habitats of microorganisms are complex. They navigate in a fluid, often through complex and confined environments. In such surroundings, long ranged hydrodynamic interactions of microorganisms with confining boundaries may play a crucial role and they decide the trajectories of these swimming bodies \cite{confinement_review}. Therefore important parameters that affect the trajectory of a microorganism, which are studied in the literature, are (i) its size and shape,  \cite{spheroidal,pedley_spherical,swimmershape}, (ii) strength of the confinement \cite{Ahana,confinement_review}, (iii) rheological properties of the suspending fluid (viscosity, elasticity etc.) \cite{powerlaw,viscoelastic,Brinkman_fluid,viscosity_gradients}, (iv) presence of a background flow (linear, quadratic flow etc.) \cite{shearflow,poiseulle_flow}, and (v) external forces acting on the microorganism (such as gravity) \cite{gravity1,gravity2}. On the other hand, far fewer studies have considered the effect of confining geometry on the dynamics of a microswimmer (or a swimming microorganism) and thus a systematic analysis of the effects of curved boundaries on the dynamics of different microswimmers are not well recorded in the literature. In this work, we address this question and fill the gap by presenting a detailed study of the effect of confining geometry on the dynamics of a microswimmer.
        
        Based on geometry, the works that have addressed the effect of confining geometry \cite{three_geometries,dario_gareth} on the dynamics of a microswimmer can be broadly classified into three categories: effects due to (i) a flat wall \cite{flatwall,Ahana, crowdywall}, (ii) a convex boundary \cite{convex,manyconvex,convex_expts,convex_simulations}, and (iii) a concave boundary \cite{Collective_concave}. Following are some of the interesting behaviours observed in these geometries. \textit{Escherichia coli} (\textit{E. coli}) and other similar bacteria are attracted to flat walls and get trapped. Experiments showed that trapped \textit{E. coli} rotates in a clockwise direction (when viewed from above the wall)  \cite{wall_exp1,wall_exp2,wall_clockwise}, but the sense of rotation reverses when it is near a flat fluid--fluid interface  \cite{interface_exp1,interfac_exp2,cou_clock_rot_swimmer_interface}. Convex boundaries also trap microswimmers, for example chemically propelled Janus microrods are captured by spherical obstacles \cite{convex_expts}. The capture occurs only when the size of the obstacle is larger than a critical size \cite{point_particle_convex}, and the trapping ability of the spherical obstacle is enhanced by the viscoelasticity of the suspending fluid \cite{three_geometries}. Similar trapping ability is exhibited by concave surfaces. \textit{Chlamydomonas reinhardtii} (\textit{C. reinhardtii}) cells confined in circular and elliptical enclosures get trapped near concave boundaries and this trapping ability is observed to increase with concavity \cite{concave_expts} but reduces with increase in activity of the microswimmer and fluid viscoelasticity \cite{three_geometries}.
        
        The experimental observations are often successfully explained by simple theoretical models and numerical simulations based on the hydrodynamic interactions between the microswimmer and the boundary. For example, a point particle model by \citet{point_particle_convex} qualitatively captured the trapping behavior of microswimmers by spherical obstacles (convex curvature). Building on this concept, both theoretical and numerical investigations have been performed later on, to include finite sized microswimmers and walls with arbitrary curvatures \cite{arbitrary_curvature, convex_simulations}. \kvsc{\citet{paolo1} analyzed the dynamics of a dilute suspension of microswimmers in channels of varying cross section. They reported that the accumulation of microswimmers at channel walls is sensitive to the swimming mechanism and the geometry of the channels.} \citet{point_particle_convex_drop} analyzed spherical, surfactant laden drops and found that the trapping ability of surfactant laden drops is larger than that of rigid spherical obstacles. Recently \citet{point_cave} analytically studied the dynamics of a point swimmer confined inside a drop. However, the point particle based theoretical models are valid only if the separation distance between the microswimmer and the boundary is much larger than the size of the microswimmer itself \cite{farfield_lauga, point_particle_convex, point_particle_convex_drop, point_cave}. \kvsc{Using fully resolved numerical simulations, several studies have also investigated the microswimmer dynamics near a solid--fluid \cite{Ahana,convex_simulations,arbitrary_curvature,confinement_LBM,powerlaw} or a fluid--fluid \cite{paolo2,lauga_drop} interface.}
        
As evident from the above discussion, most of the experimental and theoretical studies have focused the analysis only on a particular type of microswimmer or a particular type of wall curvature (flat, concave or convex). To the best of our knowledge, there are no studies that compare and contrast the way in which the concave and convex curvatures affect the microswimmer dynamics and relate the observations to that of a flat wall. Therefore, in this work we analyze the dynamics of various types of microswimmers dictated by different geometrical confinements using the theory developed by \citet{dario_gareth} and fully resolved numerical simulations based on the lattice Boltzmann method. We present our results in a unified framework illustrating the combined role of microswimmer type and wall curvature in determining the behaviour of microswimmers in confinements. Further, the  combined affect  of convex and concave curvatures in governing the dynamics of microswimmers are revealed by studying the dynamics of a microswimmer in an annular confinement.  
        
        In this work, we restrict the analysis to two dimensions. Such an approach is justified since (i) the motion of the microswimmer is restricted to a plane in most of the experimental studies, (ii) previous studies \cite{crowdy_2011,Or_3d} have shown that the two dimensional models offer physical insights and qualitatively capture the dynamics shown by the three dimensional microswimmers, and  (iii) curvature is a second rank tensor and thus the analysis of a three dimensional microswimmer near an arbitrary two dimensional surface is cumbersome due to the large parametric space involved. 
        
        This paper is organized as follows. In section~\ref{sec:Mathematical_details}, we give the details of two dimensional squirmer model for the microswimmer, and then present the exact solutions derived by \citet{dario_gareth} for the squirmer dynamics near a curved boundary. We then outline the algorithm of the lattice Boltzmann numerical scheme for numerical simulations. In section~\ref{sec:results}, we present our results by characterising the dynamics of a microswimmer near convex and concave boundaries. The instantaneous dynamics is characterized in terms of the velocity field of the fluid, and the translational and angular velocities of the microswimmer, whereas the long time dynamics is characterized \kvsc{by constructing trajectories in physical space and dynamical space. The trajectories are analyzed} in terms of \kvsc{fixed points and} \kvsc{experimentally relevant} quantities such as proximity parameter, retention time, average orientation and average tangential velocity of the microswimmer near the boundary, and scattering angle resulting from the wall collision. Then, in section~\ref{sec:annular_confinement}, we present the analysis of the microswimmer confined in an annulus in terms of fluid velocity fields and the microswimmer trajectories.
        
        \section{\label{sec:Mathematical_details}Theoretical and numerical details}
        
        In this section, we discuss the squirmer model for the microswimmer, the exact expressions for the squirmer dynamics near a curved boundary, and then outline the lattice Boltzmann method for numerical simulations.
        
        \subsection{Squirmer model for the microswimmer}

        We model the microswimmer as a two dimensional circular squirmer proposed by \citet{blake_2d_squirmer}, according to which, ciliary motion on a circular microswimmer is modeled by prescribing a slip velocity ($\mathbf{u}^{s}$) on its surface,
        \begin{equation}
        \mathbf{u}^{s} = \sum_{n=0}^\infty\Big[A_n \cos(n \theta_c) \boldsymbol{i}_r+B_n\sin(n \theta_c)\boldsymbol{i}_{\theta}\Big].
        \label{eqn:slipvel}
        \end{equation}
        Here $A_n$ and $B_n$ are the strengths of the $n^{th}$ radial and tangential modes respectively, $\theta_c$ is the polar angle on the squirmer surface relative to its orientation, and $\boldsymbol{i}_r$ and $\boldsymbol{i}_{\theta}$ are the radial and azimuthal unit vectors in the polar coordinate system respectively.
        
        \begin{figure}
        \centering
          \subfigure[]{
          \includegraphics[height=5.0cm]{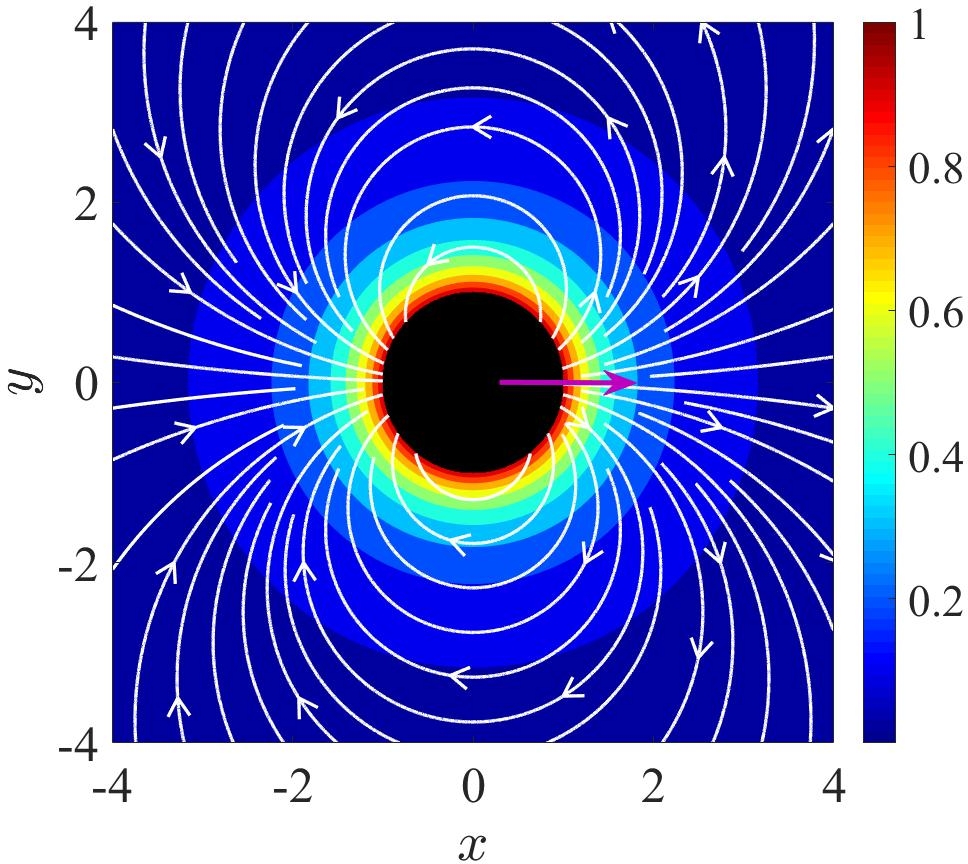}
              \label{fig:unbounded_B1_ff}}\quad
        \subfigure[]{
          \includegraphics[height=5.0cm]{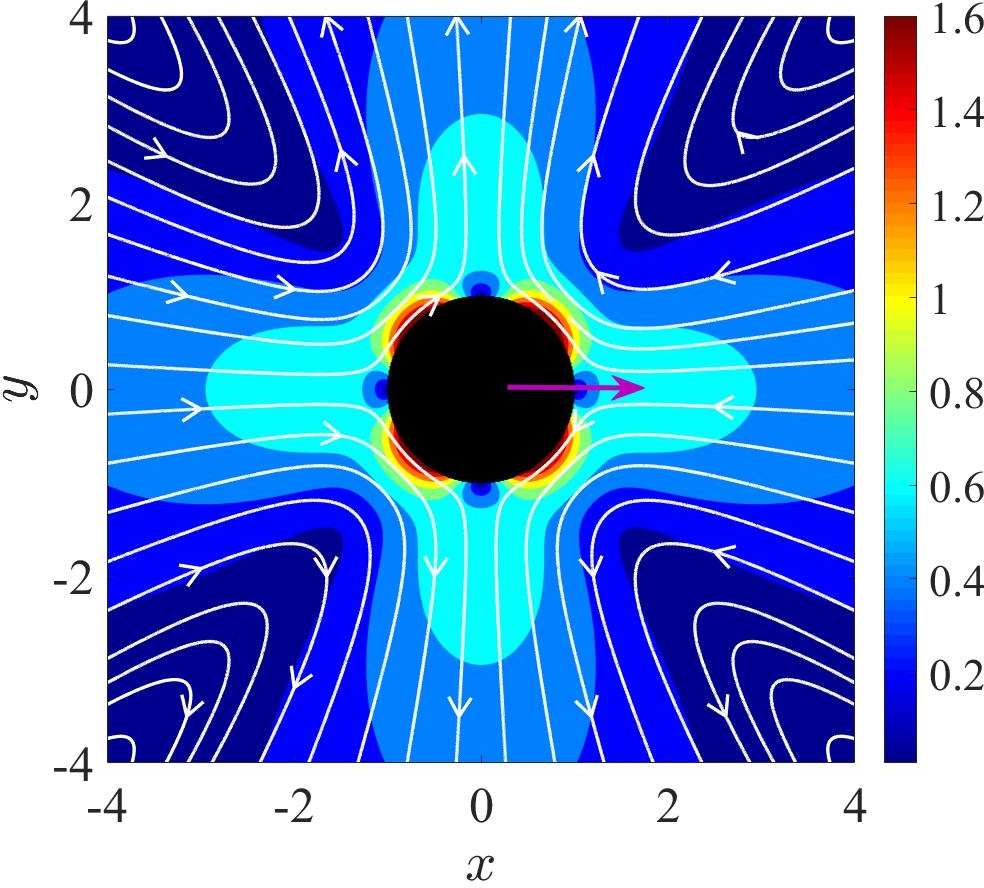}
          \label{fig:unbounded_B2_ff}}\quad 
           \caption{Velocity field  generated by an unconfined squirmer: (a) neutral swimmer ($B_1>0$, $B_2 = 0$) and (b) shaker ($B_1=0$, $B_2 > 0$). The black colored circle in the center represents the squirmer, and the thick horizontal arrow shows its orientation. Continuous lines are streamlines, and the color field shows the magnitude of velocity (normalised by $B_n/2$).} 
             \label{fig:unbounded_vel_field}
        \end{figure}
        
        In the far field, each squirmer mode ($A_n, B_n$) corresponds to a different fundamental solution of Stokes' equations \cite{farfield_lauga}. For example, the velocity field generated by $B_1$ mode corresponds to a source dipole which  decays as $1/r^2$ (see Fig.~\ref{fig:unbounded_B1_ff}), and that by $B_2$ mode corresponds to a force dipole (stresslet) which decays as $1/r$ (see Fig.~\ref{fig:unbounded_B2_ff}). All other modes produce faster decaying velocity fields that the first two modes remain strongest in the far field \cite{blake_2d_squirmer}.
        Therefore, we consider only the effect of first two tangential modes in this work, and consider $A_n = 0$ for all $n$ and $B_n = 0$ for $n>2$. Usually, the type and strength of the microswimmer is indicated by the non-dimensional number $\beta = B_2/B_1$, which is referred as activity hereafter. The limit $\beta\to0$ corresponds to a squirmer with dominant source dipole mode, and $\beta\to \infty$ corresponds to a squirmer with dominant force dipole mode.
        
        Unlike $B_1$ mode, self propulsion cannot be achieved by $B_2$ mode in an unconfined medium or in symmetric environments \cite{lauga_drop,Ahana,KVS_2020}, since the flow field produced by $B_2$ mode has a mirror symmetry. However, in the presence of boundaries, the symmetry in the velocity field is broken which, then, can significantly affect the propulsion velocity of the squirmer.
        
                \begin{figure}
        \centering
          \subfigure[]{
          \includegraphics[height=4.0cm]{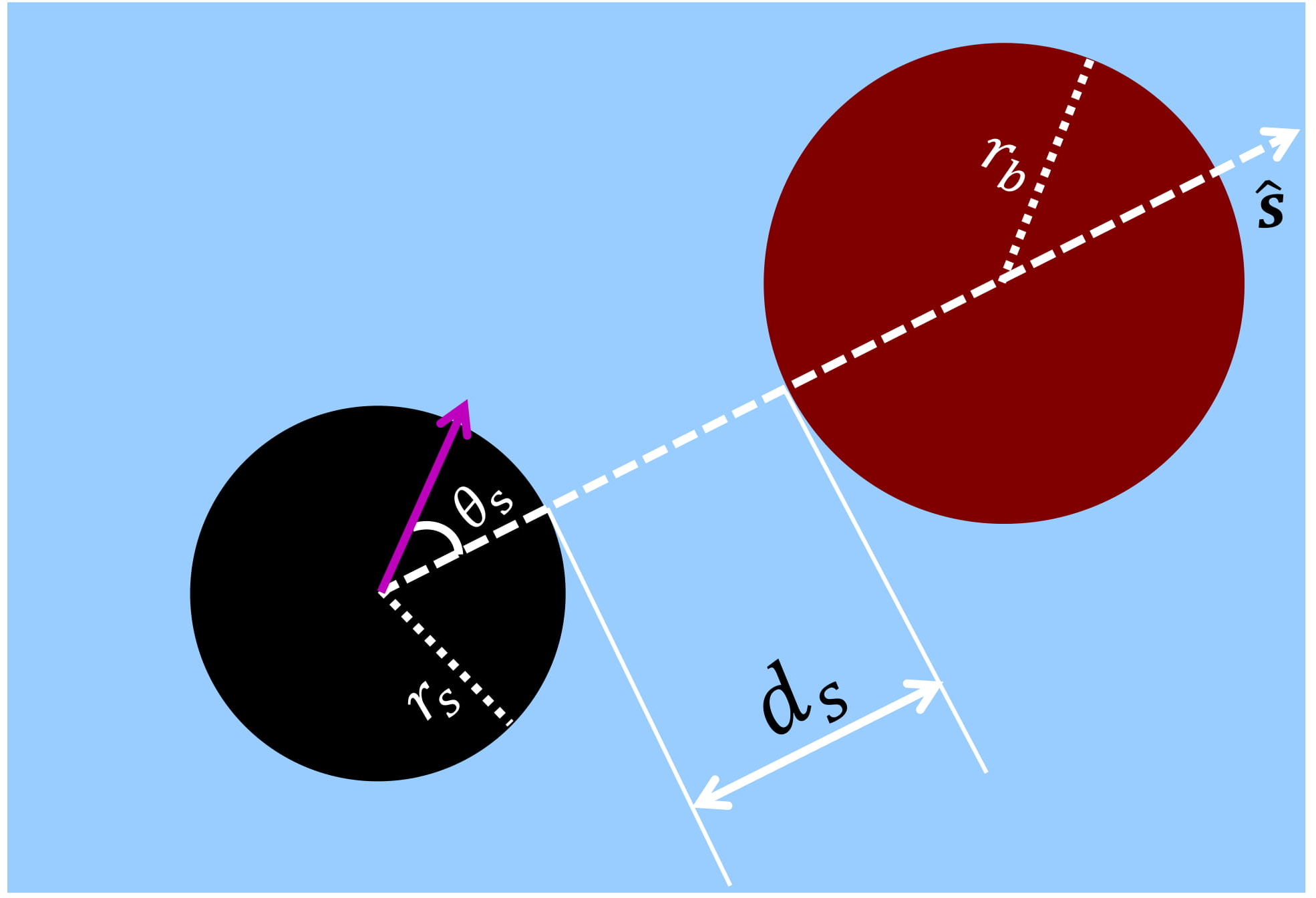}
              \label{fig:schematic}}\quad
        \subfigure[]{
          \includegraphics[height=3.7cm]{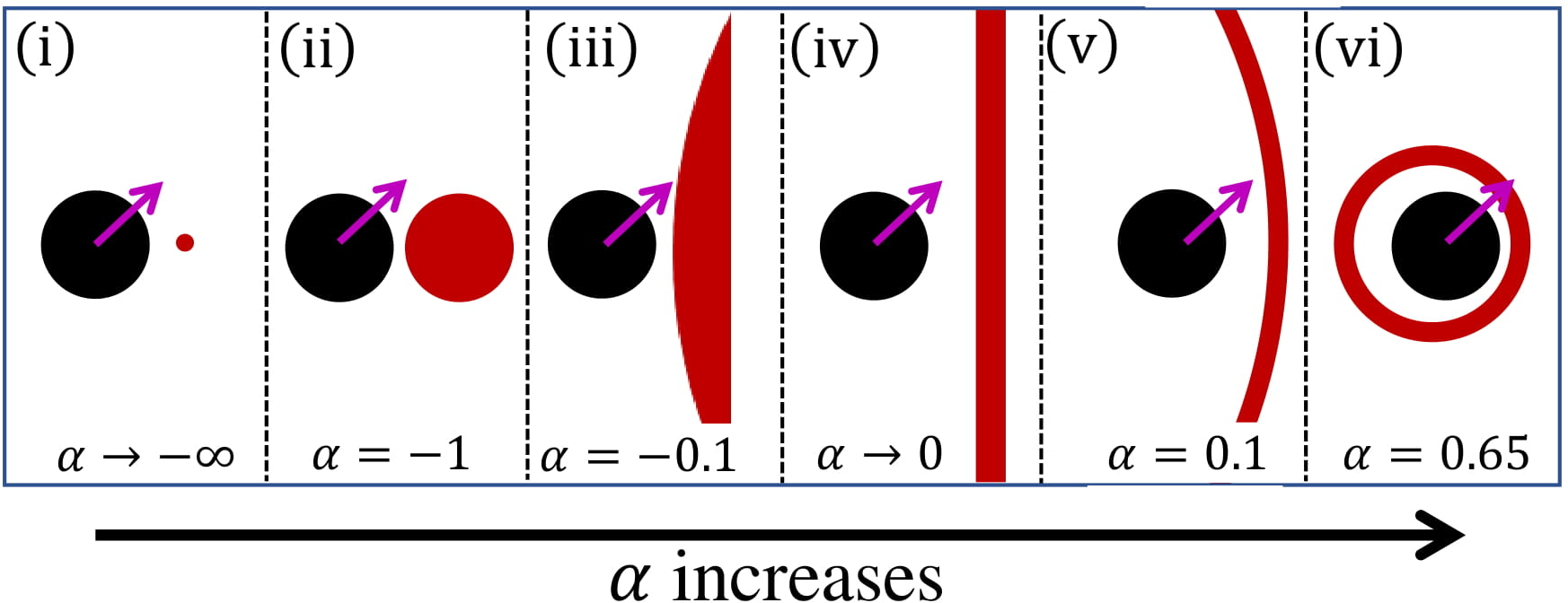}
         \label{fig:curv_schematic}}\quad 
          \caption{Schematic illustration of the system considered. (a) A squirmer of radius $r_s$ is located near a circular post of radius $r_b$ at a distance of $d = d_s+r_s+r_b$, where $d_s$ is the shortest distance between the surface of the swimmer and the surface of the boundary. 
           The orientation vector of the squirmer makes an angle $\theta_s$ with the separation vector ($\hat{\bm{s}}$). (b) Change in the curvature $\alpha = {r_s}/{r_b}$ corresponds to different configurations (i) $\alpha \to -\infty$, squirmer near an infinitesimally small post, (ii) $\alpha = -1$, post has the same radius as the squirmer, (iii) $\alpha = -0.1$, squirmer near a convex boundary, (iv) $\alpha = 0$, squirmer near a flat wall, (v) $\alpha = 0.1$, squirmer near a concave boundary, and (vi) $\alpha = 0.65$, squirmer in a strong circular confinement.} 
             \label{fig:schematicfigures}
        \end{figure}

        Based on the far field flows generated, microswimmers are broadly classified into three types \cite{pedley_spherical}, namely (i) neutral swimmers such as Janus swimmers \cite{janus_swimmers} and molecular motors \cite{molecular_motors}, pullers such as \textit{C. reinhardtii}  \cite{Chlamy1,Chlamy2}, and pushers such as \textit{E. coli} \cite{Ecoli1,Ecoli2}. A squirmer with only $B_1$ mode ($B_2 = 0$ or $\beta = 0$),  a positive $B_2$ (or $\beta > 0$) and a negative $B_2$ (or $\beta < 0$)  mimic the flow field generated by neutral swimmers, pullers, and pushers respectively.

        \subsection{Exact solution for squirmer dynamics near a curved boundary}
        \label{sec:sec2B}
        
        We consider a squirmer of radius $r_s$ placed near a circular boundary of radius $r_b$. They are separated by a distance $d = d_s+r_b+r_s$, where $d_s$ is the shortest distance between the surface of the squirmer and the surface of the boundary. The polarity of the squirmer is indicated by the orientation vector that makes an angle $\theta_s$ with the separation vector $\hat{\bm{s}}$ as shown in the Fig.~\ref{fig:schematic}.

        In order to unify the results related to convex, concave, and flat walls, we define a non-dimensional curvature, as the ratio of curvature of the boundary  ($1/r_b$) to the curvature of the squirmer ($1/r_s$), $i.e.$, $\alpha = r_s/r_b$ with the following sign convention, $\alpha < 0$ for a convex and $\alpha > 0$ for a concave  boundary. The variation of $\alpha$ with change in the radius of curvature of the neighbouring boundary for a squirmer of fixed size is shown in Fig.~\ref{fig:curv_schematic}, in the complete range $-\infty<\alpha<1$. A squirmer near a circular post of infinitesimally small radius corresponds to $\alpha \to -\infty$. Increase in the value of $\alpha$ corresponds to increase in the size of the convex post. At $\alpha = -1$, the squirmer and the post have the same radius. Further increase in $\alpha$ corresponds to posts of larger size with the limit being $\alpha = 0$, which corresponds to a flat wall. $\alpha > 0$ corresponds to a squirmer confined in concave boundaries and the confinement becomes stronger as $\alpha$ increases with the limit being $\alpha = 1$, when the radius of the confinement is equal to that of the squirmer.  Thus, a single parameter $\alpha$ can be used to represent the variation in the curvature of the curved boundary.

        \citet{crowdywall} solved for the dynamics of a 2D squirmer near a flat wall using the complex variables and the conformal mapping approach in conjunction with the reciprocal theorem. Later \citet{dario_gareth} extended this analysis for a curved boundary, and derived the following expressions for squirmer dynamics,
        \begin{subequations}
        \begin{align}   
        	V_{\parallel} &= \Bigg[-\frac{(1-k_1^2 k_2^2)(1-k_2^2)}{1+k_1^2k_2^2} B_1 \cos(\theta_s-c_1) + 2 k_2\frac{(1-k_1^2 k_2^2)(1-k_2^2)}{1+k_1^2k_2^2} B_2 \cos2(\theta_s-c_1)\Bigg],  \\
        	V_{\perp} &= \Bigg[\frac{(1-k_1^2 k_2^2)(1+k_2^2)}{(1+k_1^2k_2^2)} B_1 \sin(\theta_s-c_1)- 2 k_2\frac{(1-k_1^2 k_2^2)(1+k_2^2)}{(1+k_1^2k_2^2)} B_2 \sin2(\theta_s-c_1) \Bigg], \\
            \Omega_z &=  \Bigg[\frac{2k_2}{r_s}\Big(1-\frac{(1-k_1^2 k_2^2)}{(1+k_1^2k_2^2)}\Big) B_1 \sin(\theta_s-c_1)- \frac{2 k_2^2}{r_s}\Big(1-2\frac{(1-k_1^2 k_2^2)}{(1+k_1^2k_2^2)}\Big) B_2 \sin2(\theta_s-c_1)\Bigg].
        \end{align}
        \label{eqn:exact_exprs}
        \end{subequations}
       
        \noindent Here, $V_{\parallel}$ and $V_{\perp}$ are the components of velocity along and perpendicular to the separation vector respectively. $\Omega_z$ is the angular velocity of the squirmer. (Note that, the expression (Eq.~16) given in \cite{dario_gareth} has a typo and that is corrected as per \cite{Dario_thesis}.)  $c_1$ is a constant, and the variables $k_1$ and $k_2$ are functions of separation distance, radius of the squirmer, and radius of curvature of the boundary as follows:
        \begin{enumerate}
        \item \textbf{Convex curvature ($\alpha < 0$):}
        \begin{equation}
          c_1 = \pi; \quad k_{1,2} =\frac{d^2+r_{b,s}^2-r_{s,b}^2-\sqrt{m}}{2dr_{b,s}},
        \end{equation}
        where, $m = d^4+r_b^4+r_s^4-2(d^2 r_b^2+d^2 r_s^2+r_b^2r_s^2)$. Here, $k_1<1$ and $0\leq k_2 \leq 1$.
        \item \textbf{Concave curvature ($\alpha > 0$):}
        \begin{equation}
           c_1 = 0; \quad k_{1,2} =  \pm \frac{d^2+r_{b,s}^2-r_{s,b}^2+\sqrt{m}}{2dr_{b,s}}.
        \end{equation}
          Here, $k_1\geq1$ and $0\leq k_2 \leq 1$.
        \end{enumerate}
       
        In the limit of $\alpha \to 0$, the above expressions for  $k_{1,2}$ of both convex and concave curvatures reduce to same form,
        \begin{align}
            k_1 = 1, \quad
            k_2 = \frac{d_s+r_s-\sqrt{d_s^2+2 d_s r_s}}{r_s}.
        \end{align}
        This is expected as $\alpha = 0$ corresponds to that of a flat wall, whether approached from the expressions for a concave or convex boundary. It may also be noted that the above limit is same as the solution derived by \citet{crowdywall} for a squirmer near a flat wall.
        
        Therefore, the equations of motion of the squirmer near any curved boundary are given by,
        \begin{equation}
            \dot{d}_s = V_{\parallel}, \ \dot{d}_{\perp} = V_{\perp}, \ \text{and} \  \dot{\theta_s} = \Omega_z,
            \label{eqn:newtons_eqn}
        \end{equation}
        where, the dot on the variable represents the time derivative, and $d_{\perp}$ is the coordinate defined perpendicular to $\hat{\bm{s}}$. These expressions were integrated using forward Euler method to construct the trajectory of a squirmer in the presence of a curved boundary.
        
        In addition to hydrodynamics, for the purpose of studying the role of repulsive interactions between the microswimmer and the boundary, a hard potential of the form,
        \begin{equation}
            V_{\parallel} = 0,\ \text{for}\ d_s \leq \delta r_s \\
        \label{eqn:repulsion}
        \end{equation}
        was also implemented on the solid boundary while constructing the squirmer trajectories. In the Eq.~\ref{eqn:repulsion}, $\delta r_s$ specifies the distance over which the repulsive potential acts. It may be noted that the hard sphere potential only modifies the component of velocity parallel ($V_{\parallel}$) to the separation vector ($\hat{\bm{s}}$) while the perpendicular component and the angular velocity remain unmodified.
        
        Eq.~\ref{eqn:exact_exprs} that govern the dynamics of the squirmer were derived by solving Stokes' equations in a  complex plane and then using the reciprocal theorem by \citet{dario_gareth}. They are derived in a context to show that the reciprocal theorem is a powerful mathematical tool to analyze the low Reynolds number hydrodynamics problems. Hence, the equations or their consequences are not  analyzed in detail in the literature. 
        Since the above expressions were obtained using reciprocal theorem, the exact flow fields associated with these solutions are not known analytically, which makes it difficult to explain the observations. Therefore, we employ a complementary approach wherein lattice Boltzmann method based numerical simulations are used to construct the flow field generated by the squirmer in the presence of a curved boundary.

        \subsection{Numerical method}
        \label{sec:LBM}
        Lattice Boltzmann Method (LBM) \cite{succi} is a numerical method to solve the fluid flow problems. Unlike the conventional numerical techniques like finite difference, finite volume etc.,\cite{cfd}, where the governing equation is directly discretized and solved, in LBM the Boltzmann equation is discretized \cite{Timms_LBM_book}. Hence, it is often referred to as a mesoscopic technique. LBM is widely popular to simulate fluid flows containing particles of different shapes \cite{diff_particles_lbm} and in complex confinements \cite{confinement_LBM}. Below, we give a brief outline of the algorithm, and the method of coupling the squirmer dynamics with the surrounding fluid flow dynamics.
        
        Fluid motion is resolved on a Cartesian mesh by solving the finite difference discretized, BGK approximated Boltzmann equation.  
        Spatial and temporal resolutions are denoted as $\Delta x$ and $\Delta t$ respectively.
        The two main steps in the solution procedure are collision and streaming \cite{Timms_LBM_book,succi}, which are given as,
        \begin{align}
            &f_{i}^{*}(\mathbf{x},t) = f_{i}(\mathbf{x},t)-\frac{\Delta t}{\tau}\Big(f_{i}(\mathbf{x},t)-f_{i}^{eq}(\mathbf{x},t)\Big),\label{eq:lb_collsion}\\
            &f_{i}(\mathbf{x}+\mathbf{e}_i\Delta t,t+\Delta t) = f_{i}^{*}(\mathbf{x},t).
        \end{align}
        Here $f_{i}(\mathbf{x},t)$ represents the discrete distribution function in the direction of the lattice velocity vector $\mathbf{e}_{i}$ at position $\mathbf{x}$ and time $t$.  $f_{i}^{*}$ is the post collision discrete distribution function, and $f_{i}^{eq}$ is the equilibrium  distribution function which is given as,
        \begin{equation}
            f_{i}^{eq} = \rho w_{i}\Big[1+3\frac{\mathbf{u}\cdot\mathbf{e}_{i}}{c_s^2}-\frac{3}{2}\frac{\mathbf{u}\cdot\mathbf{u}}{c_s^2}+\frac{3}{2}\frac{\mathbf{(u}\cdot\mathbf{e}_{i})^2}{c_s^4}\Big],
        \end{equation}
        where $w_{i}$ is the weight factor, $c_s$ is the speed of the sound, $\rho$ is the fluid density, and $\mathbf{u}$ is the fluid velocity. Lattice velocities $\mathbf{e}_{i}$ and weight factors $w_i$ depend on the particular lattice model used. In this work, we use the $D_{2}Q_{9}$ model \cite{LBM_D2Q9} which has nine velocity directions ($i = 0$ to $8$) at a given lattice point. The relaxation time ($\tau$) in Eq.~(\ref{eq:lb_collsion}) is related to the kinematic viscosity $\nu$ of the fluid via the relationship $\nu = c_s^{2} (\tau-\frac{\Delta t}{2})$ \cite{Timms_LBM_book}.  The macroscopic variables can be calculated from the solution as zeroth and first moments of the discrete distribution function $i.e.,$ $\rho = \sum_{i} f_{i}$ and $\rho \mathbf{u} = \sum_{i}f_{i}\mathbf{e}_{i}$.

        \begin{figure}
        	\centering
        		\includegraphics[height=7cm]{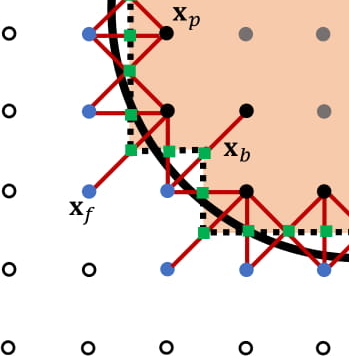}
        	\caption{The curved boundary of the solid particle (continuous black line) is approximated with a stair case construction (dotted line) in the simulations. There are four types of nodes: (i) $\mathbf{x}_p$ - solid nodes that interact with the fluid nodes (solid black circles), (ii) $\mathbf{x}_f$ - the fluid nodes that interact with the solid nodes (blue filled circles), (iii) solid nodes that don't interact with the fluid nodes (grey filled circles), and (iv) fluid nodes that don't interact with the solid nodes (hollow circles). Lattice Boltzmann populations moving along the links (brown lines) from $\mathbf{x}_f$ to $\mathbf{x}_p$ (or from $\mathbf{x}_p$ to $\mathbf{x}_f$) are bounced back at boundary nodes $\mathbf{x}_b$ (green squares).
        	 }	\label{fig:bb_schematic}
        \end{figure}

        \noindent\textbf{Implementation of boundary conditions - bounce back scheme:} In order to impose the boundary conditions on the surface of the solid (squirmer and the curved boundary) the discrete distribution function streaming in the direction of the boundary nodes are made to bounce back in the opposite direction with a modification based on the boundary node velocity \cite{ladd1,ladd2}. In this work, we use a mid grid bounce back scheme, \textit{i.e.}, boundary nodes ($\mathbf{x}_b$) are located half way between the nodes inside the solid surface ($\mathbf{x}_p$) and the external fluid nodes ($\mathbf{x}_f$) as shown in Fig.~\ref{fig:bb_schematic}. Then the bounce back scheme can be implemented as, 
         \begin{equation}
          f_{i^{-}} = f_{i}^{*}-2\rho_s w_{i}\frac{\mathbf{u}_{b}\cdot\mathbf{e}_{i}}{c_s^2},
          \label{eq:bounceback}
          \end{equation} 
         where  $\rho_s$ is the density of the solid particle and $\mathbf{u}_{b}$ is the velocity of the boundary node. For the fixed boundary, $\mathbf{u}_b = 0$, and for the squirmer, 
         \begin{equation}
          \mathbf{u}_{b} = \mathbf{u}^s+\mathbf{V}+\boldsymbol{\Omega}_z \wedge (\mathbf{x}_b-\mathbf{x}_s),
          \label{eqn:boundarynodevel}
          \end{equation}
          where $\mathbf{V}$ and $\boldsymbol{\Omega}_z$ are the translational and rotational velocities of the squirmer respectively, and $\mathbf{x}_s$ is the centre of mass of the squirmer.  \\
          
        \noindent\textbf{Squirmer dynamics:}
          Squirmer motion is governed by the Newton's second law of motion. The total momentum exchange ($\Delta \mathbf{P}$) and the total angular momentum exchange ($\Delta \mathbf{L}$) between the fluid and the squirmer during one time step are given by,
          \begin{align}
              \Delta \mathbf{P} &= \Delta x^3 \sum_{\mathbf{x}_b,\ i}\big(f_{i}^{*}(\mathbf{x}_f,t)+f_{i^{-}}(\mathbf{x}_f,t+\Delta t)\big)\mathbf{e}_{i},\\
            \Delta \mathbf{L} &= \Delta x^3 \sum_{\mathbf{x}_b, \ i}\big(f_{i}^{*}(\mathbf{x}_f,t)+f_{i^{-}}(\mathbf{x}_f,t+\Delta t)\big)(\mathbf{x}_b-\mathbf{x}_s) \wedge \mathbf{e}_{i},
          \end{align}
        where the sum runs over all the boundary links pointing from fluid nodes ($\mathbf{x}_f$) to solid nodes ($\mathbf{x}_p$). 
          
         The equations of motion that govern the squirmer dynamics are,
          \begin{align}
              \frac{d\mathbf{V}}{dt} &= \frac{\Delta\mathbf{P}}{M},      \label{eqn:force}\\
                \frac{d\mathbf{\Omega_z}}{dt} &= \frac{\Delta \mathbf{L}}{I},
              \label{eqn:torque}
          \end{align}
        where $M =  \rho_s \pi r_s^2$ is the mass, $I = \frac{1}{2} M r_s^2$ is the moment of inertia.
        
        Eq.~(\ref{eqn:force})--(\ref{eqn:torque}) are numerically integrated to track the dynamics of the squirmer using a modified implicit scheme. While an explicit scheme (\textit{e.g.,} forward Euler method) which calculates the velocities at time $t$ using velocities and forces at time $t-\Delta t$ is simpler to implement, it is known to be inaccurate, in particular for small angular velocities. Therefore, an implicit scheme was necessary to integrate Eq.~(\ref{eqn:force}) and (\ref{eqn:torque}).  However, no such schemes were available in the literature that could be directly used for an active particle. Hence, we suitably modified the implicit scheme that was originally proposed for a passive particle in the literature \cite{Frenkel} to apply for the squirmer. The details of these modifications are given in the Appendix~\ref{appendix:appendixA}. Using this modified implicit scheme, the time evolution of the position and the orientation of the squirmer are determined.
        
        Thus, the momentum exchange between the fluid and the squirmer is completely taken into account which ensures a simultaneous and coupled evolution of the dynamics of the fluid flow and the squirmer.
        
        \subsection{Simulation details}
        
       In the simulations, spatial ($\Delta x$) and temporal ($\Delta t$) resolutions are chosen to be unity. The density ($\rho$) and viscosity ($\mu$) of the fluid are taken as $1$ and $1/6$  ($\tau = 1$) lattice units respectively. In this work, we consider a neutrally buoyant swimmer ($\rho_s = \rho$) of size $r_s = 30$ lattice units, and size of the boundary is varied according to $r_b = r_s/\alpha$ for the simulations with the concave boundary. Domain size of $20r_b \times 20r_b$ is used for the simulations with the convex boundary. Simulations are performed at a Reynolds number ($=\frac{r_s(B_n/2)\rho}{\mu}$) $\approx0.075$.
        
        The trajectory of the squirmer near a curved surface is constructed using the algorithm described in the section \ref{sec:sec2B}. 
         However in the results shown below, namely the instantaneous fluid velocity field, and the translational and angular velocities attained by the squirmer are determined by a simpler, quasi-steady approach \cite{Ahana}.
        Numerical method has also been validated by comparing the results with that from the analytical calculations, as shown in Fig.~\ref{fig:inst_velocities_ds}. In this figure, markers in each plot are obtained from the simulations for the case of $\alpha = 0.33$. A good agreement with the analytical solution can be observed in all cases, validating the proposed numerical scheme. \kvsc{The small deviations from the analytical solution is due to the staircase construction of curved surfaces (of both squirmer and the boundary) in the simulations, which suggests that accuracy of numerical simulations decreases with decrease in the separation distance between curved surfaces. }
        
          \section{\label{sec:results}Results and discussion}
        In this section, we first discuss the fluid flow around the squirmer when it is located near a wall having either convex or concave curvature. Following this, the instantaneous and then the long term dynamics of the microswimmer in response to the curved boundaries will be discussed. In most of the analysis that follows, the case of $B_1$ and $B_2$ modes are dealt separately to understand the effect of each mode. Since the linear Stokes' equation governs the dynamics of fluid flow which in turn determines the dynamics of the squirmer, the superposition of  individual solutions of $B_1$ and $B_2$ modes can be performed to obtain full solution and analyze the effect of activity $\beta$, as we do in the later sections. A squirmer with only $B_1$ mode and only $B_2$ mode are referred to as a neutral swimmer and a shaker respectively.
        
        \begin{figure*}
        \centering
            \subfigure[]{
          \includegraphics[height=3.1cm]{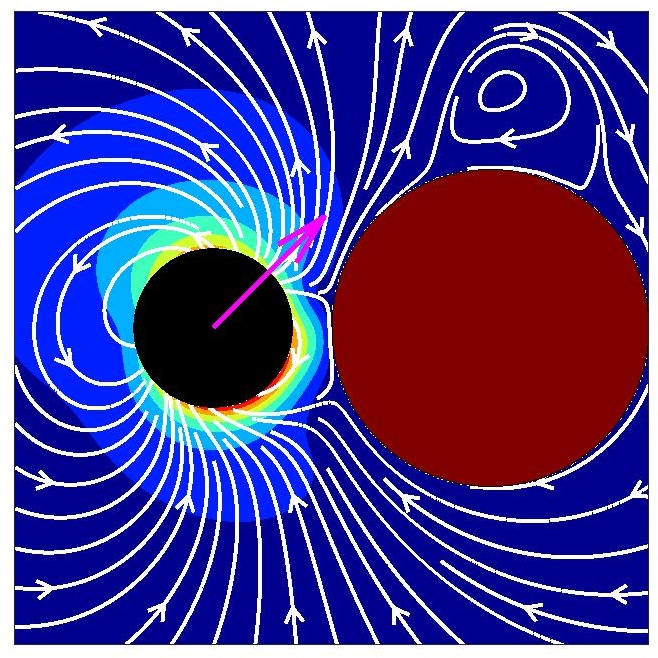}
              \label{fig:vex_B1_ff_0p5}}\quad
            \subfigure[]{
          \includegraphics[height=3.1cm]{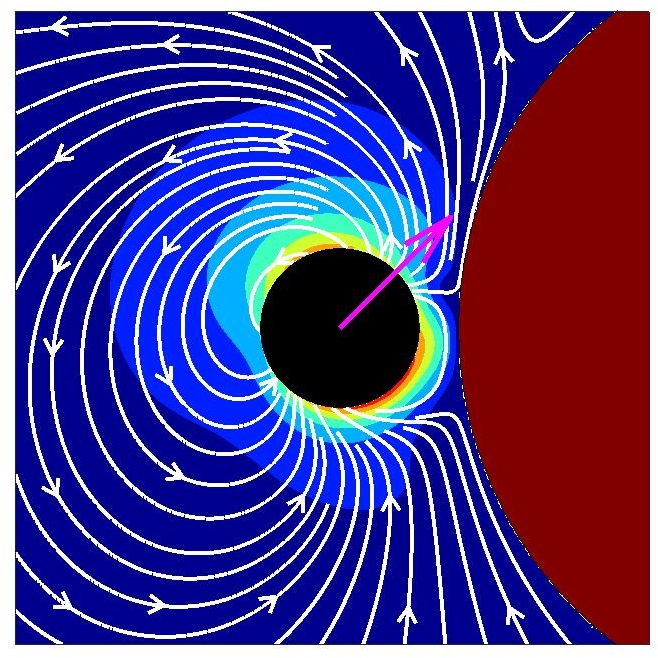}
              \label{fig:vex_B1_ff_0p2}}\quad
                  \subfigure[]{
          \includegraphics[height=3.1cm]{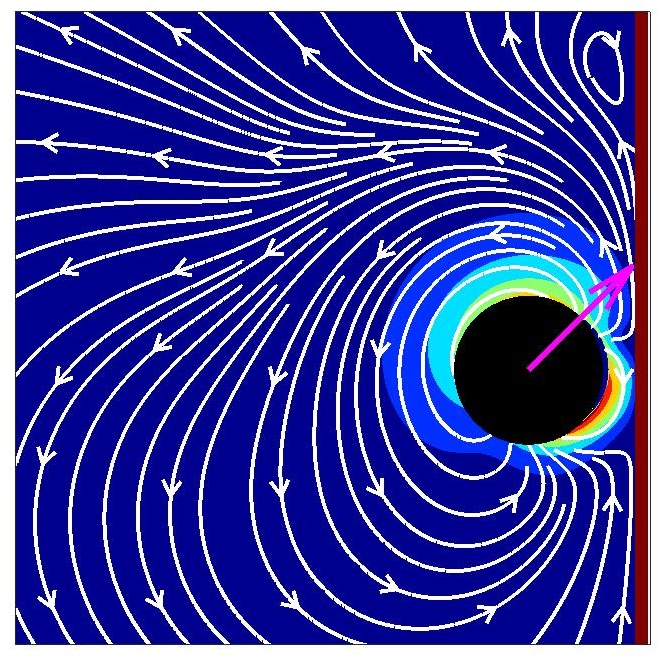}
          \label{fig:wall_B1_ff_0p5}}\quad 
                \subfigure[]{
          \includegraphics[height=3.1cm]{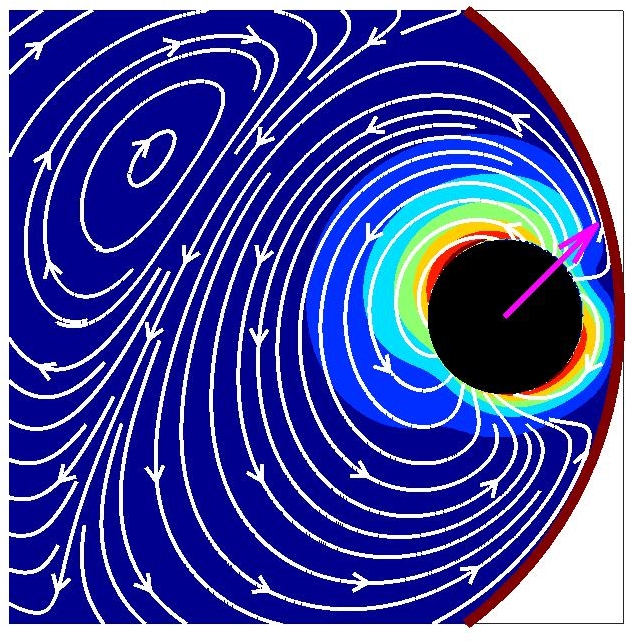}
          \label{fig:cave_B1_ff_0p2}}\quad
          \subfigure[]{
          \includegraphics[height=3.1cm]{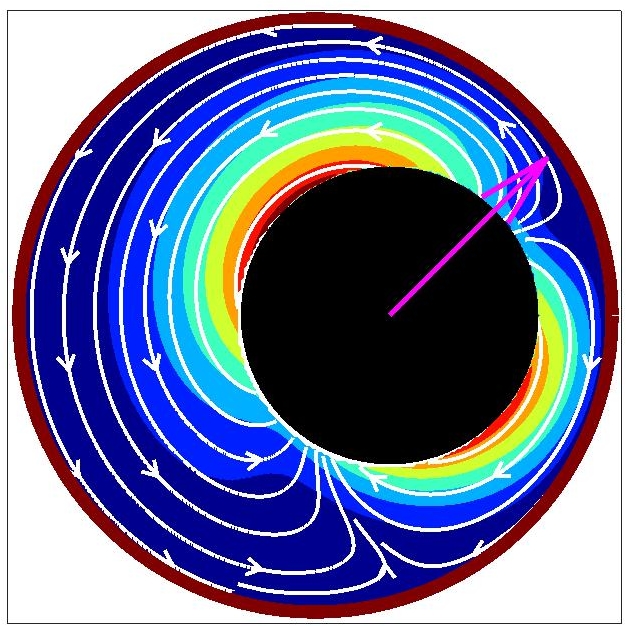}
          \label{fig:cave_B1_ff_0p5}}\quad 
        \subfigure[]{
          \includegraphics[height=3.1cm]{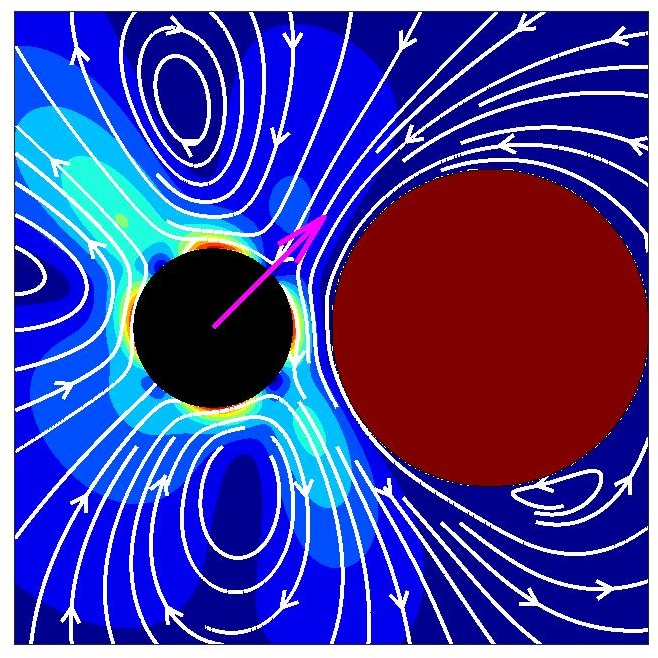}
          \label{fig:vex_B2_ff_0p5}}\quad  
        \subfigure[]{
          \includegraphics[height=3.1cm]{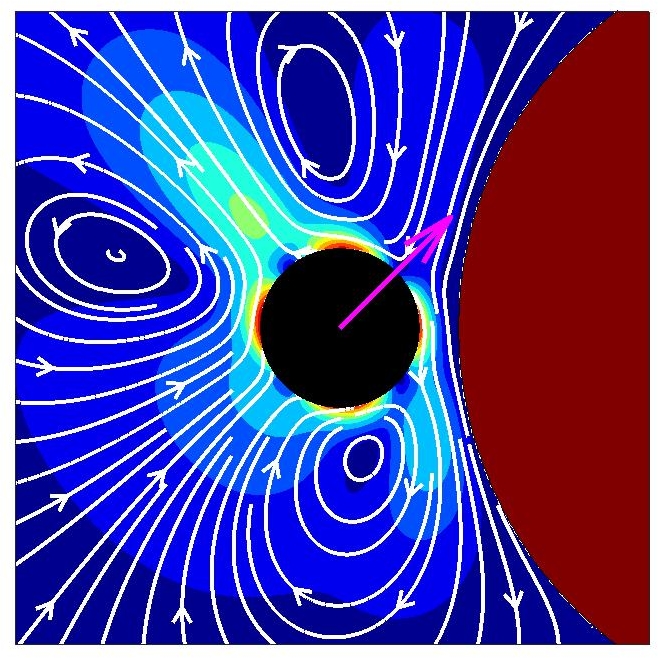}
          \label{fig:vex_B2_ff_0p2}}\quad 
                  \subfigure[]{
          \includegraphics[height=3.1cm]{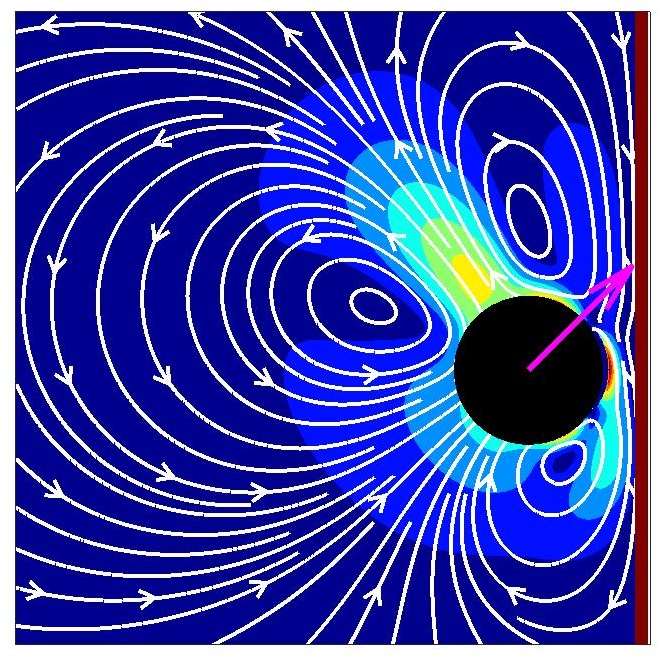}
          \label{fig:wall_B2_ff_0p5}}\quad 
          \subfigure[]{
          \includegraphics[height=3.1cm]{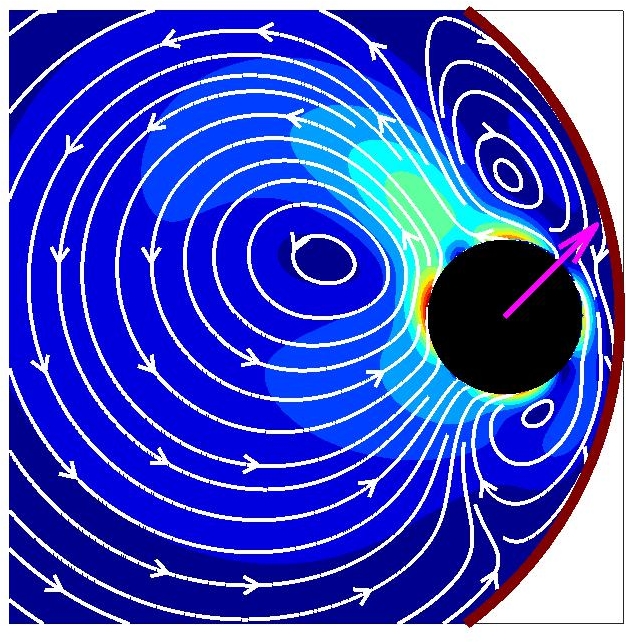}
          \label{fig:cave_B2_ff_0p2}}\quad
        \subfigure[]{
          \includegraphics[height=3.1cm]{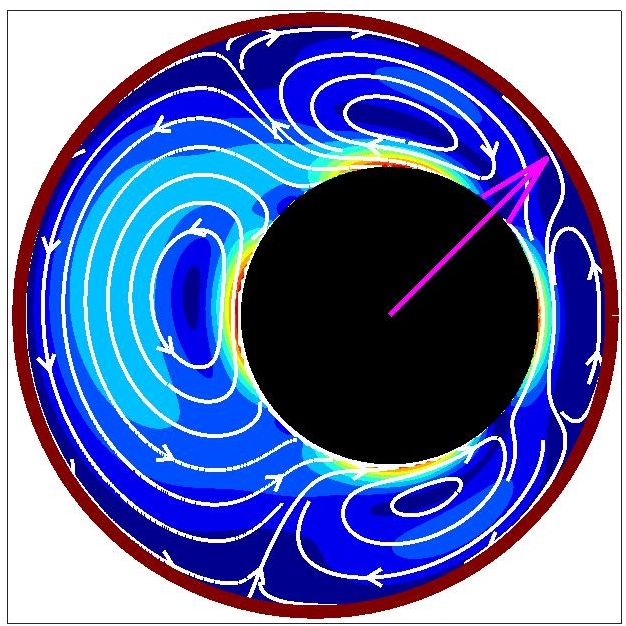}
          \label{fig:cave_B2_ff_0p5}}\quad
           \caption{Instantaneous velocity fields of the fluid around the squirmer are obtained from the lattice Boltzmann simulations for $\theta_s = 45^{\circ}$ and $d_s/r_s = 0.5$. The first row is for a neutral swimmer ($B_1>0$, $B_2 = 0$): (a) $\alpha = -0.5$, (b) $\alpha = -0.2$, (c) $\alpha = 0$, (d) $\alpha = 0.2$, and (e) $\alpha = 0.5$. The continuous lines are streamlines, and the background color corresponds to the magnitude of the normalized velocity field (red - highest, blue - lowest). The second row is for a shaker ($B_1=0$, $B_2 > 0$), and for the same $\alpha$ as in first row.
         } 
             \label{fig:inst_simu_velocity_fields}
        \end{figure*}

        \subsection{Fluid dynamics around the squirmer near a curved boundary}
        \label{sec:fluid_dynamics}
          
        The dynamics of a microswimmer in the neighbourhood of a curved wall is governed by four parameters namely, the curvature of the boundary ($\alpha$), the activity ($\beta$), and its position \big($x$, $y$\big) and orientation ($\theta_s$) with respect to the curved boundary. In this section, we analyze the velocity fields generated by the squirmer near the curved boundaries at a given location and orientation but as $\alpha$ varies.
        
        The steady state velocity fields of the fluid generated by a squirmer that is oriented at an angle with respect to the nearby curved boundary, obtained from lattice Boltzmann simulations, are shown in  Fig.~\ref{fig:inst_simu_velocity_fields}. The first row corresponds to neutral swimmers ($B_1>0$, $B_2 = 0$), and the second row corresponds to shakers ($B_1=0$, $B_2 > 0$). These flow fields are to be contrasted with those in the unbounded fluid domain shown in Fig.~\ref{fig:unbounded_vel_field}. For example, $B_1$ mode generates a velocity field corresponding to a source dipole in an unconfined domain (Fig.~\ref{fig:unbounded_B1_ff}). This field is altered near a convex boundary with streamlines on the source side (in this case) curving and going around the convex object as shown in the Fig.~\ref{fig:vex_B1_ff_0p5}--\ref{fig:vex_B1_ff_0p2}. As size of the convex boundary increases, the streamlines also deflect by a large extent to go around the convex object. In the limit of a flat wall, the streamlines cannot bend around the object any more (Fig.~\ref{fig:wall_B1_ff_0p5}). The streamlines bend in the opposite direction as $\alpha$ increases further, aligning with the concave boundary (Fig.~\ref{fig:cave_B1_ff_0p2}--\ref{fig:cave_B1_ff_0p5}). This change in the streamline patterns (compared to that of an unbounded squirmer) increases viscous dissipation in the system, and therefore, the dissipation can be expected to increase with increase in $\alpha$ for a given value of $B_1$. 
        
        \kvsc{Another consequence of the presence of a nearby boundary is the formation of eddies (circulating flow patterns) see for e.g., clockwise eddy ahead of the squirmer} next to the convex object in Fig.~\ref{fig:vex_B1_ff_0p5} and the clockwise eddy developed on the left side of the squirmer in concave confinement in Fig.~\ref{fig:cave_B1_ff_0p2}.
        These eddies are formed far from the squirmer surface. They are generally weak compared to the velocity of the squirmer, so they have less effect on the squirmer dynamics. As the boundary effects become stronger ($\alpha$ increases) these  eddies disappear, see for \textit{e.g.,} the clockwise eddy ahead of the squirmer in Fig.~\ref{fig:vex_B1_ff_0p5}--\ref{fig:wall_B1_ff_0p5} reduces its size as $\alpha$ increases. Similarly, with increase in $\alpha$ in concave confinements, the clockwise eddy on the left side of the squirmer in Fig.~\ref{fig:cave_B1_ff_0p2} vanishes and the streamlines are more aligned with the confining boundary as shown in Fig.~\ref{fig:cave_B1_ff_0p5}. It may also be noted that, for the same $|\alpha|$, the strength of the velocity field is weaker in the case of a concave boundary compared to a convex boundary. This can be expected as the fraction of the boundary to which the microswimmer is exposed increases with increase in $\alpha$ and is maximum in the case of a strong concave confinement ($\alpha \to 1$). This increased frictional resistance reduces the strength of the fluid velocity field.
        \par We now consider the case of a shaker, a squirmer with only $B_2$ mode, shown in the second row of  Fig.~\ref{fig:inst_simu_velocity_fields}. The velocity field generated by a $B_2$ mode in an unbounded fluid domain is that of a stresslet, and it is mirror symmetric as shown in Fig.~\ref{fig:unbounded_B2_ff}. However, this symmetry breaks down near a boundary as shown in Fig.~\ref{fig:vex_B2_ff_0p5}--\ref{fig:cave_B2_ff_0p5}. 
        Considering the first case of a smallest convex object (see Fig.~\ref{fig:vex_B2_ff_0p5}) it can be noticed that the streamlines curve and go around the convex object similar to the flow generated by the neutral swimmer. However, unlike the neutral swimmer, flow around the shaker on all other sides are also affected. A series of eddies appear around the shaker. The case shown in  Fig.~\ref{fig:vex_B2_ff_0p5} has two clockwise eddies closer to the squirmer, and one anticlockwise eddy slightly far from the squirmer and on the opposite side of the convex object. It is interesting to note that the occurrence of these eddies themselves (closed streamlines)  are in sharp contrast with the open streamlines found in the unconfined case. As the size of the convex object increases, the eddy opposite to the curved boundary gets larger compared to the other two (see Fig.~\ref{fig:vex_B2_ff_0p2}). This continues to be the case for a flat wall ($\alpha = 0$, Fig.~\ref{fig:wall_B2_ff_0p5}) and even for weak concave curvatures ($\alpha = 0.2$, Fig.~\ref{fig:cave_B2_ff_0p2}). However in the case of a strong concave confinement ($\alpha = 0.5$, Fig.~\ref{fig:cave_B2_ff_0p5}), another eddy sandwiched between the squirmer and the nearby concave boundary appears and the flow field becomes more symmetric. Thus, the asymmetry in the velocity field varies non-monotonically with the curvature, $i.e.,$ asymmetry being minimum both in the limit $\alpha \to -\infty$ and  in the limit $\alpha \to 1$.
      
          \subsection{Instantaneous dynamics}
          \label{sec:3B}
        
        In this section, we discuss the instantaneous dynamics of the squirmer which is located close to a convex or a concave boundary. The instantaneous translational and angular velocities are analyzed as a function of squirmer orientation and location  with respect to the neighbouring solid surface. The translational velocity is reported as its components parallel ($V_{\parallel}$) and perpendicular ($V_{\perp}$) to the separation vector pointing from the centre of the squirmer to the surface of the nearest neighbouring boundary ($\hat{\mathbf{s}}$). 
           
          \subsubsection{Effect of orientation of the squirmer with respect to the curved boundary}
          \label{sec:orio_inst_dynamics}
          
         \begin{figure*}
        \centering
          \subfigure[]{
          \includegraphics[height=5.0cm]{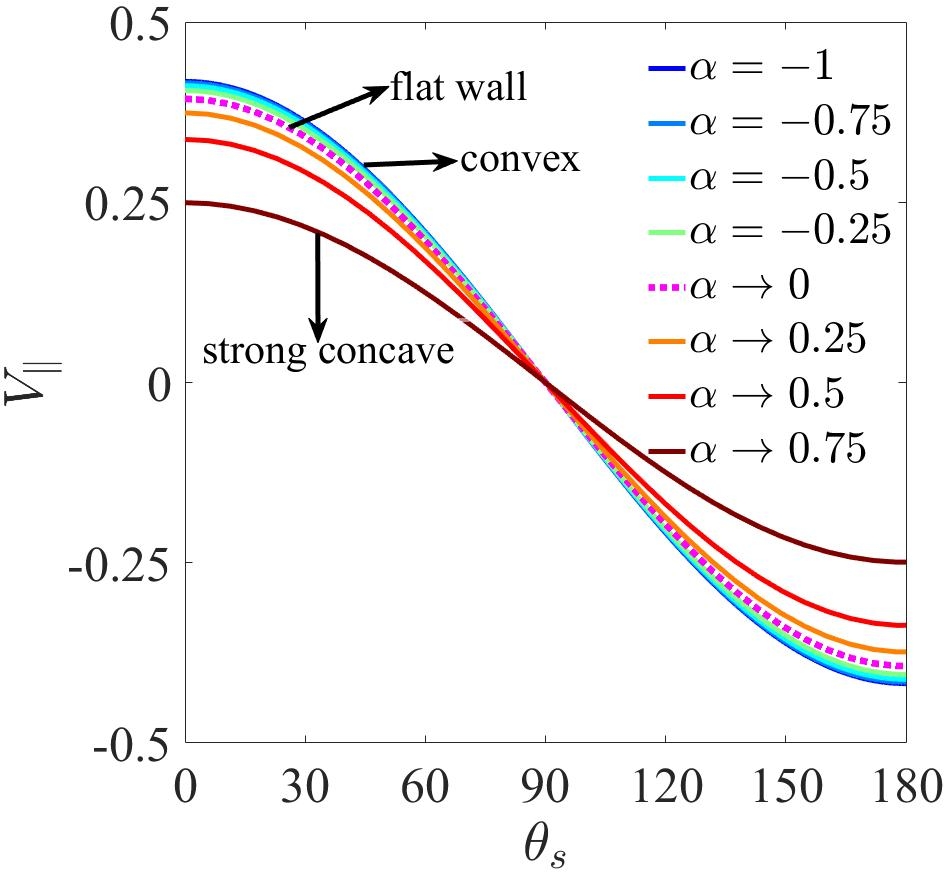}
              \label{fig:B1_mode_Vr}}\quad
        \subfigure[]{
          \includegraphics[height=5.0cm]{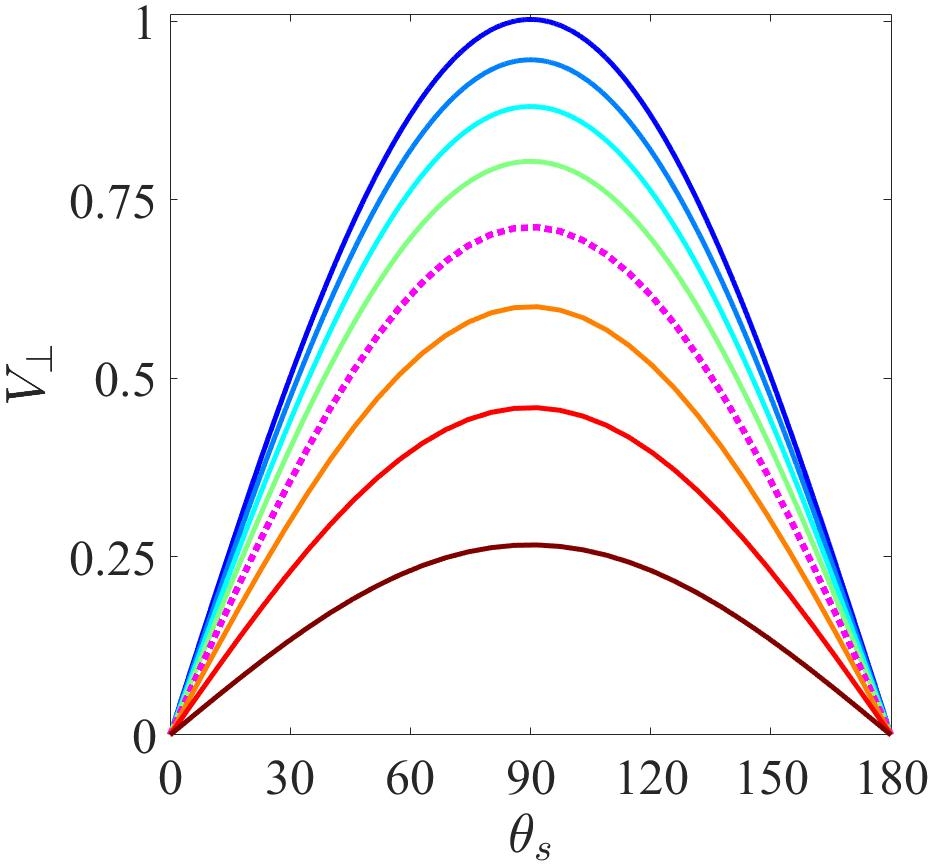}
          \label{fig:B1_mode_Vtheta}}\quad 
          \subfigure[]{
          \includegraphics[height=5.0cm]{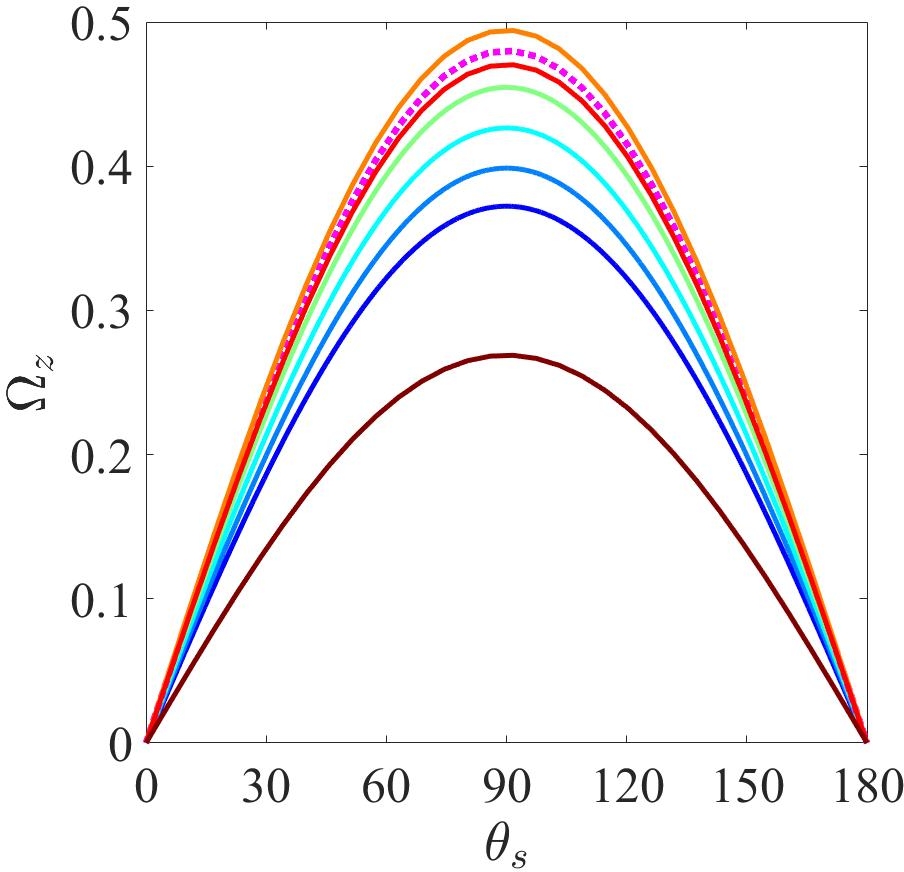}
          \label{fig:B1_mode_omega}}\\
        \subfigure[]{
          \includegraphics[height=5.0cm]{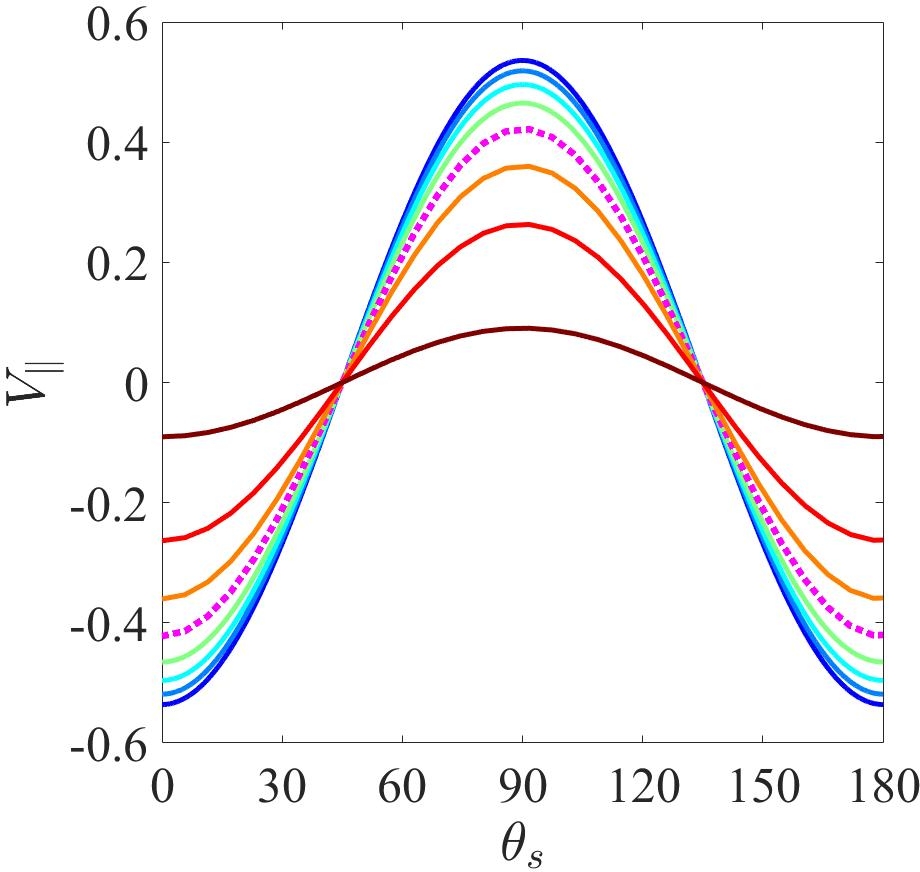}
          \label{fig:B2_mode_Vr}}\quad
        \subfigure[]{
          \includegraphics[height=5.0cm]{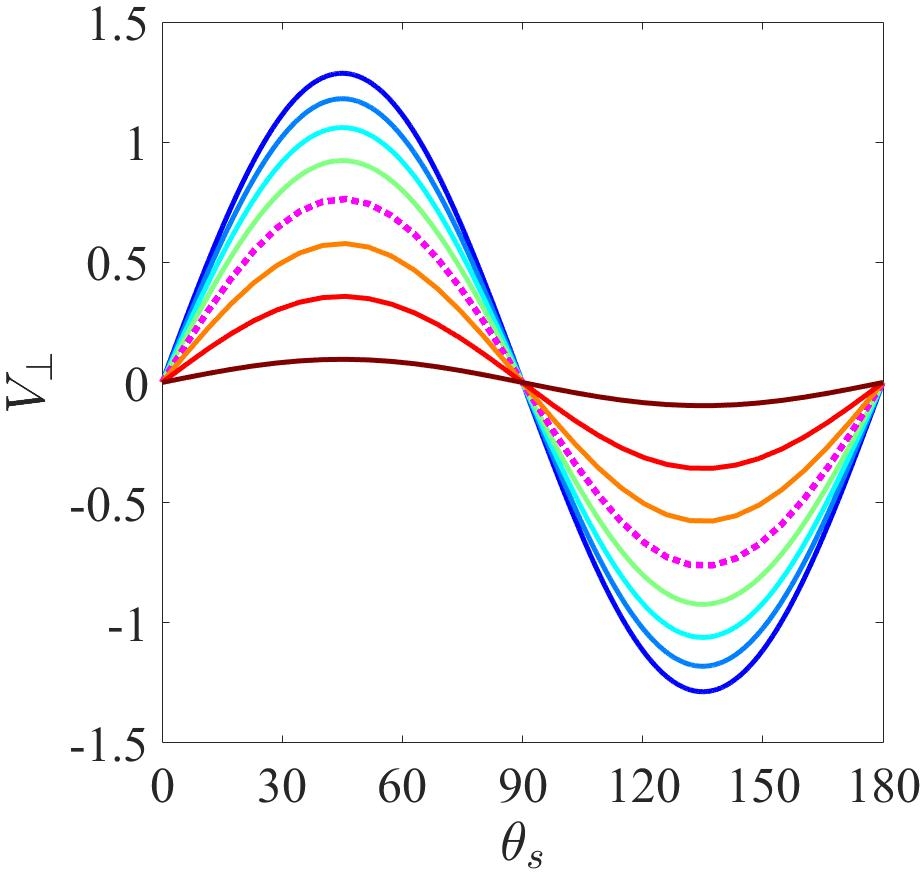}
          \label{fig:B2_mode_Vtheta}}
        \subfigure[]{
          \includegraphics[height=5.0cm]{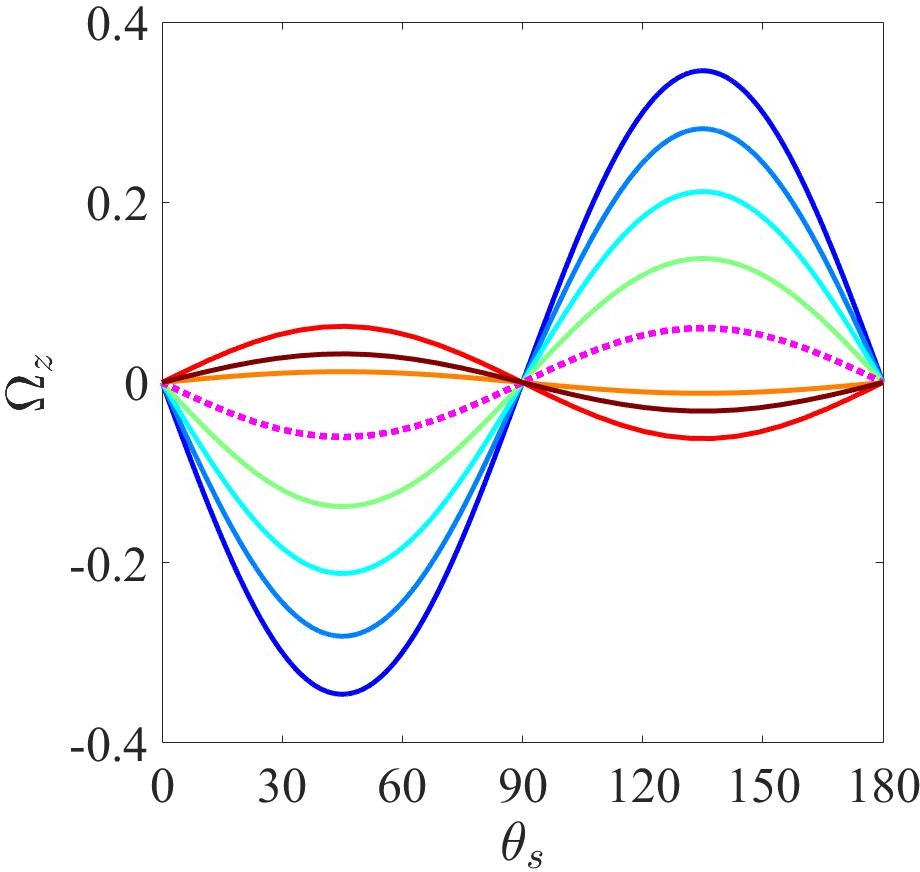}
        +  \label{fig:B2_mode_omega}}\quad
           \caption{Effect of curvature of a neighbouring boundary ($\alpha$)  on the instantaneous dynamics of a squirmer. (a) $V_{\parallel}$, (b) $V_{\perp}$ (the components of instantaneous translational velocity parallel and perpendicular to the separation vector), and (c) $\Omega_z$ (the angular velocity) of a neutral swimmer ($B_1>0$, $B_2 = 0$) are plotted as a function of squirmer orientation. (d)-(f) are similar plots for a shaker ($B_1=0$, $B_2 \neq 0$). In all cases, the surface to surface separation is maintained as $d_s = 0.2 r_s$. Legends of all plots are same as that given in (a).  $V_{\parallel}$ and $V_{\perp}$ are normalized by the $U_0$, and $\Omega_z$ is normalized by $r_s U_0$, where $U_0 = B_1/2$ for top row and $B_2/2$ for bottom row.} 
             \label{fig:inst_velocities}
        \end{figure*}
        
        Figure~\ref{fig:inst_velocities} shows the normalized components of instantaneous translational velocity ($V_{\parallel}$ and $V_{\perp}$), and the angular velocity ($\Omega_z$) of a squirmer as a function of its orientation ($\theta_s$): Fig.~\ref{fig:B1_mode_Vr}--\ref{fig:B1_mode_omega} correspond to a neutral swimmer ($B_1>0$, $B_2 = 0$), and  Fig.~\ref{fig:B2_mode_Vr}--\ref{fig:B2_mode_omega} correspond to a shaker ($B_1 = 0$, $B_2  \neq 0$). Different curves in each plot are for different values of $\alpha$.  It may be observed that, irrespective of the curvature, the magnitude of instantaneous velocities are symmetric about $\theta_s = 90^{\circ}$ (swimmer oriented parallel to the boundary) for both neutral swimmer and shaker. However, the sign (direction of velocity) depends on the squirmer orientation. 

        Figure~\ref{fig:B1_mode_Vr} shows that 
        $V_{\parallel}$ of the neutral swimmer continuously decreases with increase in $\theta_s$, reaches a value of zero before increasing further in the opposite direction, \textit{i.e.,} as the squirmer which is oriented towards the boundary ($\theta_s = 0^\circ$) turns and orients away from the boundary ($\theta_s = 180^\circ$). However, both perpendicular component $V_{\perp}$ and angular velocity $\Omega_z$ are maximum when the squirmer is orientated parallel to the boundary ($\theta_s = 90^{\circ}$) and they reduce as the neutral swimmer turns in either direction. Similarly, $V_{\parallel}$ is maximum for  a shaker when it is oriented parallel the boundary ($\theta_s = 90^{\circ}$) and reduces as the swimmer rotates in either direction. In contrast, the perpendicular component and the angular velocity are zero when the shaker is aligned parallel or perpendicular to the boundary but show a maximum at the intermediate angle $\theta_s = 45^{\circ}$. These results remain valid irrespective of the size and nature of curvature of the boundary.
      
        Now lets look into the effect of curvature of the neighbouring boundary on the instantaneous dynamics of the microswimmer. As shown in Fig.~\ref{fig:B1_mode_Vr}--\ref{fig:B1_mode_Vtheta} and Fig.~\ref{fig:B2_mode_Vr}--\ref{fig:B2_mode_Vtheta}, the translational velocity (both $V_{\parallel}$ and $V_{\perp}$) decreases with increase in the curvature. In other words, squirmer has a larger translational velocity near a convex surface compared to a concave surface. As the curvature of the neighbouring boundary changes from a highly convex ($\alpha \rightarrow -\infty$) to a highly concave ($\alpha \rightarrow 1)$ surface, squirmer (specified by a fixed $B_1$ and $B_2$) slows down. This slow down is due to the fact that the larger fraction of the surrounding fluid of the microswimmer is exposed to no-slip boundaries as the curvature of the neighbouring surface changes from convex to concave. The result is that more recirculating regions of the fluid appear as seen in the Fig.~\ref{fig:inst_simu_velocity_fields} and viscous dissipation in the fluid increases. This increase in frictional resistance slows down the microswimmer. The squirmer dynamics near a flat wall ($\alpha = 0$) is well studied in the literature \cite{crowdywall,Ahana}. Our study shows that, compared to a flat wall ($\alpha = 0$), a convex surface ($\alpha < 0 $) enhances the translational velocity of the squirmer but a concave surface ($\alpha > 0$) diminishes it irrespective of the orientation of the squirmer.
        
        However the angular (spin) velocity of the squirmer shows a non-monotonic variation with the change in curvature of the neighbouring boundary. And this variation is also different for $B_1$ and $B_2$ modes. As shown in Fig.~\ref{fig:B1_mode_omega},  the angular velocity of a neutral swimmer increases as the curvature of the boundary changes from convex ($\alpha < 0$) to concave ($\alpha > 0$) but this trend reverses beyond a critical $\alpha$. When $\alpha > 0.276$, further increase in $\alpha$ (or concavity of the wall) reduces the angular velocity of the neutral swimmer. This reduction may be due to the fact that, as $\alpha \to 1$ the squirmer occupies a symmetric position in the concave confinement. Thus, compared to a flat wall, both concave and convex boundaries reduce the angular velocity of the neutral swimmer except when for weak concave curvatures, namely $0 < \alpha < 0.276$ when the angular velocity is increased. Two features that are worth noticing are that: (i)  the boundary induced angular velocity always rotates the neutral swimmer away from the boundary,  (ii) the angular velocity of the squirmer does not go to zero as $\alpha \rightarrow -\infty$, \textit{i.e.,} in the limit of a point convex object because the point object  generates a Stokeslet disturbance velocity field, thus rotating the squirmer even in the limiting case of $\alpha \rightarrow -\infty$.
        
        A shaker shows a more subtle behaviour in terms of its angular velocity. As $\alpha$ increases from that corresponding to a convex curvature to that of a concave curvature, the magnitude of the angular velocity decreases. It becomes zero at $\alpha = 0.206$, a weak concave curvature. Further increase in the concavity of the wall increases the angular velocity of the shaker in the opposite direction, but this behaviour again breaks down beyond a critical $\alpha$. When $\alpha > 0.573$, further increase in the concavity of the boundary reduces the angular velocity of the shaker. This reduction is similar to that of a neutral swimmer where it attains a symmetric position in strong confinements and slows down significantly. Unlike neutral swimmer, the boundary induced direction of rotation of the shaker depends on both $\alpha$ and the orientation of the squirmer. Near a convex object, a shaker oriented towards the boundary rotates further towards it but one rotated away from the boundary rotates further away from it. Interpreting the results in another way - compared to that near a flat plate, a shaker near a convex object rotates faster, and near a concave object rotates slower. Moreover, concavity of the boundary may also change the direction of rotation of the shaker compared to that near a flat wall. The change in the sign of the angular velocity is due to the non-monotonic variation of asymmetry in the velocity field with the curvature (refer Fig.~\ref{fig:vex_B2_ff_0p5}--\ref{fig:cave_B1_ff_0p2}). 
        
        The instantaneous dynamics of a microswimmer can be calculated by a scaled sum of the two solutions (that of $B_1$ and $B_2$ modes) discussed above, and clearly, this behaviour can be complicated based on the individual contributions from each mode.
         \begin{figure*}
        \centering
          \subfigure[]{
          \includegraphics[height=5.0cm]{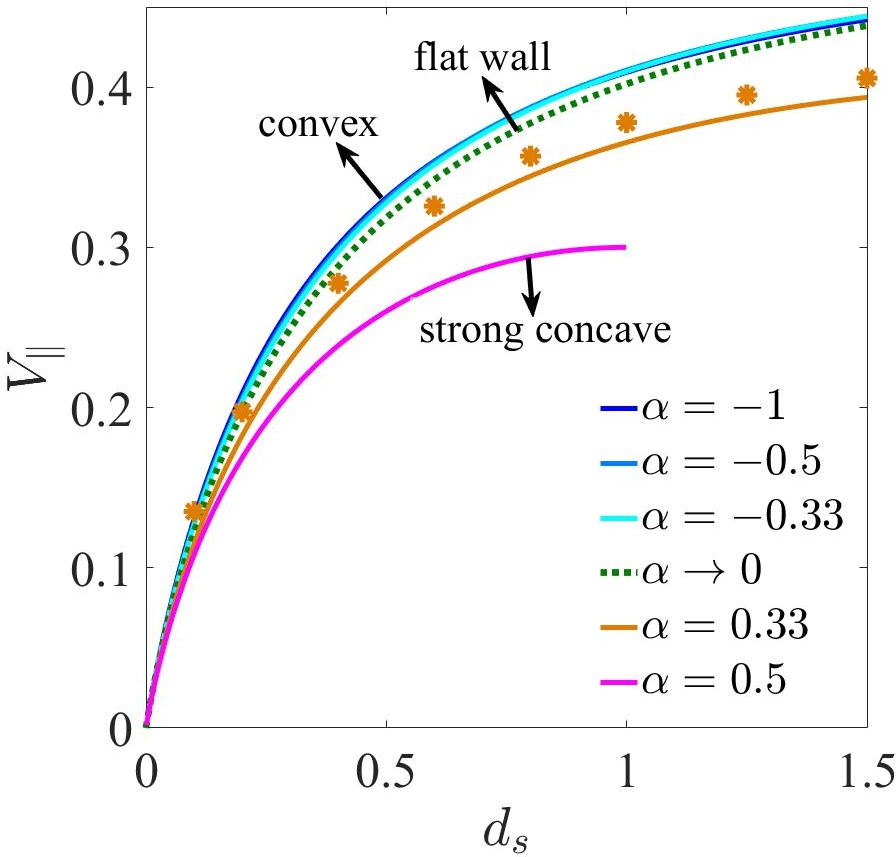}
              \label{fig:ds_B1_mode_Vr}}\quad
        \subfigure[]{
          \includegraphics[height=5.0cm]{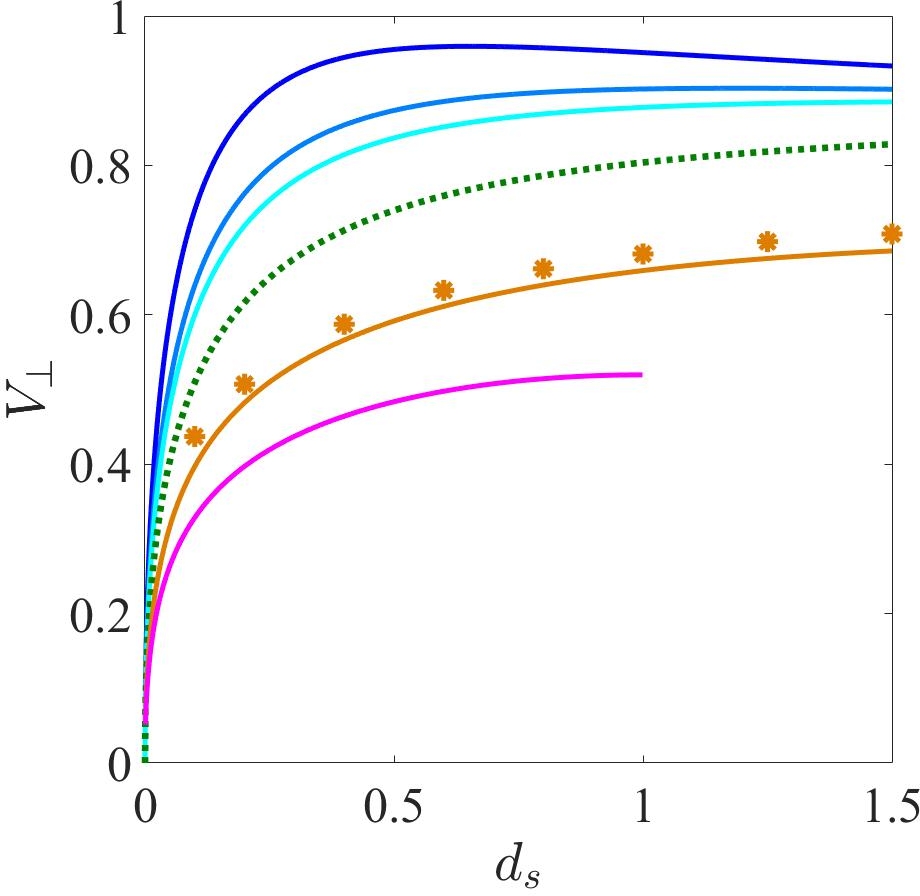}
          \label{fig:ds_B1_mode_Vtheta}}\quad 
          \subfigure[]{
          \includegraphics[height=5.0cm]{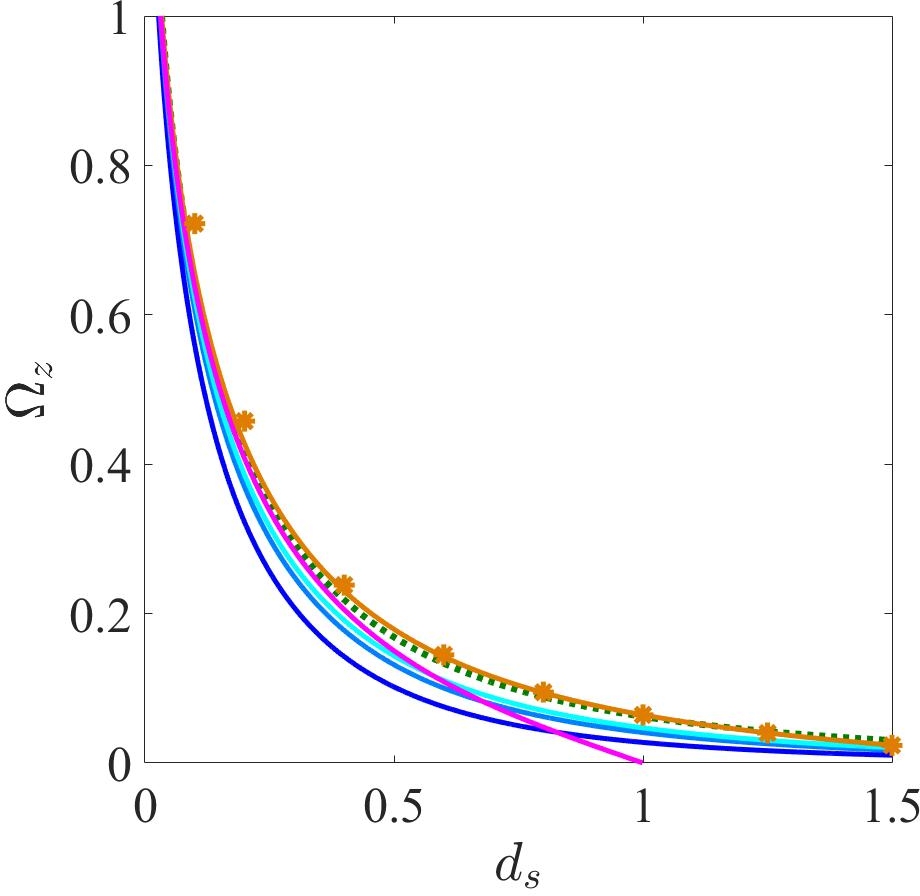}
          \label{fig:ds_B1_mode_omega}}\\
        \subfigure[]{
          \includegraphics[height=5.0cm]{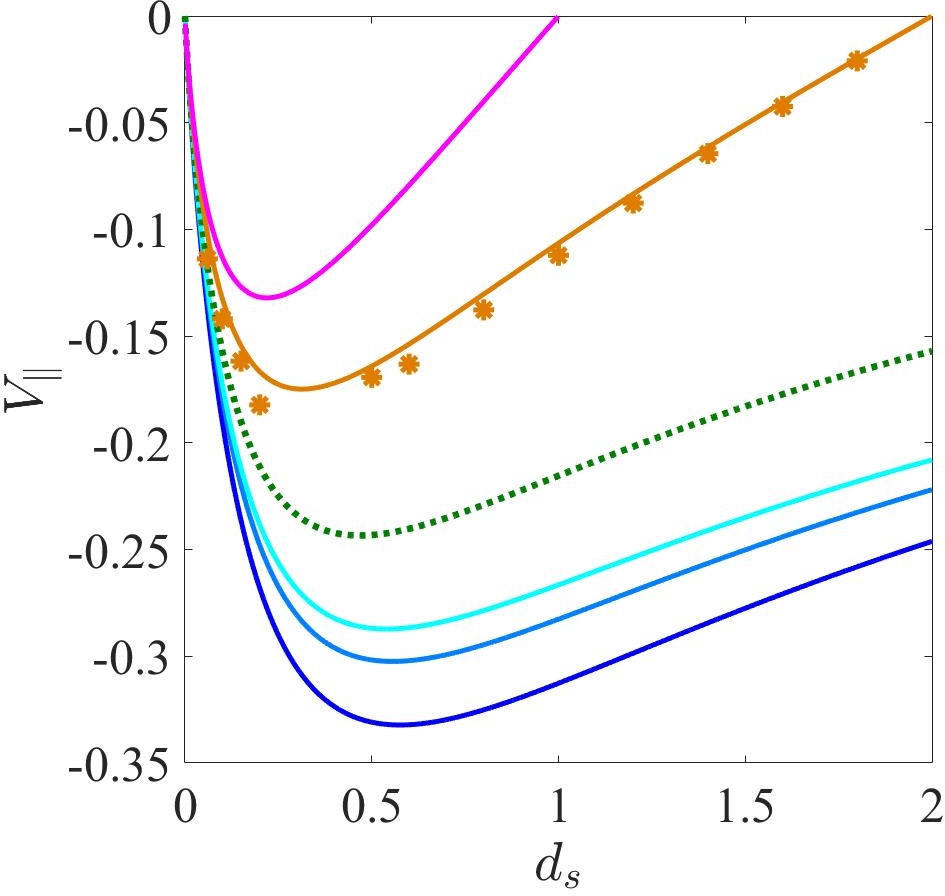}
          \label{fig:ds_B2_mode_Vr}}\quad
        \subfigure[]{
          \includegraphics[height=5.0cm]{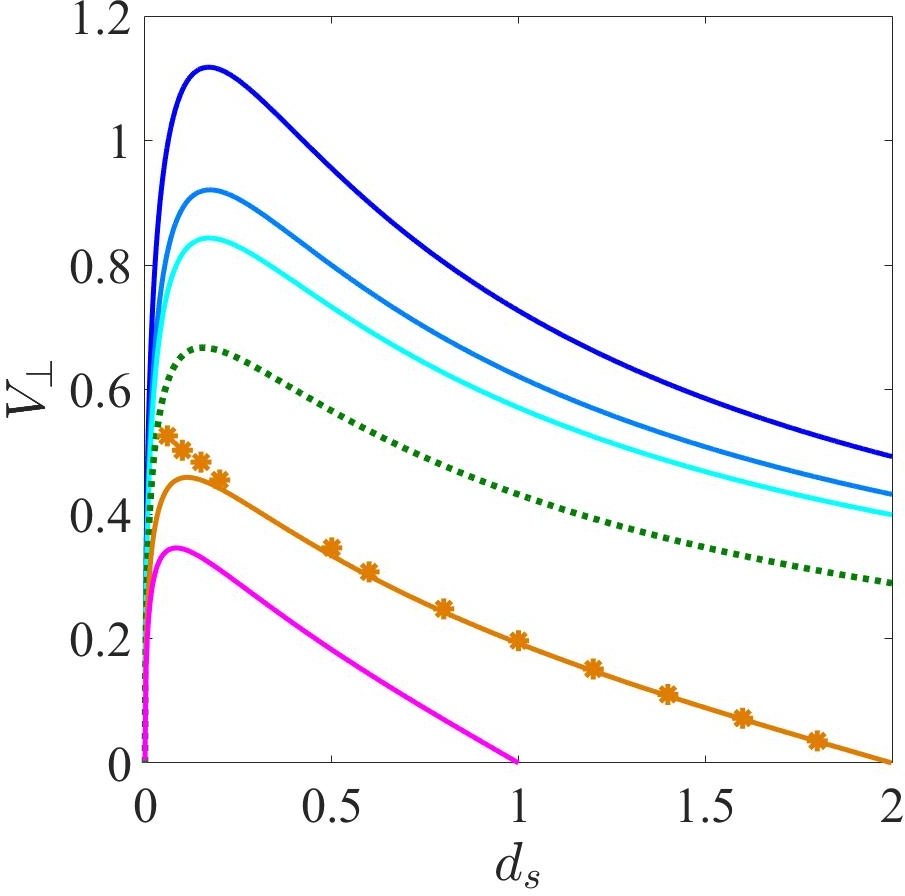}
          \label{fig:ds_B2_mode_Vtheta}}
        \subfigure[]{
          \includegraphics[height=5.0cm]{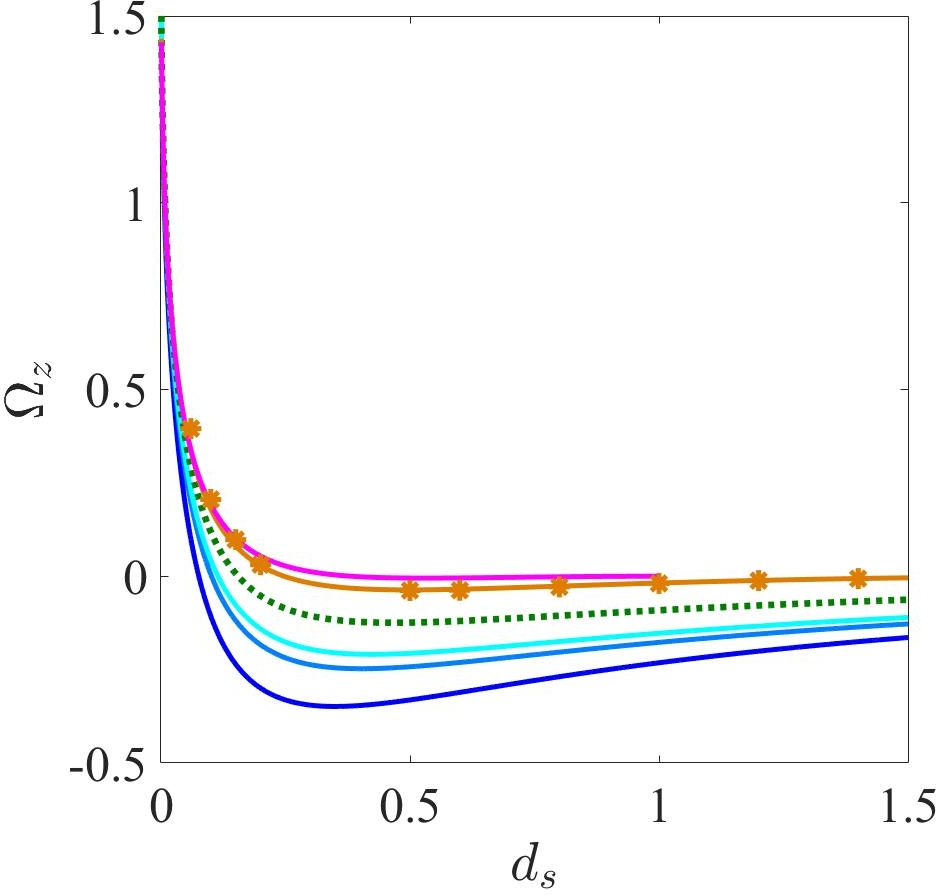}
        +  \label{fig:ds_B2_mode_omega}}\quad
           \caption{Effect of curvature of a neighbouring boundary ($\alpha$)  on the instantaneous dynamics of a squirmer. The components of instantaneous velocity (a) $V_\parallel$ and (b) $V_{\perp}$, and (c) the angular velocity $\Omega_z$ of a neutral swimmer are plotted as a function of surface to surface separation distance $d_s$. (d)--(f) are similar plots for a shaker.  In all cases, the orientation of the squirmer is chosen as $\theta_s = 60^{\circ}$. Legends of all plots are same and is given in (a). Both $V_\parallel$ and $V_{\perp}$ are normalized by $U_0$, and $\Omega_z$ is normalized by $r_s U_0$, where $U_0 = B_1/2$ for top row and $B_2/2$ for bottom row. The markers in each plot are obtained from lattice Boltzmann simulations for the case of $\alpha = 0.33$.} 
             \label{fig:inst_velocities_ds}
        \end{figure*}
        
        \subsubsection{Effect of location of the squirmer with respect to the curved boundary}
        
        In the subsection (\ref{sec:orio_inst_dynamics}), we analyzed the effect of orientation  on the instantaneous dynamics of the squirmer for a constant separation distance. Here, we analyze the effect of separation distance ($d_s$) on the squirmer dynamics when it is in the neighbourhood of a curved boundary.
        
        The instantaneous velocities $V_{\parallel}$, $V_{\perp}$, and $\Omega_z$ as a function of separation distance between the surface of the squirmer and the curved boundary ($d_s$) are shown in the Fig.~\ref{fig:inst_velocities_ds}. The minimum separation distance $d_s = 0$ corresponds to a physical contact between the squirmer and the boundary. For a concave boundary, the maximum separation distance is $d_s = r_b-r_s$, which corresponds to a squirmer located at the center of the circular confinement. There is no such maximum $d_s$ for a convex boundary. As earlier, the analysis for $B_1$ and $B_2$ modes are done separately: Fig.~\ref{fig:ds_B1_mode_Vr}--\ref{fig:ds_B1_mode_omega} correspond to a neutral swimmer, and Fig.~\ref{fig:ds_B2_mode_Vr}--\ref{fig:ds_B2_mode_omega} correspond to a shaker. In all cases, at $d_s = 0$ the translational velocities vanish while angular velocities exhibit a maximum value.
        
        Let us consider the case of a neutral swimmer first, shown in  Fig.~\ref{fig:ds_B1_mode_Vr}--\ref{fig:ds_B1_mode_Vtheta}. As $d_s$ increases both $V_{\parallel}$ and $V_{\perp}$ increase irrespective of whether it is near a convex or a concave boundary. However, for a given $d_s$ both $V_{\parallel}$ and $V_{\perp}$ decrease with increase in $\alpha$. In other words, the translational velocity of a neutral swimmer is smaller near a concave boundary compared to a convex boundary. This difference between convex and concave boundaries increases with increase in $d_s$. For large $d_s$, the translational velocity approaches that of a squirmer in the unbounded domain when it is located in the neighbourhood of a convex boundary, but it approaches that of a concentrically placed squirmer in a circular confinement exhibiting a smaller translational velocity when it is in the neighbourhood of a concave boundary.
        As shown in Fig.~\ref{fig:ds_B1_mode_omega}, the angular velocity of the squirmer decreases monotonically with separation distance. However the variation of angular velocity with $\alpha$ is non monotonic at a given $d_s$. The angular velocity of the squirmer slightly increases with increase in $\alpha$ (compare a convex wall with a flat wall), but as $\alpha\to 1$ (strongly concave boundary), it decreases again as squirmer occupies a symmetric position in the strong confinement.
        
        Unlike that of a neutral swimmer, the induced velocities on the shaker is a non-monotonic function of $d_s$. At $d_s = 0$, both $V_{\parallel}$ and $V_{\perp}$ are zero since the squirmer is in contact with the boundary. 
        On the other hand, at large $d_s$, a squirmer located near a convex boundary approaches the dynamics of an unconfined shaker which will have $V_{\parallel} = 0$ and $V_{\perp} = 0$. Similarly a squirmer in the neighbourhood of a concave surface, when $d_s$ is large, corresponds to a concentric squirmer in circular confinement with a symmetric flow field around it and therefore, $V_{\parallel} = 0$ and $V_{\perp} = 0$.
        Thus, the translational velocity of a shaker is zero in either limit (small and large $d_s$) and it exhibits a maximum velocity for some intermediate values of $d_s$ as shown in Fig.~\ref{fig:ds_B2_mode_Vr}--\ref{fig:ds_B2_mode_Vtheta}. Similar to $B_1$ mode, at a fixed value of $d_s$ the components of induced velocity $V_{\parallel}$ and $V_{\perp}$ also decrease with increase in $\alpha$. It may be noted that, as shown in Fig.~\ref{fig:ds_B2_mode_omega} the variation in the angular velocity of a shaker is different from its translational counterpart: $\Omega_z$ is maximum when $d_s = 0$ and it decreases to zero as $d_s$ increases. This trend itself is not monotonic. For a squirmer located near a convex wall the angular velocity decreases as $d_s$ increases, changes sign, reaches a maximum value before reducing further to reach zero. On the other hand, for a squirmer in the neighbourhood of a concave surface the reduction in angular velocity with $d_s$ is monotonic. Despite this complex behaviour the following two points may be noticed in Fig.~\ref{fig:ds_B2_mode_omega}: (i) when $d_s$ is large, at a given $d_s$ the induced angular velocity is larger due to a neighbouring convex boundary than a concave boundary, and (ii) when $d_s \to 0$ the trend reverses, and at a given $d_s$ the angular velocity due to a neighbouring concave boundary is more than that due to a convex boundary.
        
        To summarise this section, for a given $d_s$ the translational velocity of both neutral swimmer and shaker are always largest due to a convex boundary and smallest due to a concave boundary, and the case of a flat wall lies in between. Thus, microswimmers have larger velocity near convex boundaries compared to concave boundaries. The signs of $V_{\parallel}$ and $V_{\perp}$ depend upon the orientation of the squirmer and the sign of $B_2$ mode. No such general statements can be made with regard to the variations of angular velocity as its dependence on $d_s$ and $\alpha$ is more complex. This diversity in the variation of instantaneous velocities ($V_{\parallel}$, $V_{\perp}$, and $\Omega_z$) makes it difficult to predict the long time behavior of a microswimmer near a curved boundary. Therefore we now calculate the trajectory of a squirmer near curved boundaries.

        \subsection{Long time dynamics}
         \begin{figure*}
        \centering
          \subfigure[]{
          \includegraphics[height=3.4cm]{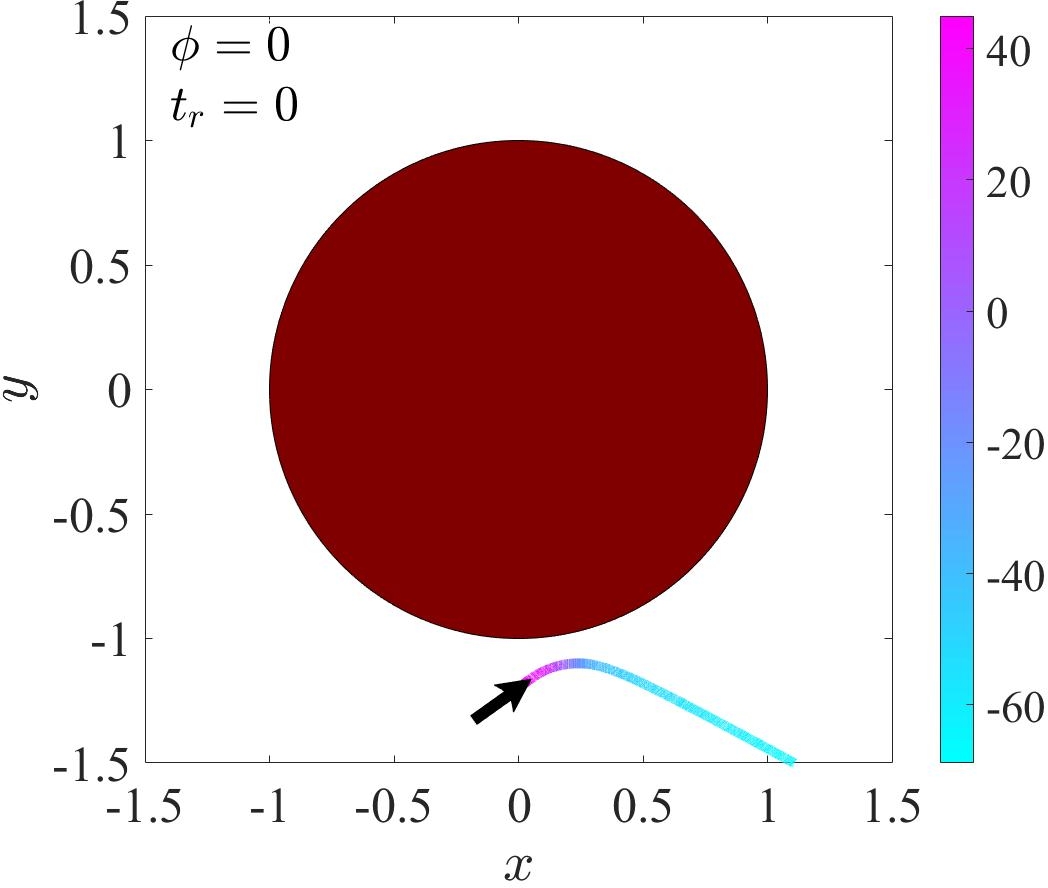}
              \label{fig:vex_zero_rep_neutral}}\quad
        \subfigure[]{
          \includegraphics[height=3.4cm]{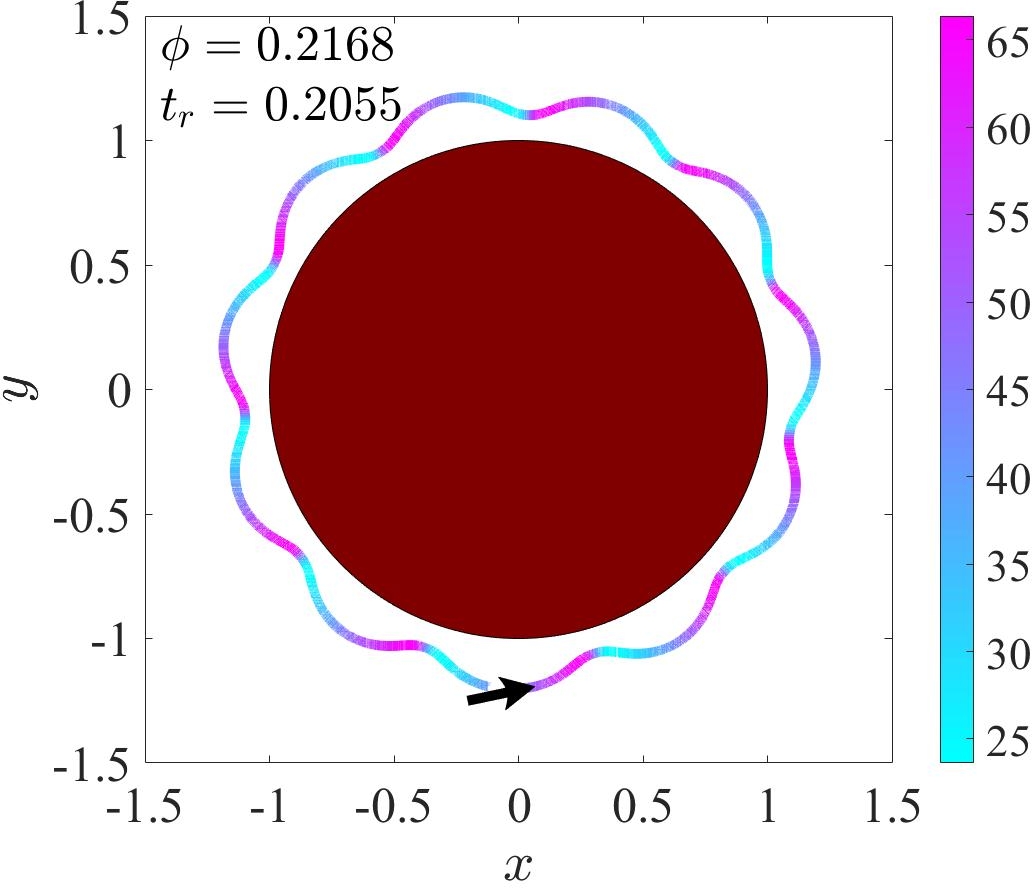}
          \label{fig:vex_zero_rep_shaker}}\quad 
          \subfigure[]{
          \includegraphics[height=3.4cm]{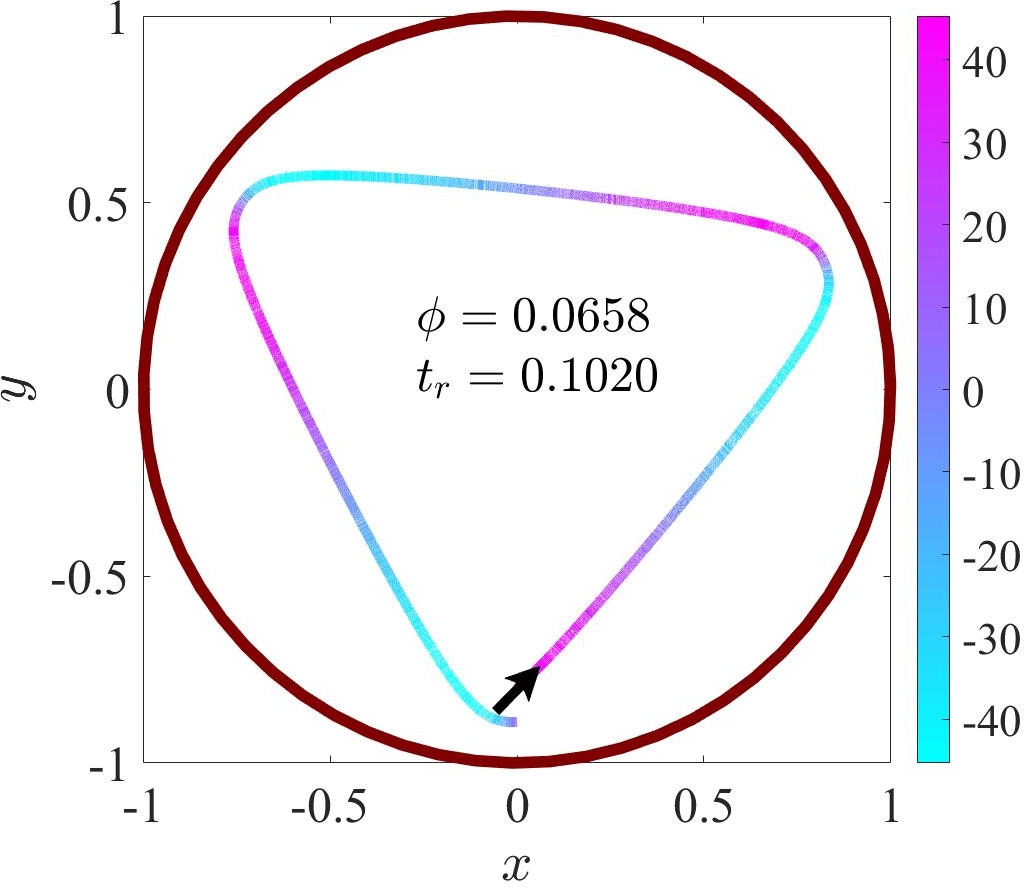}
          \label{fig:cave_zero_rep_neutral}}
        \subfigure[]{
          \includegraphics[height=3.4cm]{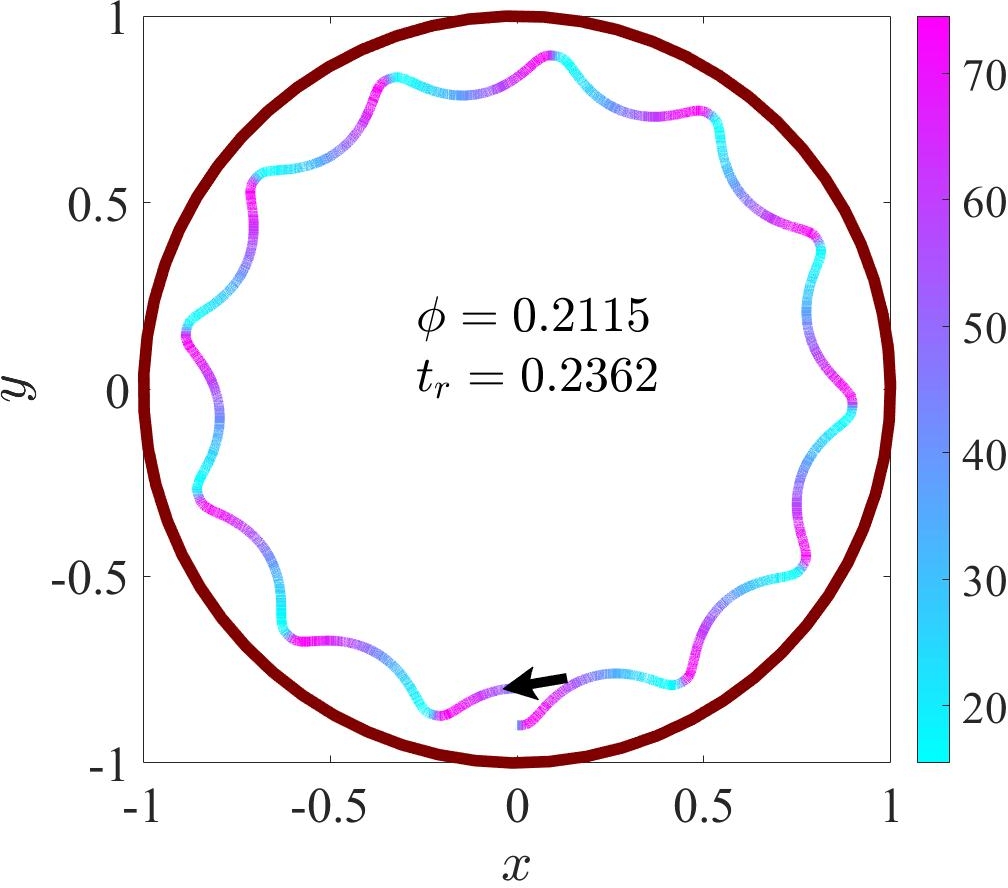}
          \label{fig:cave_zero_rep_shaker}}\quad
        \subfigure[]{
          \includegraphics[height=3.4cm]{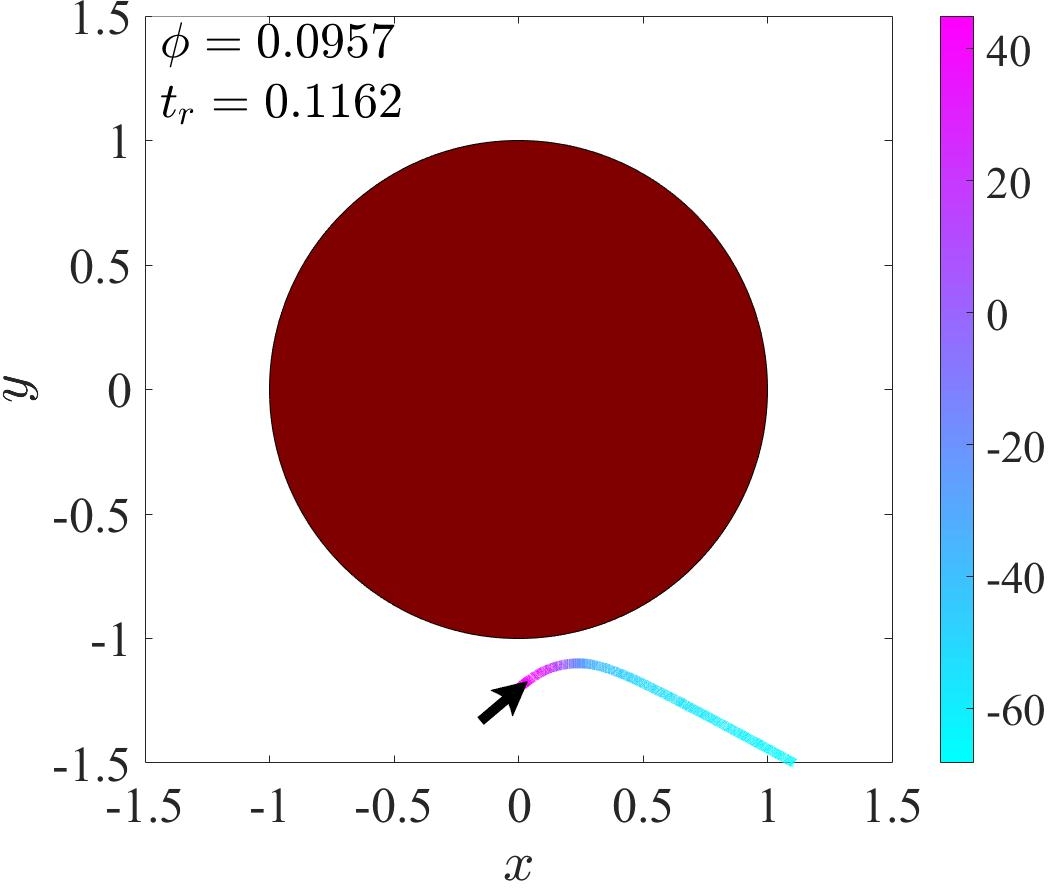}
          \label{fig:vex_0p05_rep_neutral}}
        \subfigure[]{
          \includegraphics[height=3.4cm]{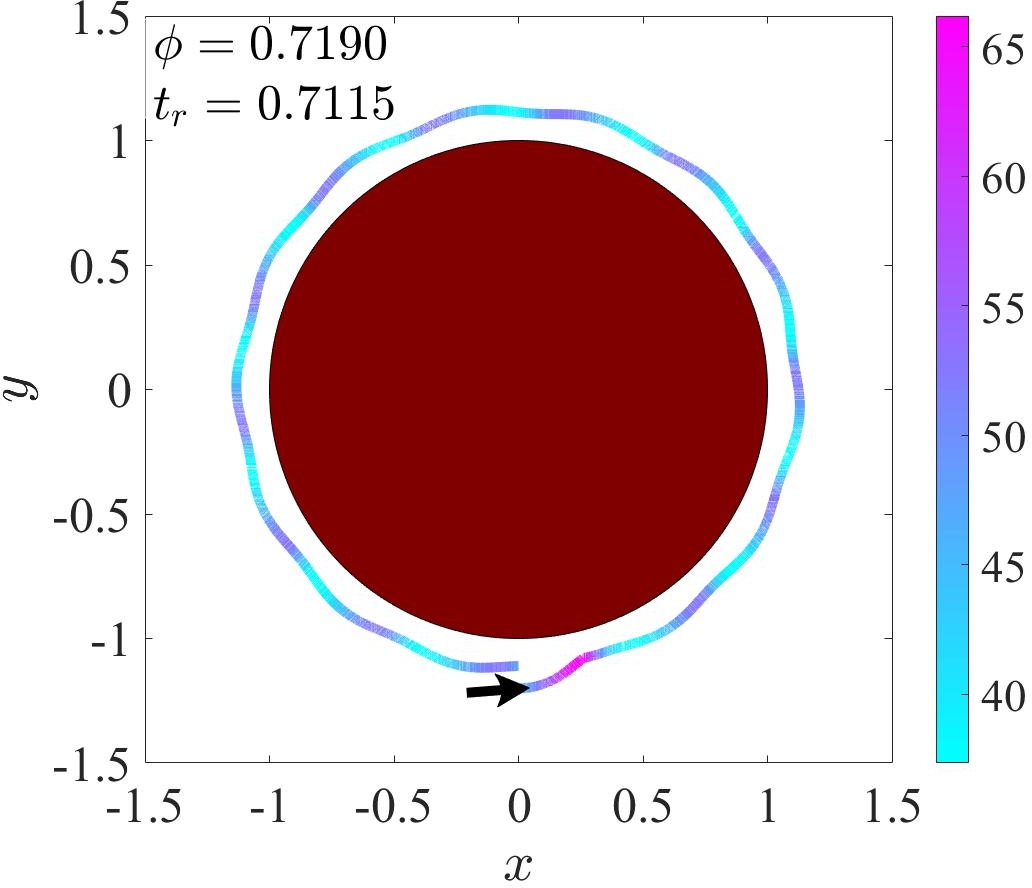}
        -  \label{fig:vex_0p05_rep_shaker}}\quad
        \subfigure[]{
          \includegraphics[height=3.4cm]{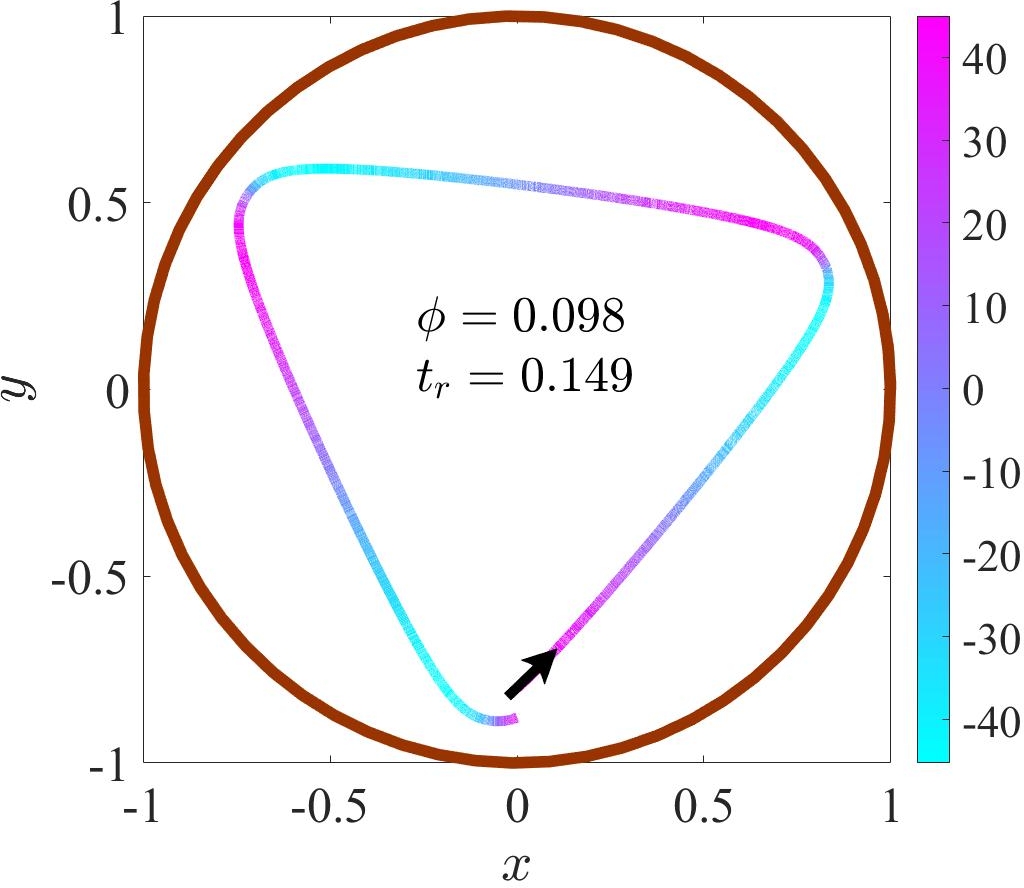}
          \label{fig:cave_0p05_rep_neutral}}
        \subfigure[]{
          \includegraphics[height=3.4cm]{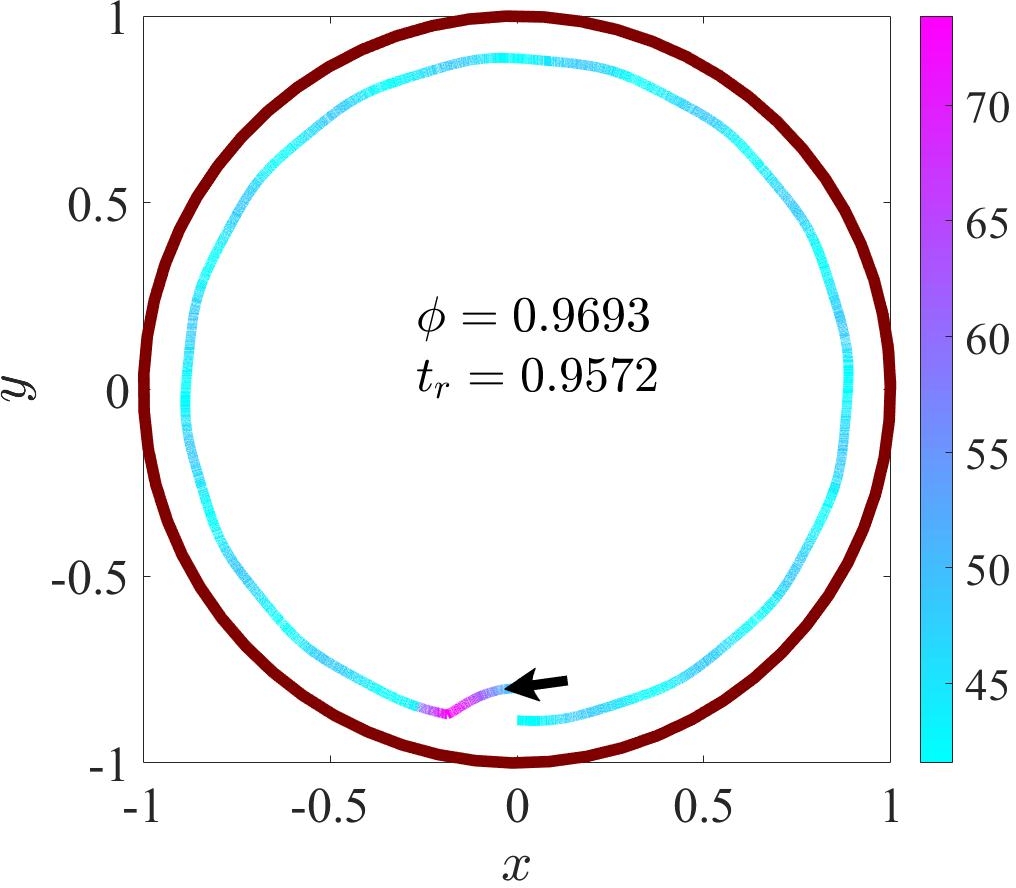}
          \label{fig:cave_0p05_rep_shaker}}\quad
           \caption{Trajectories of a squirmer near a curved boundary ($|\alpha| = 0.1$).   (a) A neutral swimmer and (b) a shaker near a convex boundary with an initial orientation $\theta_s^i$ $= 45^{\circ}$ (shown by the black arrow). (c) A neutral swimmer and (d) a shaker near a concave boundary with $\theta_s^i$ $= 135^{\circ}$. Here, ($x, y$) are normalized by $r_s$.
           Color-bar indicates the orientation of the squirmer with the surface tangent, $90^{\circ}-\theta_s$. (e)-(f) are altered trajectories corresponding to cases (a)-(d) but when the boundary has a repulsive potential. }
             \label{fig:trajectories}
        \end{figure*}
        The analysis on the instantaneous dynamics in the subsection \ref{sec:3B} showed that the dynamics of the microswimmer depends upon (i) its location and orientation with respect to the curved boundary, (ii) the size and nature of curvature of the boundary, and (iii) the strength and nature of the squirmer. In order to comprehensively understand the resulting dynamics due to these multiple factors and competing effects we now follow the squirmers over longer times as they come in contact with a curved surface and construct their trajectories \kvsc{in physical space ($x,y$) and in dynamical space ($\theta_s,d_s$)}. \kvsc{The trajectories are first analyzed and then different measures that may have experimental relevance are introduced to quantify the behaviour of microswimmers near curved surfaces.}
        
        \subsubsection{Trajectory of a squirmer near the curved boundary}
\noindent \kvsc{\textbf{Physical space}:} Figure~\ref{fig:trajectories} illustrates the typical trajectories of a squirmer \kvsc{in the vicinity of the curved boundary} when initialised at $d_s^{i} = r_s$. In all cases the trajectories are colored by $90^{\circ}-\theta_s$, the instantaneous angle that the squirmer makes with the tangent drawn on the boundary surface.  Figure~\ref{fig:vex_zero_rep_neutral} shows that a neutral swimmer bounces off from the convex object after the hydrodynamic collision. On the other hand, a shaker which bounces off from the convex surface is attracted towards the surface again and thus exhibiting an oscillatory trajectory on the convex object as shown in Fig.~\ref{fig:vex_zero_rep_shaker}. Similar bouncing and oscillatory trajectories are shown by neutral swimmers (Fig.~\ref{fig:cave_zero_rep_neutral}) and shakers (Fig.~\ref{fig:cave_zero_rep_shaker}) on concave surfaces. \citet{dario_gareth} have reported similar trajectories for a squirmer in a concave confinement.
        
The presence of a repulsive potential on the boundary surface (Eq.~\ref{eqn:repulsion}) restricts the distance of closest approach of the squirmer. Therefore, the trajectory of the squirmer is affected by the repulsive potential of the curved walls, as has been demonstrated earlier for flat walls \cite{repulsive_walls}. The altered trajectories corresponding to four different situations described above, due to the repulsive potential of the curved surface, are shown in  Fig.~\ref{fig:vex_0p05_rep_neutral}--\ref{fig:cave_0p05_rep_shaker}. It may be noticed that the presence of a repulsive potential does not change the qualitative nature of the trajectories in these four cases, and both bouncing off and oscillatory trajectories are seen. However, the amplitude of the oscillations are smaller in case of the repulsive boundary, which arises from the restricted motion of the squirmer near the surface. In some cases, the oscillations may completely disappear and consequently the squirmer crawls on the curved surface, as shown in Fig.~\ref{fig:cave_0p05_rep_shaker}
        

\noindent \kvsc{\textbf{Dynamical space}:}
        \begin{figure}[h!]
 \centering
 \subfigure[]{
 \includegraphics[height = 5.7cm]{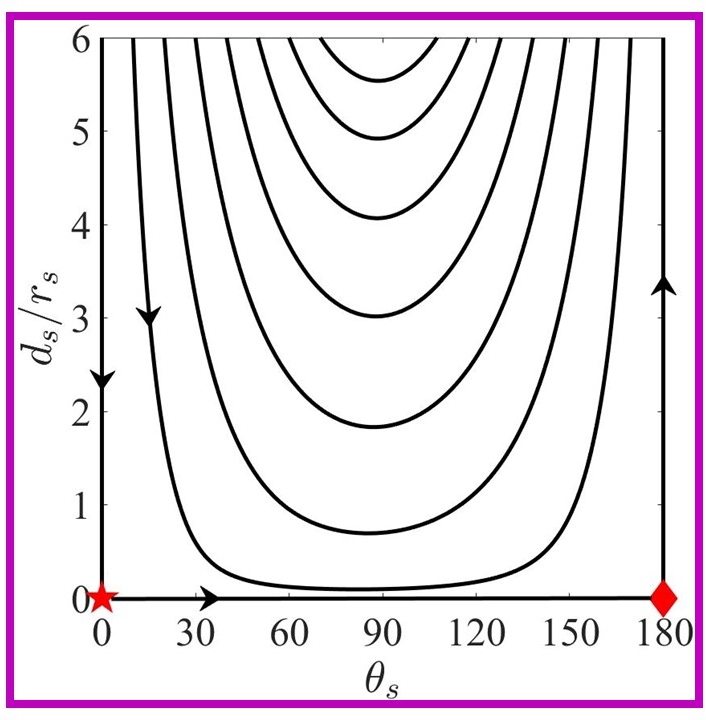}
  \label{fig:m1}}
 \subfigure[]{
 \includegraphics[height = 5.7cm]{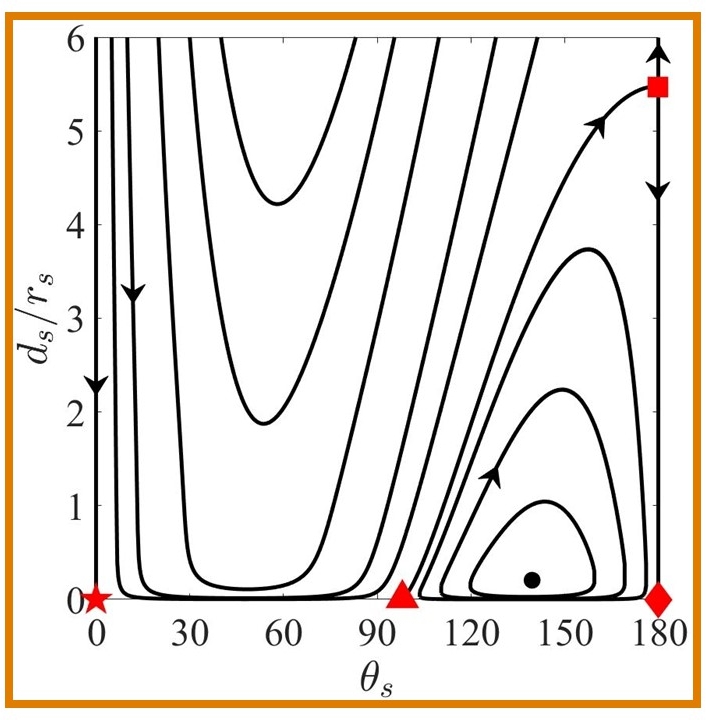}
  \label{fig:m2}}
   \subfigure[]{
  \includegraphics[height=5.7cm]{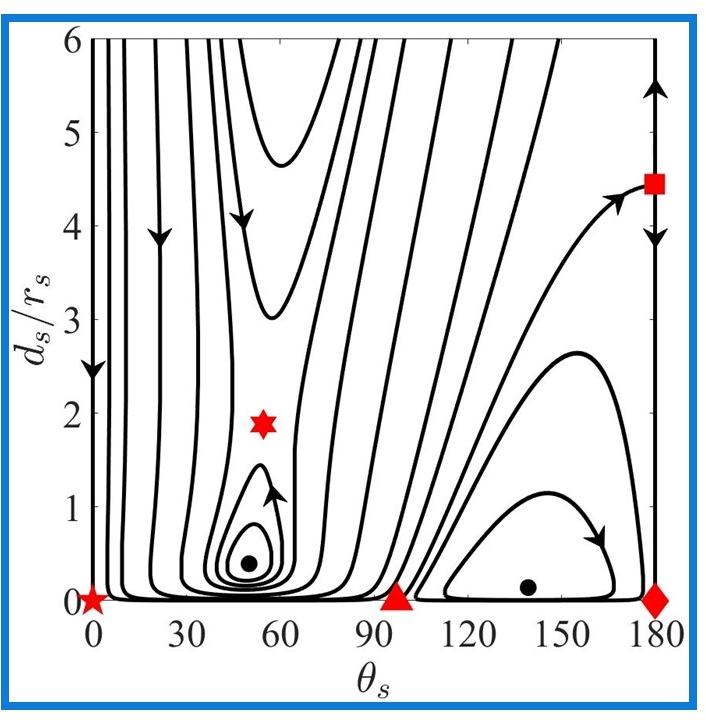}
  \label{fig:m3}}\\
       \subfigure[]{
  \includegraphics[height=5.7cm]{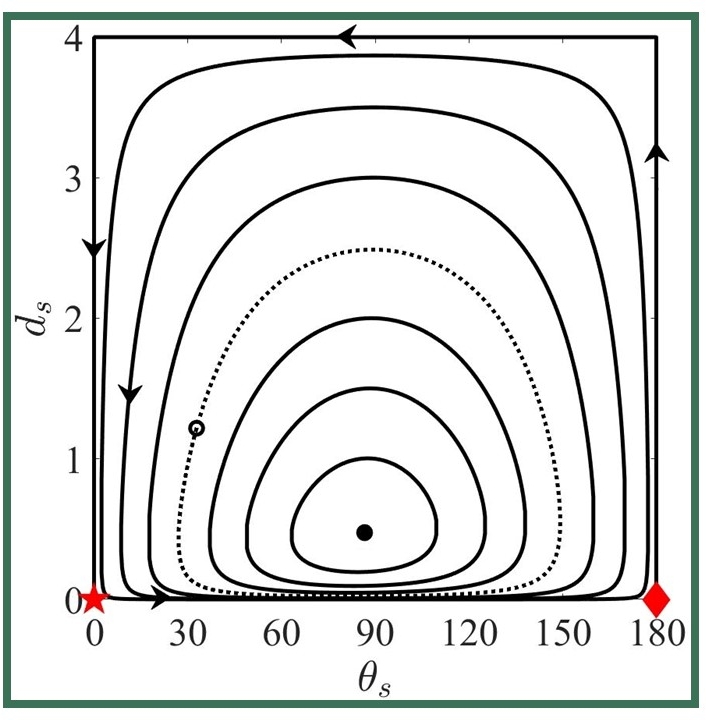}
  \label{fig:m4}}
   \subfigure[]{
  \includegraphics[height=5.7cm]{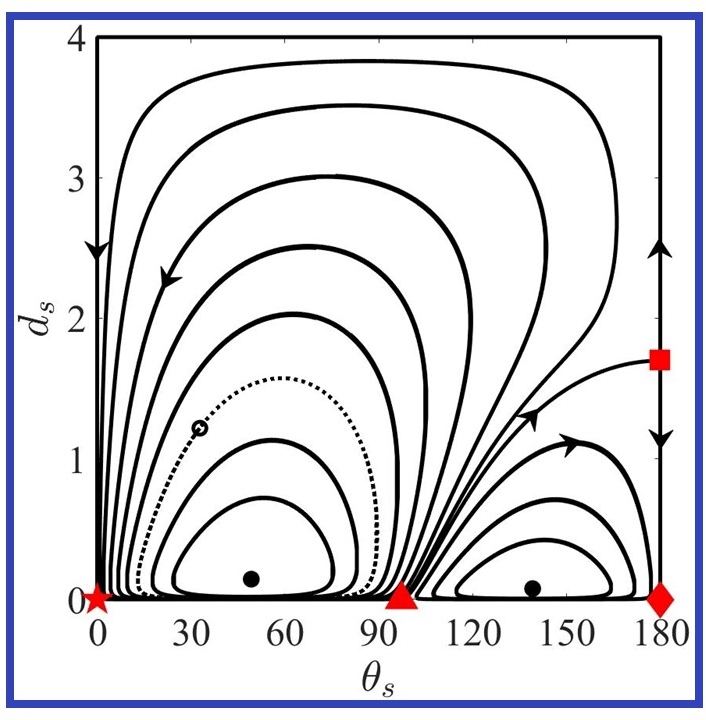}
  \label{fig:m5}}
  \subfigure[]{
  \includegraphics[height=5.7cm]{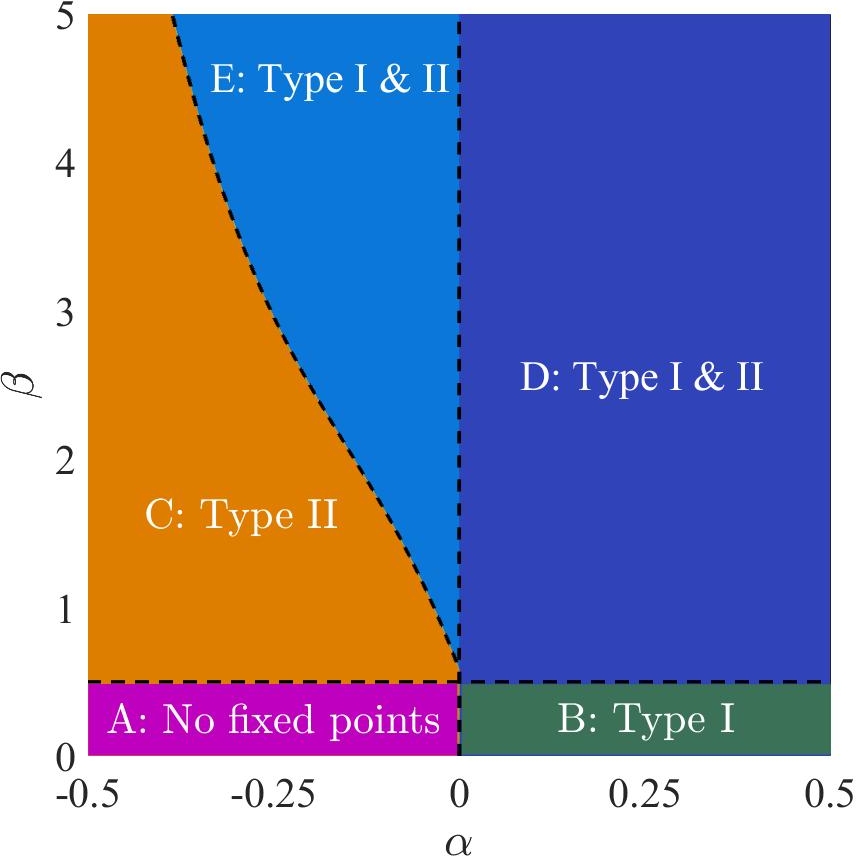}
  \label{fig:phaseplane}}
    \caption {The trajectories of a squirmer in a dynamical space ($d_s$, $\theta_s$) for various values of size ratio and activity: (a) $\alpha = -0.5$, $\beta = 0.1$, (b) $\alpha = -0.5$, $\beta = 4$, (c) $\alpha = -0.2$, $\beta = 4$,  (d) $\alpha = 0.2$, $\beta = 0.1$, and  (e) $\alpha = 0.2$, $\beta = 4$. Fixed points corresponding to crawling trajectories, marked as thick black dots, with $\theta_s<90^\circ$ and  $\theta_s>90^\circ$ are referred to as Type I and Type II respectively. The regions of existence of Type I and Type II fixed points in $\alpha-\beta$ space is illustrated in (f). Labels A--E indicate different regions which are also colored differently. Note that plots (a)--(e) are enclosed in boxes colored according to the color coding of the regions in (f).}
\label{fig:mdynamical_traj}
 \end{figure} 
 \kvsc{The trajectories depicted in Fig.~\ref{fig:trajectories} correspond to a particular initial condition defined by $d_s^{i}$ and $\theta_s^{i}$. However, the change in the initial condition may completely change the nature of the trajectory: bouncing, oscillatory, and crawling.} 
\kvsc{Analyzing the sensitivity of the microswimmer trajectories for different initial conditions in physical space is cumbersome and is often accomplished in a dynamical space. In Fig.~\ref{fig:mdynamical_traj}, we plot the trajectory of a microswimmer in the dynamical space ($\theta_s,d_s$) for different values of size ratio ($\alpha$) and activity ($\beta$).}

\kvsc{\textbf{Bouncing and oscillatory trajectories:} Figure~\ref{fig:m1} shows the dynamical trajectory of a swimmer with dominant source dipole mode ($\beta<<1$) near a convex boundary ($\alpha = -0.5$). The trajectories are open, indicating that such swimmers exhibit a bouncing trajectory on a convex boundary irrespective of the initial location and orientation as shown in Fig.~\ref{fig:vex_zero_rep_neutral}. Figure~\ref{fig:m2} shows the dynamical trajectories for a squirmer with dominant force dipole mode, $\beta = 4$, but near  the same convex object (same $\alpha$) as in Fig.~\ref{fig:m1}. Unlike the previous case, both open and closed trajectories are observed indicating that the swimmer will exhibit either bouncing or oscillatory trajectory based on its initial location and orientation. The open trajectories occur when squirmer is oriented towards the boundary ($\theta_s<90^{\circ}$) while the closed trajectories are possible when squirmer is oriented away from the boundary ($\theta_s > 90^{\circ}$).  Figure~\ref{fig:m3} shows the dynamical trajectories for $\alpha = -0.2$ but at $\beta$ same as that in Fig.~\ref{fig:m2}. In this case, we observe that some of the open trajectories have turned to closed trajectories even for $\theta_s<90^{\circ}$ (see Fig.~\ref{fig:vex_zero_rep_shaker}), thus illustrating that increase in the size ratio can also result in closed trajectories. In other words, Figure~\ref{fig:m1}--\ref{fig:m3} show that the transition from bouncing trajectory to an oscillatory trajectory may happen either with increase in $\alpha$ or with increase in $\beta$ near a convex boundary.}
 
\kvsc{In the case of a concave confinement, irrespective of size ratio ($\alpha$) and activity ($\beta$), we always observe closed trajectories in the dynamical space (Fig.~\ref{fig:m4} and \ref{fig:m5}). The closed trajectories indicate that the swimmers approach the concave surface in a periodic fashion and the maximum possible amplitude of the oscillations is equal to maximum value of the $d_s$ in the dynamical trajectory. Smaller the size of the closed loop, smaller is the amplitude of the oscillations. For a given initial condition, the amplitude of the oscillations is much larger for a swimmer with dominant source dipole mode (Fig.~\ref{fig:cave_zero_rep_neutral}) compared to that of a swimmer with dominant force dipole mode (Fig.~\ref{fig:cave_zero_rep_shaker}). This can be inferred by comparing Fig.~\ref{fig:m4} and \ref{fig:m5}, for example compare the dotted--line trajectory from the initial point marked by the hollow circle.
Hence all trajectories in the dynamical space are either open which extend upto infinity or are closed. It illustrates that there are no attractors, instead the squirmers exhibit trajectories that are dependent on initial conditions even at long times.\\
\noindent\textbf{Crawling and hovering trajectories:} In addition to bouncing and oscillatory trajectories squirmers may also exhibit crawling and hovering trajectories. Such behaviour will correspond to fixed points in the dynamical space. As discussed above, increase in $\alpha$ or $\beta$ introduces closed orbits in the dynamical space thus introducing fixed points as the limiting cases of these closed orbits (crawling trajectory) or the points that separate out closed and open orbits (hovering state). Below we discuss these fixed points and their dynamics in the phase space when the size ratio or activity is changed.
\subsubsection{Dynamics of fixed points}
\noindent\textbf{Hovering fixed points}: These are fixed points characterized by $V_{\parallel} = V_{\perp} = \Omega_z = 0$. Utmost, five such fixed points are possible as shown in Fig.~\ref{fig:m3}, and they will be referred as $H_{1-5}$. Three of them are located on the boundary surface \textit{i.e.,} $d_s = 0$ and at (i) $\theta_s = 0^{\circ}$ ($H_1$), (ii) $\theta_s = 180^{\circ}$ ($H_2$), and (iii) $0^\circ<\theta_s<180^\circ$ ($H_3$). The other two are located in the domain, \textit{i.e.,} $0<d_s<\infty$ and are at (iv) $\theta_s = 180^{\circ}$ ($H_4$) and (v) $0^\circ<\theta_s<90^{\circ}$ ($H_5$). The complex interplay of these fixed points in $\alpha$--$\beta$ space is described below.
\begin{enumerate}[i.]
\item $H_1$ ($d_s = 0$, $\theta_s = 0^\circ$): This fixed point is indicated by a pentagram in Fig.~\ref{fig:mdynamical_traj}. Existence of this fixed point indicates that, irrespective of the initial location, if a squirmer is initialized with $\theta_s = 0^{\circ}$, it moves towards the boundary surface, comes in contact with the surface and stays there, exhibiting a hovering state. This fixed point exists at all values of $\alpha$ and $\beta$ indicating that existence of this hovering state is independent of size ratio and activity. The phase space trajectories in the vicinity of $H_1$ indicate that it is stable to the perturbations in $d_s$ but unstable to the perturbations in $\theta_s$. For example, for a squirmer located at $H_1$, if slightly perturbed in $\theta_s$, will attain a non--zero angular velocity. The squirmer will then spin on the surface of the curved boundary till it attains a stable orientation with $\Omega_z = 0$. This dynamics leads to the presence of two other fixed points on the curved surface, namely, $H_2$ and $H_3$ as shown in Fig.~\ref{fig:m2}--\ref{fig:m4} and are discussed below.
\item $H_2$ ($d_s = 0$, $\theta_s = 180^\circ$): This fixed point is indicated by a rhombus in Fig.~\ref{fig:mdynamical_traj}. Similar to $H_1$, this point exists irrespective of $\alpha$ and $\beta$. Figure~\ref{fig:m1} shows that, for swimmers with a dominant source dipole, this fixed point is stable with respect to the perturbations in $\theta_s$ but unstable with respect to the perturbations in $d_s$. The case is exactly opposite if the swimmer have a dominant force dipole mode, \textit{i.e.,} the fixed point is stable (unstable) with respect to perturbations in $d_s$ ($\theta_s$) as shown in Fig.~\ref{fig:m2}. This change in behavior occurs beyond a critical $\beta$ when two other fixed points, namely, $H_3$ and $H_4$  as shown in Fig.~\ref{fig:m2}, \ref{fig:m3}, \ref{fig:m5} appear.
\item $H_3$ ($d_s = 0$, $90^\circ <\theta_s <180^\circ$): This fixed point is indicated by a triangle in Fig.~\ref{fig:mdynamical_traj}. Unlike $H_1$ and $H_2$, this fixed point appears only when activity is larger than a critical $\beta$. At large activities when $H_3$ exists, it can be seen that slight perturbations in $\theta_s$ on a swimmer located  at either $H_1$ or $H_2$ drives it towards $H_3$, however $H_3$ is also a saddle point, it is stable to the perturbations in $\theta_s$ but unstable to the perturbations in $d_s$. Irrespective of size ratio, as activity increases this fixed point approaches $\theta_s = 90^\circ$. Therefore, swimmers with dominant source dipole mode ($\beta<<1$) can hover at $\theta_s = 0^{\circ}$ and $\theta_s = 180^\circ$ while the swimmers with dominant force dipole mode ($\beta>>1$) hover at $\theta_s = 90^\circ$. Swimmers with intermediate values of $\beta$ can hover at an angle in between $90^\circ$ and $180^\circ$ as shown in Fig.~\ref{fig:m3} and \ref{fig:m4}.
\item $H_4$ ($0<d_s <\infty$, $\theta_s =180^\circ$): This point is indicated by a square in Fig.~\ref{fig:mdynamical_traj}. A positive perturbation in $d_s$ drives the swimmer towards infinity and a negative perturbation in $d_s$ drives the swimmer towards the fixed point $H_2$. The location of $H_4$ is sensitive to the values of $\alpha$ and $\beta$. $H_4$ moves towards infinity with increase in $\beta$ but moves towards the curved surface with increase in $\alpha$.
\item $H_5$ ($0<d_s<\infty$, $0^\circ<\theta_s<90^\circ$): This fixed point is indicated by a hexagram in Fig.~\ref{fig:mdynamical_traj}. It separates the closed and open trajectories in phase space as shown in Fig.~\ref{fig:m3}. Therefore, this point doesn't exist for a concave surface (see Fig.~\ref{fig:m5}), as all the trajectories are closed for a concave surface. For a convex surface, $H_5$ appears beyond a critical $\beta$, and it moves away from the curved surface with increase in either $\alpha$ or $\beta$.
\end{enumerate}
While identifying the hovering fixed points helped us to understand the phase portraits with respect to variation in the size ratio and activity, these points are of less physical significance since they are generally not stable.}\\

\kvsc{\noindent\textbf{Crawling fixed points}:  The other class of fixed points that are shown in Fig.~\ref{fig:mdynamical_traj} correspond to the limiting trajectory of closed orbits with zero radii.  Figure~\ref{fig:mdynamical_traj} confirms the existence of these fixed points near both concave and convex--curved boundaries, and utmost, two such fixed points are possible as shown in Fig.~\ref{fig:m3} and \ref{fig:m5}. The fixed point with $\theta_s<90^{\circ}$ and $0<d_s<\infty$ is referred as Type I and the fixed point with $\theta_s>90^{\circ}$ and $0<d_s<\infty$ is referred to as Type II. Both these fixed points correspond to crawling trajectory of swimmers in physical space.}

\kvsc{Crawling trajectories should satisfy the following relation
\begin{equation}\label{eqn:mfixed_points}
V_{\parallel} = 0, \quad \frac{V_{\perp}}{d} = \Omega_z.
\end{equation}
The above relations correspond to the conditions that (i) the radial velocity of the squirmer is zero and (ii) the change in the angular position of the microswimmer with respect to the curved surface caused by its tangential velocity should be equal to the change in the orientation of the squirmer resulting from its angular velocity. Thus, the first condition constraints the motion of the microswimmer only in the tangential direction with respect to the curved surface and the second constraint assures that the microswimmer moves along the tangent such that its orientation with respect to the curved surface does not change. Equation~\ref{eqn:mfixed_points} is numerically solved to determine the Type I and Type II fixed points; their behaviour in $\alpha-\beta$ plane is shown in Fig.~\ref{fig:phaseplane} and is described below.}

\kvsc{Let us consider the case of squirmers with small activity ($\beta << 1$) first.  These are swimmers with a dominant source dipole and there are no Type I or Type II fixed points when they are near a convex boundary as shown in Fig.~\ref{fig:m1}.  On the other hand, as shown in Fig.~\ref{fig:m4}, Type I fixed point exists when the squirmer is in the concave confinement even when $\beta<<1$. The change from the absence of fixed points to presence of Type I fixed points with change in curvature for swimmers of small activity ($\beta$) can be seen as a transition from region A to region B in Fig.~\ref{fig:phaseplane}. This transition occurs at $\alpha=0$ (corresponds to a flat wall boundary) irrespective of the value of $\beta$.}

\kvsc{Now consider the case of squirmers with increased activity. As shown in  Fig.~\ref{fig:m1}--\ref{fig:m2} increase in activity results in the emergence of type II fixed points for $\alpha < 0$. Similarly as shown in Fig.~\ref{fig:m3}--Fig.~\ref{fig:m4}, Type II fixed points emerge for $\alpha > 0$ with increase in activity. The critical $\beta$ beyond which the Type II fixed points appear near convex surfaces is shown as a transition from region A to region C in Fig.~\ref{fig:phaseplane}. Similarly for concave surfaces the appearance of Type II fixed points is shown as a transition from region B to region D in Fig.~\ref{fig:phaseplane}. It is clear that the transitions indicating the emergence of Type II fixed points for both convex and concave surfaces is independent of the size ratio, $\alpha$.}
 \kvsc{The third bifurcation occurs at $\beta = f(\alpha)$, beyond which Type I fixed points appear in the phase space for convex surfaces as observed in Fig.~\ref{fig:m2} and \ref{fig:m3}. Unlike the transitions discussed above, this bifurcation depends on both size ratio and activity.  As size ratio increases, the critical $\beta$ beyond which Type I fixed point appears decreases.  Thus the bifurcation leading to the formation of Type I fixed points can happen either with increase in the activity (for a fixed size ratio) or with increase in the size ratio (for a fixed activity). This change in behaviour is indicated as a transition from region C to region E in Fig.~\ref{fig:phaseplane}.}

\kvsc{To summarize, increase in either $\alpha$ or $\beta$ results in a transition from open trajectories to closed trajectories in the phase space which is reflected as an increase in the number of fixed points in Fig.~\ref{fig:phaseplane}. Open trajectories in phase space correspond to bouncing behaviour of the swimmer on the curved wall characterized by a single collision. However, the closed curves in phase space don't necessarily correspond to oscillatory trajectories. For example, the closed trajectories in concave confinement are not  really the consequence of hydrodynamic interactions but due to the motion in closed space. Thus the trajectories in phase space cannot distinguish the bouncing versus oscillatory behaviour as observed in the physical space (see Fig.~\ref{fig:cave_zero_rep_neutral} and \ref{fig:cave_zero_rep_shaker}). Therefore, below we define new measures that may have experimental relevance and can effectively distinguish different kinds of trajectories near boundaries with either curvatures.} 


        \subsubsection{Characterising the trajectories - proximity parameter and retention time}
        \label{sec:orio_avg_char}
        In order to classify different kinds of trajectories, and thus to quantify the effect of wall curvature on the behaviour of microswimmers we define two measures:  proximity parameter and retention time.
        
        Proximity parameter quantifies the affinity of the microswimmer towards the boundary based on its trajectory. It is defined as the fraction of the distance that a microswimmer traverses in close proximity of a curved boundary,
        \begin{align}
            \phi = \frac{\int_{d_s < d_c} dS}{\int dS},
        \end{align}
        where $dS$ is the differential arc length on the trajectory of the microswimmer, and $d_c$ is a cut off distance. If a squirmer bounces off from the boundary and swims away after the hydrodynamic collision (\textit{e.g.,} Fig.~\ref{fig:vex_zero_rep_neutral},~\ref{fig:cave_zero_rep_neutral}), then $\phi \to 0$, $i.e.,$ the distance travelled by the squirmer in proximity of the boundary is much smaller than the total distance that it travelled. In the other limit, $\phi \to 1$ corresponds to a crawling behavior of the squirmer where the squirmer is always in close proximity of the boundary (\textit{e.g.,} Fig.~\ref{fig:cave_0p05_rep_shaker}). Intermediate values of $\phi$ represent oscillatory trajectories of varying amplitude and frequency ( as in Fig.~\ref{fig:vex_zero_rep_shaker} and Fig.~\ref{fig:cave_zero_rep_shaker}), with increase in $\phi$ indicating more frequent oscillations with decreasing amplitude and the squirmer trajectory lying more in close proximity of the surface.
        
        Since velocity of the microswimmer is not constant along its path, it is also useful to quantify the time spent by the microswimmer in proximity of a surface, as it gives an independent measure of the affinity of the microswimmer towards the surface. Thus, a parameter, retention time is defined as,
        \begin{equation}
            t_r =  \frac{\int_{d_s < d_c} dt}{\int dt},
        \end{equation}
        where $dt$ is the differential time taken by the microswimmer on its trajectory. For a bouncing trajectory $t_r \to 0$, for a crawling behavior $t_r \to 1$, and intermediate values of $t_r$ indicates oscillatory trajectories, similar to the proximity parameter. 
        
        As discussed earlier, the trajectory and hence both $\phi$ and $t_r$ are initial condition dependent. Therefore, we integrate $\phi$ and $t_r$ over all possible initial orientations to define an average proximity parameter,
        \begin{equation}
            \langle \phi \rangle = \frac{1}{2\pi} \int_0^{2\pi} \phi(\theta_{s}^i) d\theta_{s}^i,
            \label{eqn:avg_def}
        \end{equation}
        and an average retention time
        \begin{equation}
            \langle t_r \rangle = \frac{1}{2\pi} \int_0^{2\pi} t_r(\theta_{s}^i) d\theta_{s}^i,
                \label{eqn:avg_tr}
        \end{equation}
        which eliminate the dependency of $\phi$ and $t_r$ on the initial orientation of the squirmer $\theta_{s}^i$.\\
        
             \begin{figure*}
          	\centering
          	\subfigure[]{
          		\includegraphics[height=5.5cm]{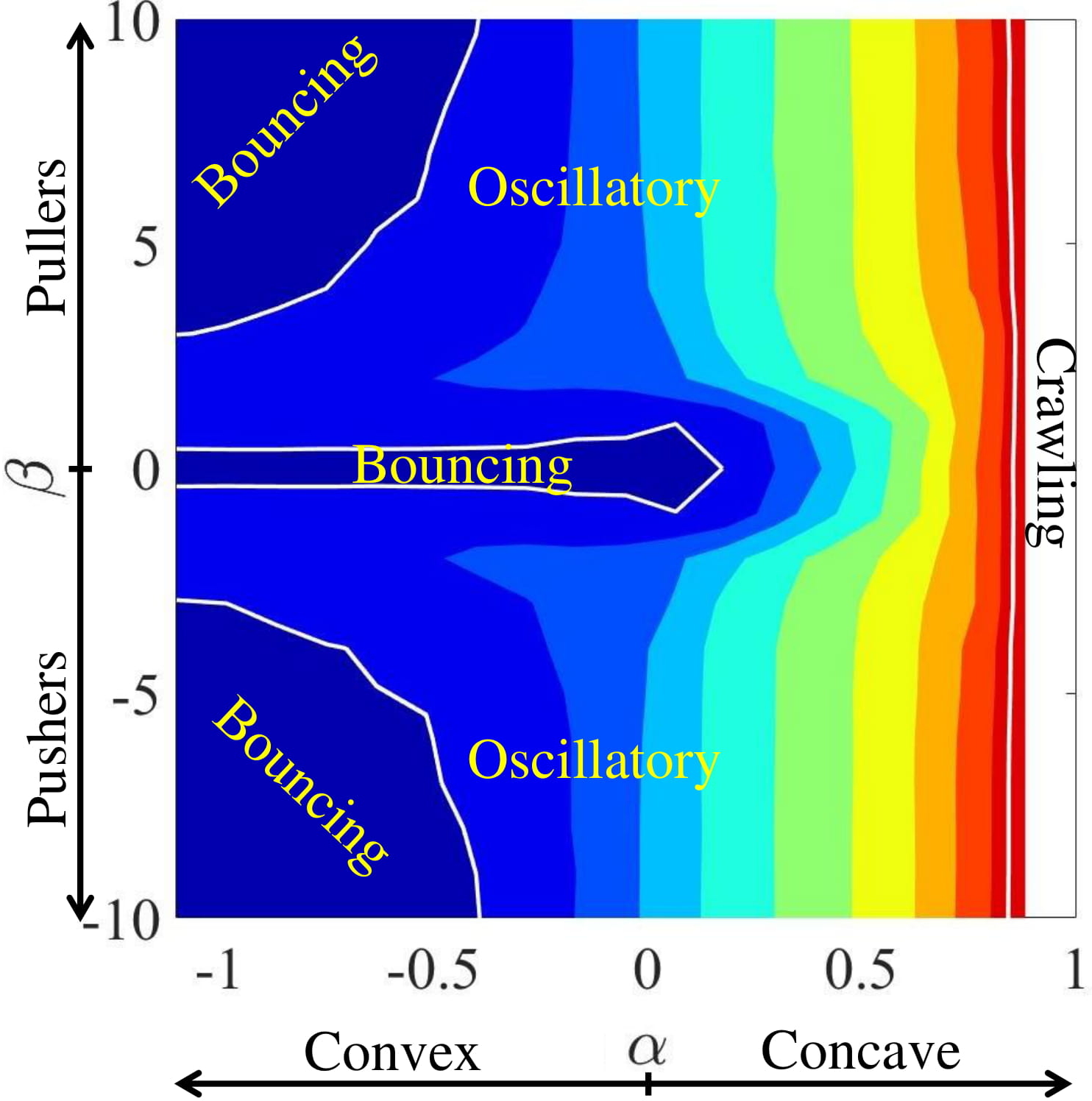}
          	\label{fig:prox_0_rep}}\hfill
          	\subfigure[]{
          		\includegraphics[height=5.5cm]{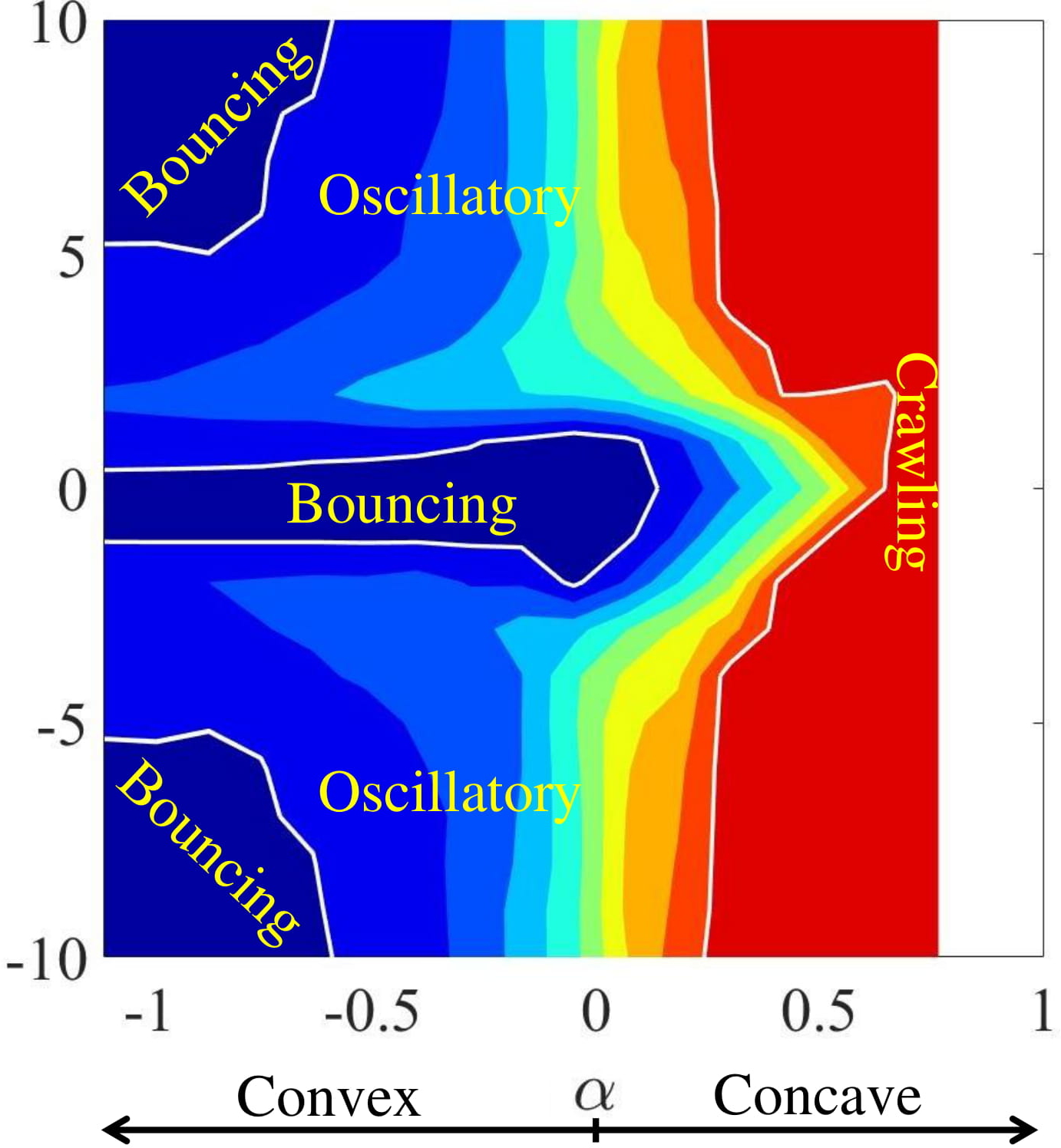}
          	\label{fig:prox_0p05_rep}}\hfill
          	  	\subfigure[]{
          		\includegraphics[height=5.5cm]{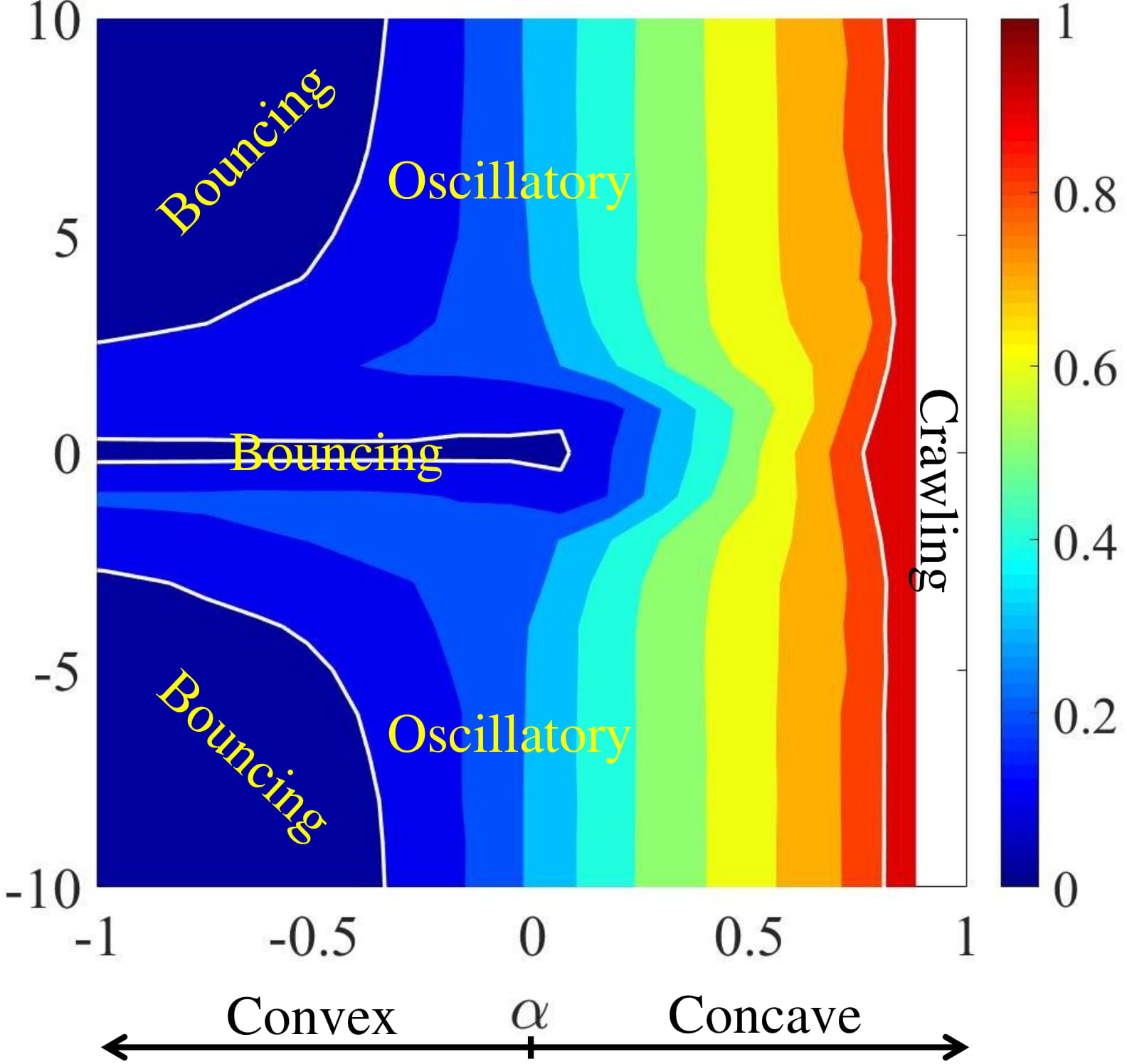}
          		\label{fig:retention_time_0_rep}}
          	\caption{The contour plot of average proximity parameter, $\langle\phi\rangle$, as a function of curvature ($\alpha$) and activity ($\beta$). (a) in the absence of repulsion ($\delta = 0$) and (b) in the presence of repulsion ($\delta$ = 0.05). In these calculations, the swimmer is considered to be in the proximity of the boundary, if the surface to surface separation distance is less than $d_c = 0.2 r_s$. \kvsc{Note that $d_c$ is an arbitrary choice and the behavior depicted here is insensitive to the value of $d_c$.} (c) The contour plot of average retention time, $\langle t_{r}\rangle$, as a function of the curvature ($\alpha$) and $\beta$, and in the absence of repulsion ($\delta = 0$). Colorbar of (a) and (b) is same as that of (c).  For the analysis, we approximate the integral in Eq.~\ref{eqn:avg_def}--\ref{eqn:avg_tr} with summation, and divide the orientational space into 18 equal size intervals.}
          	\label{fig:prox}
          \end{figure*}

        \noindent\textbf{Average proximity parameter:} Figure~\ref{fig:prox_0_rep} depicts the dependency of average proximity parameter on the curvature $\alpha$ and the activity $\beta$. It may be observed that $\langle \phi \rangle \lessapprox 0.2$, in the $\alpha-\beta$ plane as long as $\alpha<0$. This small value of $\langle \phi \rangle$ indicates minimal contact between the squirmer and convex surfaces it comes into contact with. The corresponding trajectories may be either single contact (bouncing trajectory, \kvsc{open trajectory in phase space and no fixed points}) or multiple contact (oscillatory trajectory, \kvsc{closed trajectories in phase space, presence of Type I and Type II fixed points}). Bouncing trajectories are encountered in two regions in $\alpha - \beta$ plane: (i) by squirmers with dominant $B_1$ mode ($\beta\approx0$: neutral swimmers, weak pullers and pushers) irrespective of the value of the curvature as long as $\alpha < 0$, and (ii) squirmers with dominant $B_2$ mode($|\beta| \gtrapprox 4$: strong pullers and pushers) on strongly curved convex surfaces ($\alpha \lessapprox -0.5)$. On contrary, when both $B_1$ and $B_2$ modes are significant  ($|\beta| < 4$: pullers and pushers), such squirmers show small to moderate values of $\langle\phi\rangle$ on convex surfaces indicating that they get trapped and exhibit wall bounded oscillations. It may also be noted that the average proximity parameter increases with increase in $\alpha$, irrespective of the value of $\beta$. This increase is least for the squirmers with dominant $B_1$ mode ($\beta \to 0$), so they continue to exhibit bouncing trajectory even on the flat wall. All other swimmers, ($|\beta|>1$) exhibit oscillatory trajectory on a flat wall as reported in \cite{crowdywall}. \kvsc{This behavior is reflected as the transition from region A to region C in Fig.~\ref{fig:phaseplane}.} The increase in $\langle \phi \rangle$ with increase in $\alpha$ continues even beyond the wall limit ($\alpha > 0$), \textit{i.e.,}  as the concavity of the confinement increases. \kvsc{This is consistent with the smaller values of $d_s$ associated with closed orbits near a concave surface compared to a convex surface in Fig.~\ref{fig:mdynamical_traj}.} Moreover, when $\alpha > 0$, the average proximity parameter is dependent only on $\beta$ when $|\beta|$ is approximately less than unity. \kvsc{A similar behavior can be seen in Fig.~\ref{fig:phaseplane} in terms of fixed points. When $\alpha>0$ the dynamics of fixed points is insensitive to the size ratio, and the critical $\beta$ beyond which the Type II fixed points appear is of order unity as predicted by the proximity parameter.}
        
        In other words, Fig.~\ref{fig:prox_0_rep} suggests that a microswimmer is likely to be attracted towards a concave boundary and swims close to it. The proximity of the trajectory to the surface increases with increase in concavity, irrespective of the type and strength of the microswimmer.  Thus, hydrodynamic interaction results in a squirmer to swim away from a convex boundary but to swim closer to a concave boundary. A flat plate has $\langle \phi \rangle \approx 0.4$, an intermediate behaviour representing neither a strong attraction nor a strong repulsion.

        The increase in proximity parameter with curvature $i.e.$, as the curvature changes from convex to concave can be understood as follows. Firstly, it may be noted that Fig.~\ref{fig:inst_velocities}--\ref{fig:inst_velocities_ds} show that the instantaneous translational velocities decrease with increase in $\alpha$. 
         Moreover, the angular velocity of the squirmer near a convex boundary is generally larger than that near a concave boundary (except a neutral swimmer near a weak concave surface).
        Therefore, a microswimmer located near a convex boundary can reorient easily, resulting in an escape from the neighbouring wall. On the other hand the microswimmer near a concave boundary will exhibit a weaker dynamics. Secondly, the geometry of the neighbouring wall makes a difference - an escaped microswimmer goes further away from the neighbouring part of the surface when escaping from a convex boundary, but this will not be the case with a concave boundary. An escaping microswimmer may get again influenced by the neighbouring wall due to the concavity of the surface\kvsc{, reflected as closed trajectories in the phase space (see Fig.~\ref{fig:m4} and \ref{fig:m5})}. Thus, a microswimmer near a convex boundary easily escapes while a microswimmer near a concave boundary continues to stay close to it.
       
        As mentioned earlier, $\langle \phi \rangle$ is independent of $\beta$ for sufficiently large $|\beta|$.  This may be expected as large $|\beta|$ indicates that the squirmer dynamics is primarily governed by $B_2$ mode, and contributions from $B_1$ mode is smaller. On the other hand when $|\beta| < 1$ neither $B_1$ nor $B_2$ contributions can be neglected. As $|\beta| \to 0$ squirmer behaviour is dictated only by $B_1$ mode. For these swimmers, if $\alpha \to 0$, but $\alpha>0$ (weak concave curvature) the angular velocity of a squirmer is sufficiently large that it easily escapes from the surface and thus, $\langle \phi \rangle$ remains small. 
         However, on these weakly concave surfaces, as $|\beta|$ increases there will be a transition from bouncing to oscillatory trajectory. This observation is similar to the behavior of two dimensional squirmers observed near a flat wall \cite{Ahana}. In the limit $\alpha \to 1$, the squirmer motion is highly restricted, and the angular velocities are small. Hence, in this limit, irrespective of $\beta$ the proximity parameter is close to 1 indicating a crawling behavior.

        Another characteristic feature to note in Fig.~\ref{fig:prox_0_rep} is the symmetry about $\beta = 0$ axis. This happens because a puller will show exactly the same trajectory as a pusher when initialised at the same location but at a different orientation\kvsc{, referred to as puller--pusher duality \cite{crowdywall}}. Therefore, integrating over the initial orientations, average proximity parameter doesn't distinguish pullers and pushers which leads to the symmetry about $\beta = 0$ axis.  However, this is not the case when the wall has a repulsive force. This case is illustrated in Fig.~\ref{fig:prox_0p05_rep} where the average proximity parameter is plotted in $\alpha-\beta$ phase space. The symmetry about $\beta=0$ axis is slightly broken. In the case of a strong convex boundary, pullers ($\beta > 0$) have slightly larger proximity parameter compared to pushers ($\beta < 0$) however both pullers and pushers have either a bouncing or an oscillatory trajectory. As the curvature changes from convex to concave, irrespective of $\beta$ the proximity parameter increases. The small asymmetry, namely pullers have slightly larger proximity parameter compared to equally strong pushers holds for weak concave and convex surfaces as well. As the concave confinement becomes stronger, irrespective of $\beta$ the proximity parameter is very large indicating a crawling trajectory.  
        
        More importantly, on comparing Fig.~\ref{fig:prox_0_rep} and Fig.~\ref{fig:prox_0p05_rep} it can be noted that the presence of repulsion increases the proximity parameter irrespective of $\alpha$ and $\beta$. Thus, the affinity of the microswimmer towards the boundary increases with presence of repulsive forces on the wall. For example, the region of $\alpha-\beta$ plane corresponding to bouncing trajectories exhibited by strong pushers and pullers shrinks. Similarly, the increase in affinity is quite significant for concave surfaces suggesting that the proximity of the squirmer trajectory near a concave surface is enhanced by the repulsive forces on the boundary, which occurs irrespective of the value of $\beta$. However, the enhancement due to repulsive force is not significant for neutral swimmers, and when $\alpha < 0$, they continue to exhibit bouncing trajectories. \\
        
        \noindent\textbf{Average retention time:} Now, we analyse the effect of $\alpha$ and $\beta$ on average retention time $\langle t_r \rangle$. This is illustrated in Fig.~\ref{fig:retention_time_0_rep}. Clearly, the dependency of retention time on $\alpha$ and $\beta$ is very similar to that of proximity parameter. For a fixed $\beta$, the retention time increases with increase in $\alpha$, $i.e.,$ as curvature changes from convex to concave, indicating a transition from bouncing to crawling trajectory. Similar to $\langle \phi \rangle$, the retention time is insensitive to $\beta$ unless $|\beta|$ is small. The symmetry about $\beta = 0$ axis is also noticeable and this symmetry breaks down when the curved surface exhibits a repulsive potential. In other words, $\langle t_r \rangle$ follows the same trend as $\langle \phi \rangle$ suggesting that (i) the nature of the behaviour of the microswimmer near a curved boundary can be inferred from either of these measures, and (ii) the conclusions drawn from Fig.~\ref{fig:prox_0_rep}--\ref{fig:prox_0p05_rep} are robust.

        \begin{figure*}
        	\centering
        	\subfigure[]{
        		\includegraphics[height=5cm]{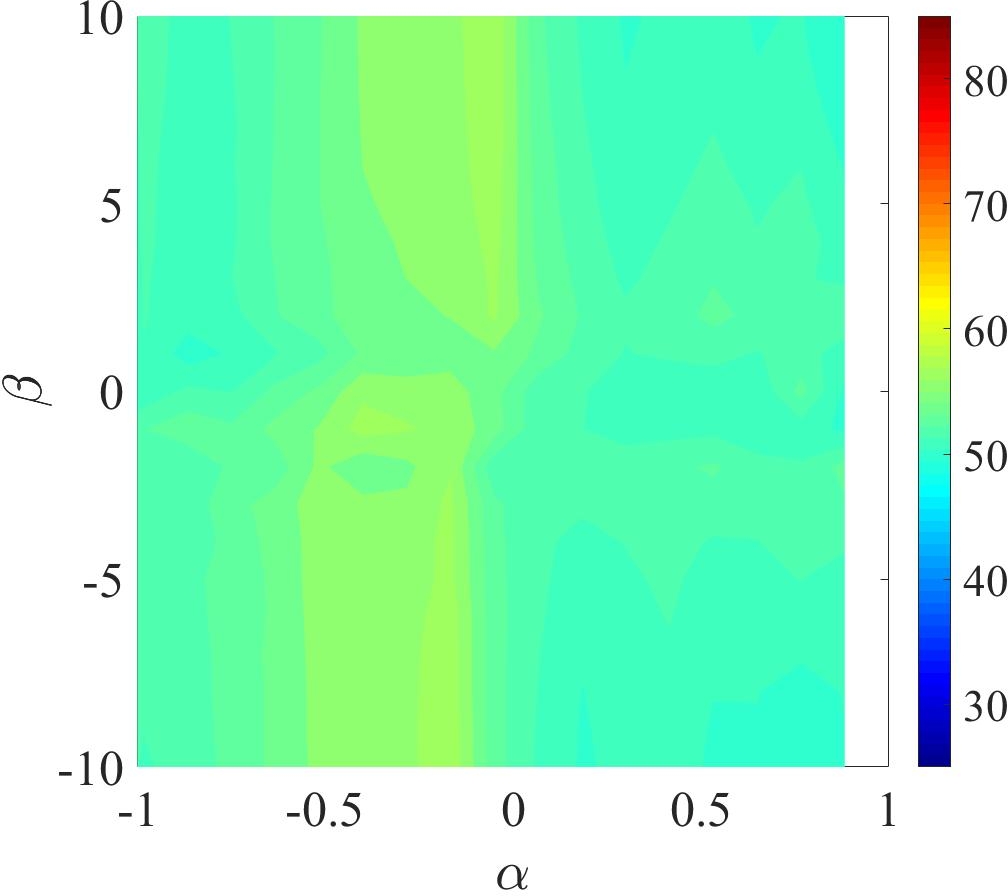}
        		\label{fig:orio_r_axis_0_rep}}
         	\subfigure[]{
         		\includegraphics[height=5cm]{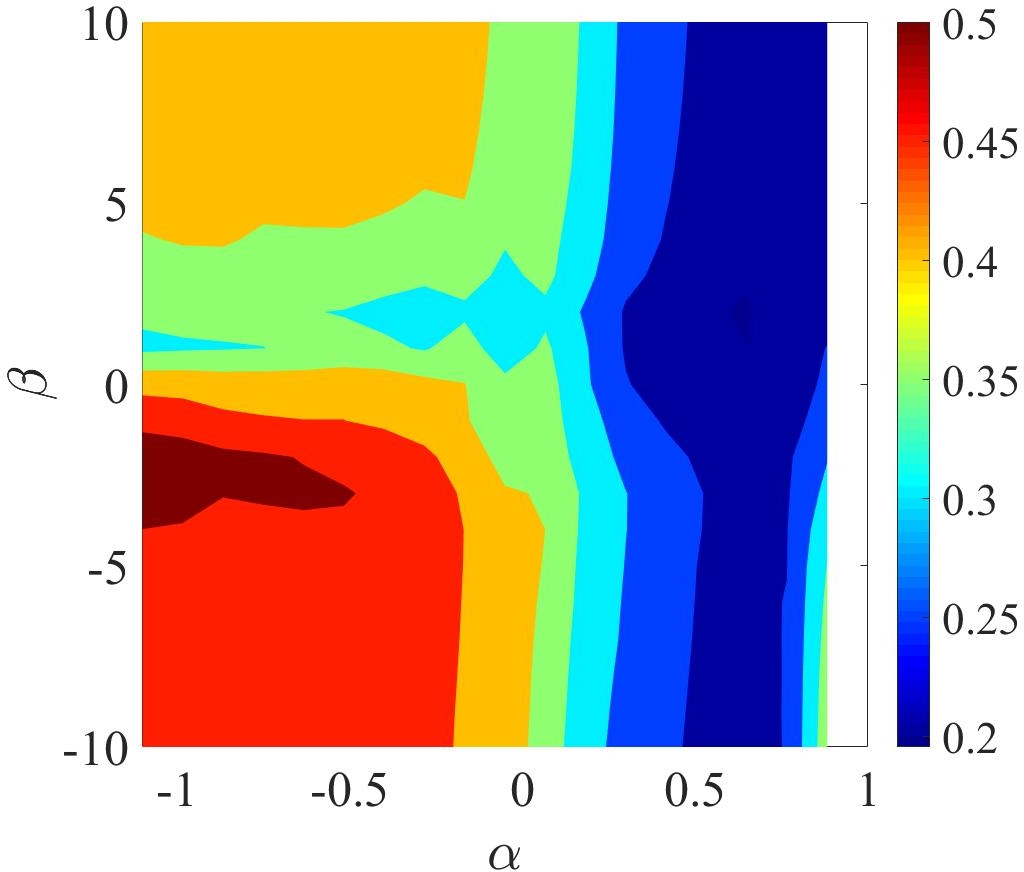}
         		\label{fig:utheta_0_rep}}
        	\subfigure[]{
        		\includegraphics[height=5cm]{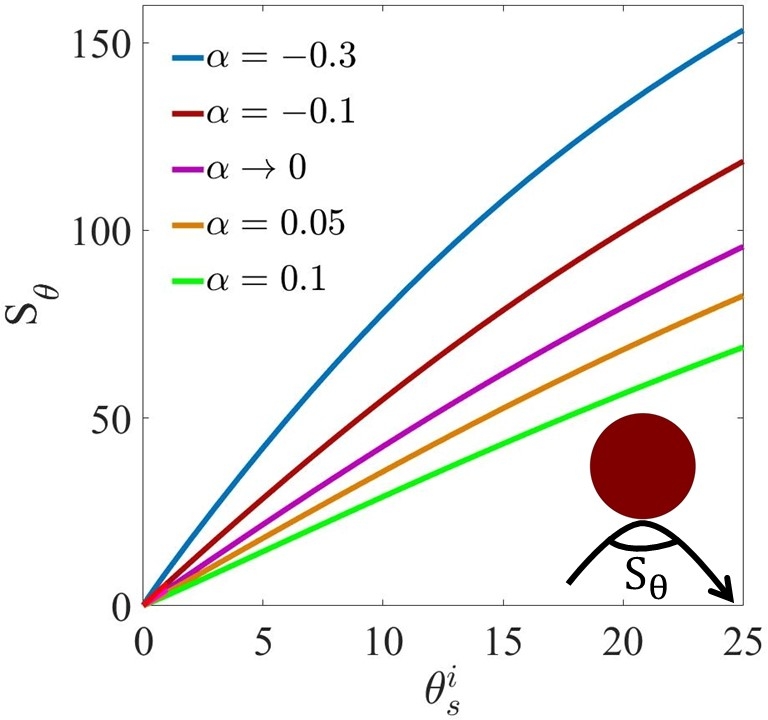}
        		\label{fig:scat_angle}}
        	\caption{The contour plot of (a) $\langle90^{\circ}-\theta_{s}\rangle$, the average orientation with respect to the surface tangent, and (b) $\langle V_{\perp}\rangle$,  the average tangential velocity near the surface are plotted as a function of $\alpha$ and $\beta$.
        \kvsc{(c) Scattering angle ($S_{\theta}$) of a neutral swimmer ($\beta = 0$) as a function of angle made by the orientation vector with the separation vector ($\theta_s^i$) at time $t = 0$ for various size ratios. The inset in (c) shows the definition of the scattering angle $S_{\theta}$.}}
        	\label{fig:orieo_r}
        \end{figure*}
        
        \subsubsection{Characterizing the dynamics close to the boundary - Squirmer orientation, tangential velocity, and scattering angle}
        \label{sec:vtheta_avg}
        
       In the subsection \ref{sec:orio_avg_char} we characterised the entire trajectory of a squirmer in terms of average proximity parameter and average retention time. Now we analyze the configurational and dynamic behaviour of the squirmer when it is in the neighbourhood of the curved boundaries.\\
       
        \noindent\textbf{Squirmer orientation near the boundary:}
        As the squirmer moves along the wall, its orientation changes continuously. We measured the angle that the squirmer makes with the neighbouring surface, \textit{i.e.,} measured as the angle between the squirmer orientation and the surface tangent, and averaged over the trajectory when the squirmer is in proximity to the boundary, $d_s \leq d_c$. Figure~\ref{fig:orio_r_axis_0_rep} illustrates that the squirmer maintains an angle close to $\approx 53^{\circ}$ with the neighbouring surface, irrespective of the curvature of the surface $\alpha$ or the activity $\beta$. However, a considerable change in the value is observed if the curved boundary possesses repulsive forces. 
        It has been found that, in this case the squirmer is more aligned with the boundary irrespective of $\alpha$ or $\beta$. However, \citet{arbitrary_curvature} reported a microswimmer dependent alignment on curved boundaries, which may be
         due to the fact that the rotational dynamics of the microswimmers in their study was affected by the thermal fluctuations.\\
         
        \noindent\textbf{Tangential velocity near the boundary:} The average velocity parallel to the boundary $\langle V_{\perp} \rangle$ is calculated when the squirmer is in the proximity of the surface ($d \leq d_c$) and is shown in Fig.~\ref{fig:utheta_0_rep} as a function of $\alpha$ and $\beta$. It may be noted that the average tangential velocity of the squirmer near a convex curvature is larger compared to that near a concave curvature. This is in agreement with our previous observations in (Fig.~\ref{fig:B1_mode_Vtheta},~\ref{fig:B2_mode_Vtheta}), where it was found that the instantaneous translational velocity decreases as the curvature changes from convex to concave. A larger tangential velocity near a convex surface indicates an easy escape of the squirmer while a smaller tangential velocity near a concave surface indicates the trapping ability of the surface. Again, these observations are consistent with our earlier findings on proximity parameter and retention time, both of which are seen to increase as the curvature changes from convex to concave. 
        
        However it is worth noting that, irrespective of the curvature the average tangential velocity of pushers is larger than pullers when $\alpha < 0$ (convex walls). This arises from the fact that instantaneous tangential velocities due to $+B_1$ and $+B_2$ modes are in opposite directions (Fig.~\ref{fig:B1_mode_Vtheta},~\ref{fig:B2_mode_Vtheta}). Hence, for a given $\alpha$, the instantaneous tangential velocity of pullers ($B_1>0, B_2 > 0$) is smaller compared to pushers ($B_1 > 0, B_2<0$). Since smaller tangential velocity indicates lesser possibility of escaping of the microswimmer from the neighbourhood of a convex surface, we may conclude that pullers have more affinity towards convex surfaces compared to pushers. Figure~\ref{fig:utheta_0_rep} shows that $\langle V_{\perp} \rangle$ is almost insensitive to $\beta$ near concave surfaces.\\
        
        \noindent\textbf{Scattering angle:}
        Here, we analyze the angle at which the squirmer scatters after its collision with the curved surface. Scattering angle ($S_{\theta}$) is defined as the angle between the incident trajectory and the reflected trajectory as shown in the inset of  Fig.~\ref{fig:scat_angle}. 
        
        
\kvsc{Figure \ref{fig:scat_angle} depicts the scattering angle ($S_\theta$) as a function of initial orientation of the squirmer ($\theta_s^i$) for various values of size ratio ($\alpha$). The scattering angle increases with $\theta_s^i$ in all cases, namely, for convex ($\alpha< 0 $) and concave--curved ($\alpha > 0 $) boundaries and planar walls. If the squirmer is initially oriented along the separation vector ($\theta_s^{i} = 0^\circ$), it exhibits a large radial velocity and zero tangential velocity. Therefore in this limit, squirmer reaches very close to the curved boundary ($d_s \to 0$). The consequence is that it experiences a large angular velocity which will reorient the squirmer immediately, resulting in smaller scattering angles. As the orientation of the squirmer deviates from the separation vector, $i.e.,$ as $\theta_s^i$ increases, the swimmer does not reach so close to the curved boundary, it deflects with a relatively smaller angular velocity thereby shows a larger scattering angle. We find that for a given orientation, scattering angle decreases with increase in the size ratio. It happens because, as $\alpha$ increases, the boundary induced angular velocity reorients the swimmer more, resulting in smaller scattering angles. In short, the microswimmers deflect with larger scattering angles from convex surfaces, and with smaller scattering angles from concave surfaces. We find that this behaviour is rather insensitive to activity.}

        Till now, we analyzed the affinity of the microswimmer towards a convex or a concave boundary \kvsc{in terms of fixed points and} by calculating various measures namely proximity parameter, retention time, squirmer orientation near the boundary, tangential velocity near the boundary, and scattering angle.  All the analysis shows that, a convex surface is less likely to trap the microswimmer hydrodynamically while a concave surface does. The  type and strength of the microswimmer does not seem to be very important in determining this behavior: compared to pushers, pullers show only a slightly greater affinity towards the convex surfaces. It has also been seen that repulsive forces on the surface enhance the affinity of the  microswimmer towards the boundary.  The extent of enhancement in the affinity can be significant for a concave surface.
        
        We now investigate the combined effects of concave and convex curvatures by analyzing the squirmer dynamics in an annular confinement. 
        
        \section{\label{sec:annular_confinement}Squirmer dynamics in an annular confinement - net effect of both convex and concave curvatures}
    
        \begin{figure}[h!]
        \centering
            \subfigure[]{
          \includegraphics[height=4cm]{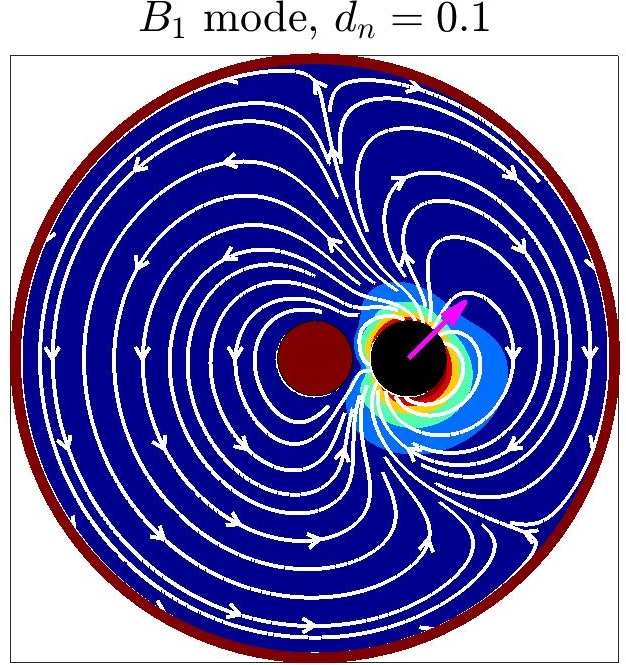}
              \label{fig:ann_B1_ff_0p1}}\quad
            \subfigure[]{
          \includegraphics[height=4cm]{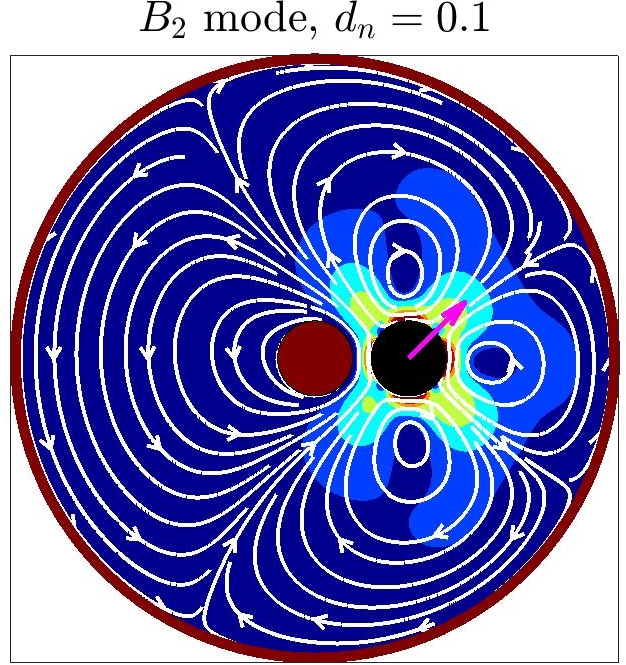}
              \label{fig:ann_B2_ff_0p1}}\quad
                \subfigure[]{
          \includegraphics[height=4cm]{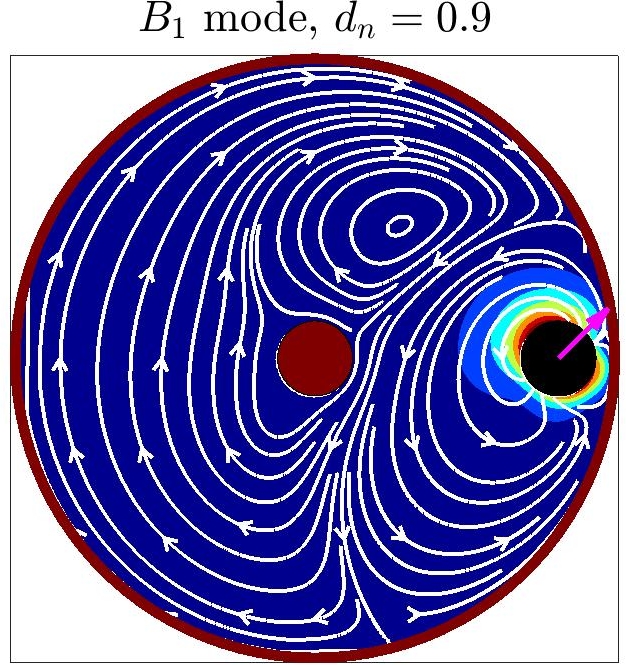}
          \label{fig:ann_B1_ff_0p9}}\quad
          \subfigure[]{
          \includegraphics[height=4cm]{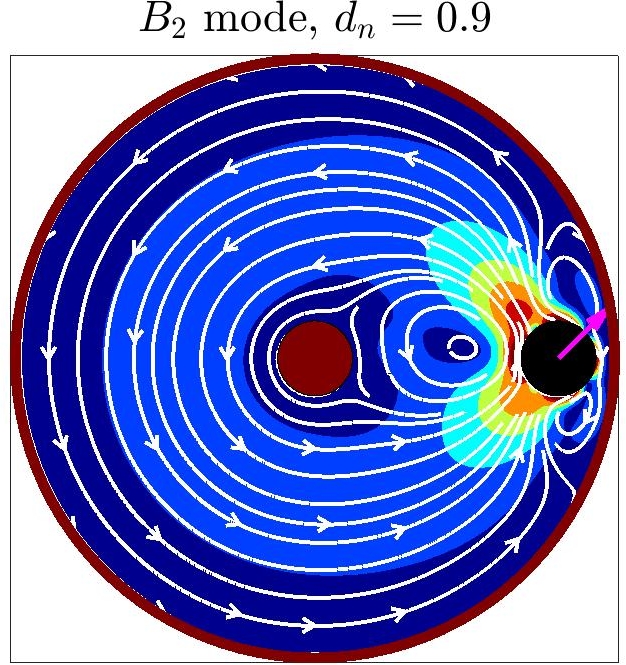}
          \label{fig:ann_B2_ff_0p9}}\quad 
          \caption{Instantaneous velocity fields  of a squirmer in annular confinement obtained from the lattice Boltzmann simulations for $\alpha_o = 0.125$, $\alpha_i = -1$, and $\theta_s = 45^{\circ}$. For $d_n = 0.1$, \kvsc{(a) a neutral swimmer ($B_1>0$, $B_2 = 0$), (b) a shaker ($B_1=0$, $B_2 > 0$)}. For $d_n = 0.9$, (c) a neutral swimmer, (d) a shaker.  Here the  continuous lines are streamlines, and the  color field corresponds to the magnitude of the velocity. Note that $\theta_s$ is defined as an angle made by the orientation vector with the separation vector joining the center of the squirmer and the surface of the concave boundary.
          } 
             \label{fig:ann_inst_simu_velocity_fields}
        \end{figure}

         \begin{figure*}
        \centering
          \subfigure[]{
          \includegraphics[height=5.0cm]{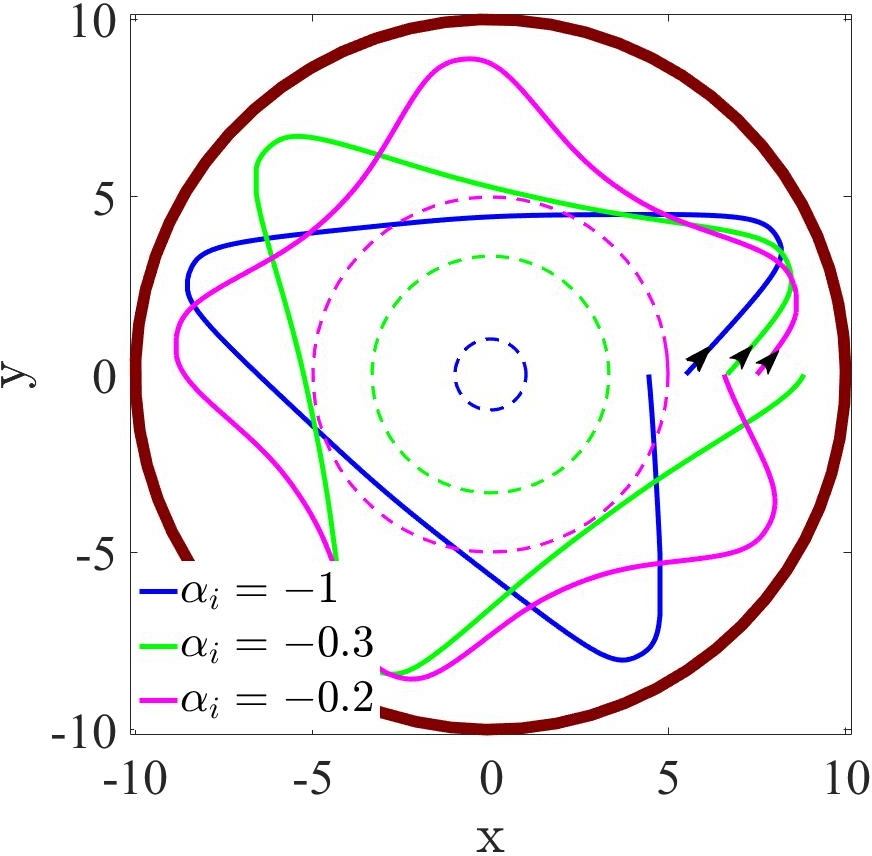}
              \label{fig:ann_neutral_alpha_o}}\quad
        \subfigure[]{
          \includegraphics[height=5.0cm]{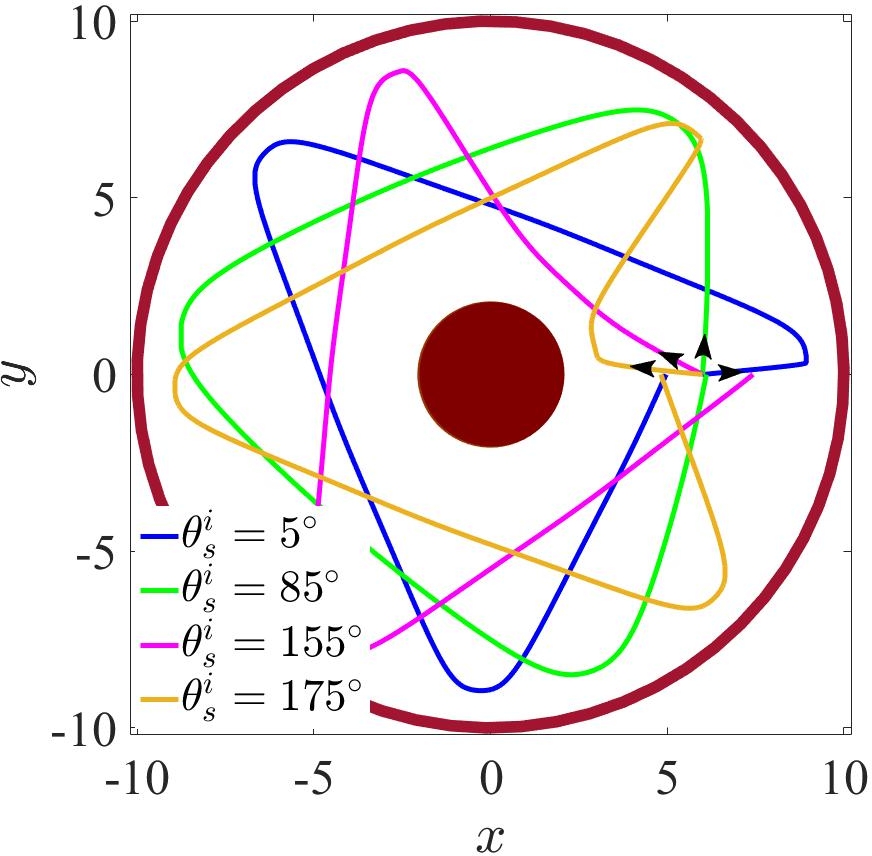}
          \label{fig:ann_neutral_orie}}\quad
              \subfigure[]{
          \includegraphics[height=5.0cm]{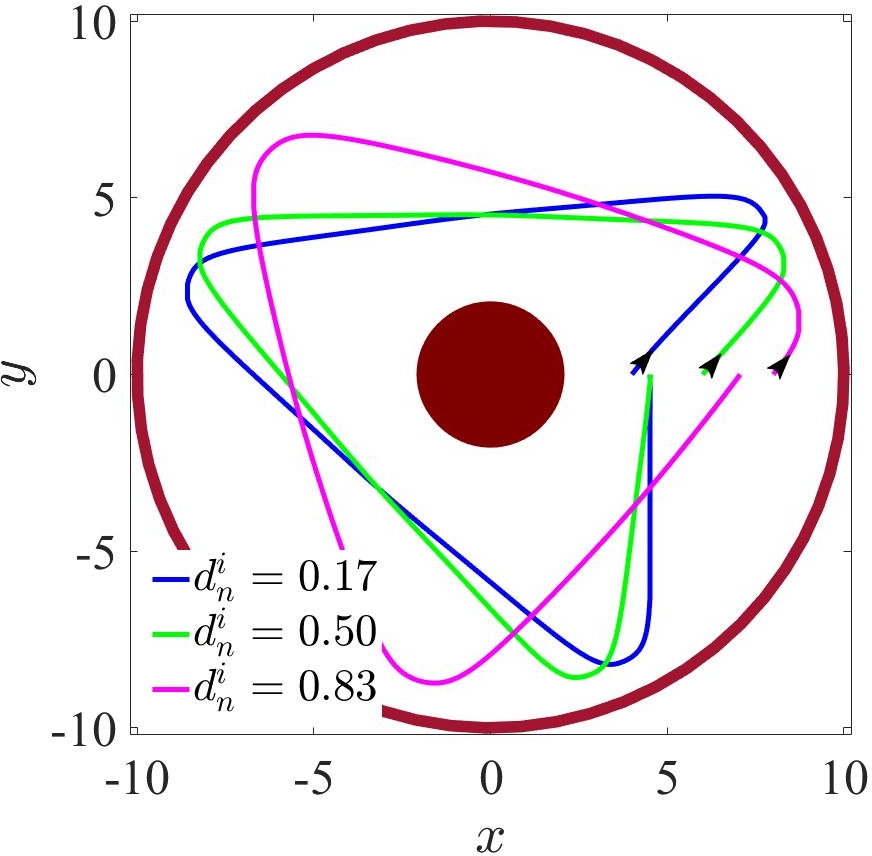}
           \label{fig:ann_neutral_initial_pos}}\quad
          \subfigure[]{
          \includegraphics[height=5.0cm]{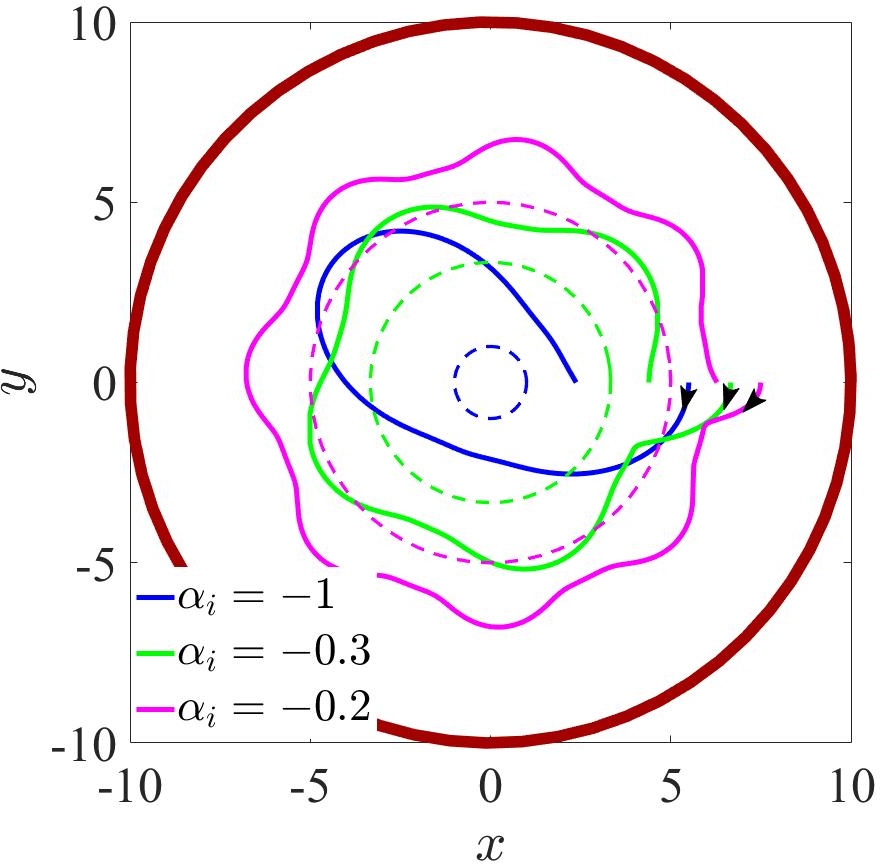}
          \label{fig:ann_shaker_alpha_o}}\quad
          \subfigure[]{
          \includegraphics[height=5.0cm]{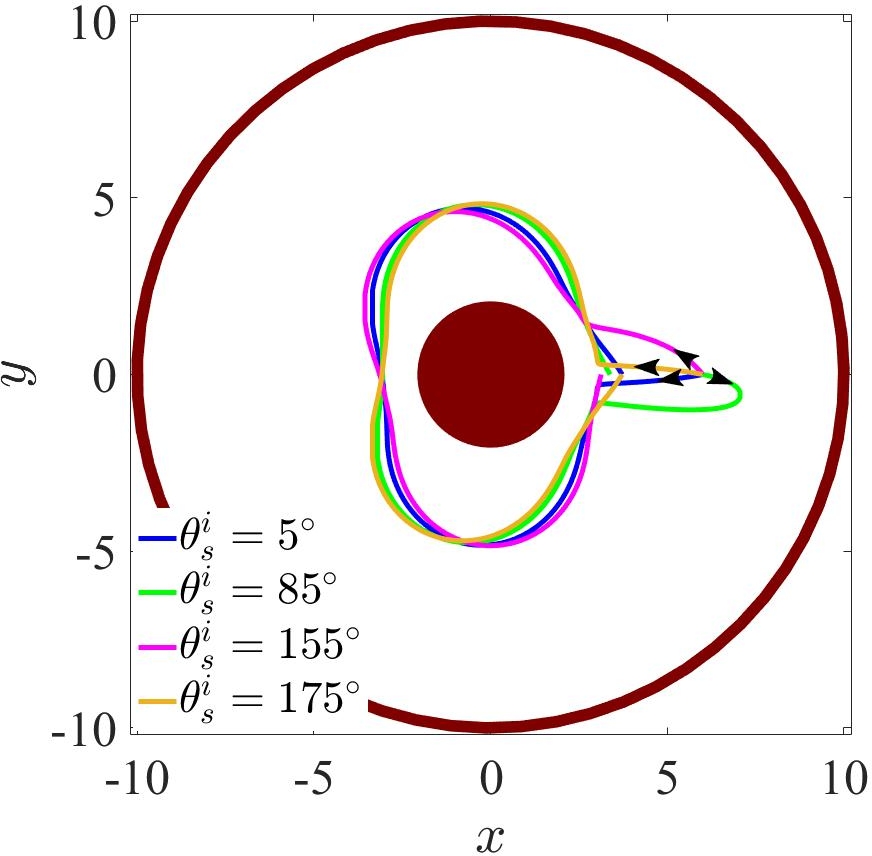}
          \label{fig:ann_shaker_orio}}\quad
            \subfigure[]{
          \includegraphics[height=5.0cm]{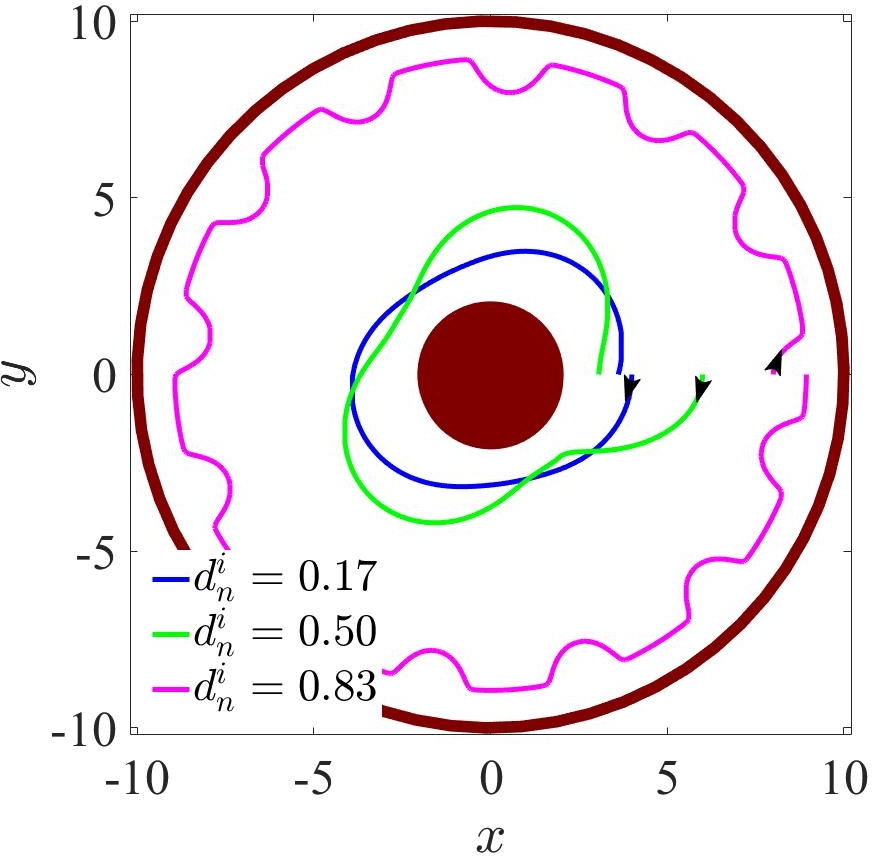}
          \label{fig:ann_shaker_initial_pos}}\quad
           \caption{Trajectories of neutral swimmers and shakers in an annular confinement. Effect of (a) $\alpha_i$ for $\alpha_o = 0.1$, $\theta_s^i = 45^{\circ}$, (b) initial orientation for $\alpha_i = -0.5$, $\alpha_o = 0.1$, and (c) initial location for $\theta_s^i = 45^{\circ}$, $\alpha_i = -0.5$, $\alpha_o = 0.1$ for a neutral swimmer. (d), (e), and (f) are corresponding plots for a shaker. In (a)-(b) and (d)-(e) the squirmer is initialised at an equal distance from both the confining surfaces. Hard repulsion is imposed at a distance $0.05r_s$ on the confining surfaces.
           } 
           \label{fig:annular_trajectories}
        \end{figure*}
        
        In earlier sections, we have analyzed the squirmer dynamics near a concave or a convex boundary. However, most often the boundaries are geometrically complex. Therefore, here we analyze the squirmer dynamics in annular confinement, where the dynamics is governed by both convex and concave curvatures of the boundary. A convex post is placed concentrically within a concave confinement, where both the convex post and the concave confinement are circular, and the motion of the squirmer in the resulting annular space is analyzed in this section. 
        
        The non-dimensional radii of curvature of the inner convex post,  $\alpha_{i} = r_s/r_{ib}$ and the outer concave confinement $\alpha_{o} = r_s/r_{ob}$ are taken to be negative and positive respectively to be consistent with the notation in the earlier sections. 
        Further, in order to characterise the separation distance between the squirmer and the surrounding surfaces, we define eccentricity
        \begin{align}
         d_n = \frac{d_{si}}{r_{ob}-r_{ib}-2r_s}   
        \end{align}
        where $d_{si}$ is the shortest distance between the surface of the inner convex post and the surface of the squirmer. $d_n = 0$ corresponds to the squirmer touching the inner convex post, and $d_n = 1$ corresponds to the squirmer touching the outer concave confinement.
        
        As earlier, two different approaches are used to analyze the squirmer dynamics:
        \begin{enumerate}
            \item In the section \ref{sec:results}, exact expressions describing the squirmer motion near a curved boundary \cite{dario_gareth} is used for the analysis. However, the exact expressions that govern the squirmer dynamics in an annular confinement is not available. Therefore, in this first approach, we construct an approximate solution by  superposition:
            \begin{align}
                \mathbf{V}^{annular} = \mathbf{V}^{convex} + \mathbf{V}^{concave}        \label{eqn:ann_V},\\
                \boldsymbol{\Omega}^{annular} = \boldsymbol{\Omega}^{convex} + \boldsymbol{\Omega}^{concave},
                \label{eqn:ann_omega}
            \end{align}
        where $\mathbf{V}^{convex}$, $\mathbf{V}^{concave}$, $\boldsymbol{\Omega}^{convex}$, and $\boldsymbol{\Omega}^{concave}$ are obtained from Eq.~\ref{eqn:exact_exprs}. Such approximate analysis based on superposition is shown to be useful in the context of a microswimmer confined between two flat walls \cite{poiseulle_flow}. To avoid the penetration of the squirmer into the boundary while constructing trajectories, a hard repulsion at a distance $\delta r_s$ is imposed from the confining surfaces. 
        \item In the second approach, no such assumptions are made, instead full numerical simulations are performed for a squirmer in an annular confinement using LBM simulations as explained in section~\ref{sec:LBM}.
        \end{enumerate}
        A comparison between two approaches, namely the instantaneous velocities obtained from the approximate expressions (first approach) and the exact values obtained from the LBM simulations (second approach) is given in Appendix~\ref{appendix:appendixB}. 
        
        Figure~\ref{fig:ann_inst_simu_velocity_fields} shows the steady state velocity fields generated by the squirmer in an annular confinement for varying eccentricities ($d_n$). These velocity fields are qualitatively similar to that observed near individual concave and convex boundaries. \kvsc{As shown in Fig.~\ref{fig:ann_B1_ff_0p1}, when the neutral swimmer ($B_1>0$, $B_2=0$) is close to the inner convex post ($d_n \to 0$) the source--sink nature of the flow is majorly affected, but no secondary eddies are observed. On the other hand, the same swimmer when close to the concave boundary ($d_n \to 1$) shows secondary eddies (Fig.~\ref{fig:ann_B1_ff_0p9}). In an annular confinement, the convex post obstructs these secondary eddies and thus the flow field is different compared to the case without the convex post (as in section \ref{sec:fluid_dynamics}). This change in the flow field results in viscous dissipation varying non--monotonically with squirmer location, with dissipation being maximum when squirmer is closer to either of the curved surfaces and minimum when it is in between.} Similarly, as shown in Fig.~\ref{fig:ann_B2_ff_0p1}--\ref{fig:ann_B2_ff_0p9}, the flow field generated by a shaker ($B_1=0$, $B_2 > 0$) in annular confinement is similar to that observed near convex and concave boundary. Again, if the squirmer is located close to a concave surface, then we can observe that the inner convex post alters the streamlines compared to the case where there is no post. The close similarity in the flow fields generated by the squirmer in the annular confinement and in the presence of individual surfaces show that the squirmer behaviour is likely to be governed \textit{solely} by the nature of nearby surface in most cases, which we verify below by constructing its trajectories. \kvsc{However, this effect of the nearby surface is dependent on the (i) relative size of the swimmer to the boundary, (ii) orientation of the swimmer, and (iii) location of the swimmer. The effect of each of these parameters is discussed below while analyzing the trajectories.}

        Figure~\ref{fig:annular_trajectories} describes the trajectories of the squirmer in an annular confinement. For a neutral swimmer, the effect of size of the convex post $\alpha_i$, the effect of initial orientation, and the effect of initial location are shown in Fig.~\ref{fig:ann_neutral_alpha_o}, Fig.~\ref{fig:ann_neutral_orie}, and Fig.~\ref{fig:ann_neutral_initial_pos} respectively. Similar plots for a shaker are shown in Fig.~\ref{fig:ann_shaker_alpha_o}, Fig.~\ref{fig:ann_shaker_orio}, and Fig.~\ref{fig:ann_shaker_initial_pos}. 
        
        It may be noticed from Fig.~\ref{fig:ann_neutral_alpha_o}--\ref{fig:ann_neutral_initial_pos} that, for a given $\alpha_o$ irrespective of the value of $\alpha_i$, initial orientation and location, the neutral swimmer bounces off from the outer concave confinement. \kvsc{Since $\alpha_o>>1$ ($r_{ob}>>r_s$), this behaviour is consistent with the earlier observations of bouncing trajectory of neutral squirmers near weak concave surfaces.} The collision with concave surface occurs multiple times thus the squirmer exhibits a bouncing trajectory on the outer concave surface, but during its path it never comes in contact with the convex post. Similar behaviour is observed even by varying $\alpha_o$ (not shown). It is interesting to note that even in cases when the squirmer is initially oriented towards the inner convex surface (Fig.~\ref{fig:ann_neutral_orie}), the squirmer is eventually attracted to the concave surface after the initial hydrodynamic collision with the convex surface. On the other hand as shown in Fig.~\ref{fig:ann_shaker_alpha_o}--\ref{fig:ann_shaker_initial_pos}, a shaker is attracted towards the inner convex surface irrespective of the size of the convex post, initial orientation or location. If the size of the convex post is small, then the trajectory is asymmetric about the post  (Fig.~\ref{fig:ann_shaker_alpha_o}). As $\alpha_i$ decreases, the size of the squirmer becomes comparable with the available annular space then the squirmer exhibits concentric but oscillatory trajectory. The exception to these observations is when the shaker is initially located very close to the outer concave surface. In this case, the squirmer remains in the attractive zone of the concave surface and exhibits an oscillatory trajectory on the concave surface (Fig.~\ref{fig:ann_shaker_initial_pos}).
        
        These observations are commensurate with earlier discussions and can be understood as follows. Proximity parameter (Fig.~\ref{fig:prox}) shows that the affinity of the neutral swimmer increases with increase in $\alpha$, $i.e.,$ neutral swimmer has higher affinity towards a concave surface compared to a convex surface of same radius. 
        Therefore, it is not surprising to see that the concave surface attracts the neutral swimmer which then exhibits a bouncing trajectory on the outer concave confinement in the annulus, similar to the neutral swimmer confined in a concave confinement without the inner post (Fig.~\ref{fig:cave_zero_rep_neutral},~\ref{fig:cave_0p05_rep_neutral}). 
        It may also be noted that the angular velocity of a neutral swimmer near a convex surface is larger than that near a concave surface. The result is that, the squirmers bounce from the convex surface with a large scattering angle even when they are initially directed towards it and then they approach the concave confinement. 
     
        The scattering at the outer concave surface is not sufficiently strong to make them move towards the convex surface back and the neutral swimmers exhibit a bouncing trajectory along the outer concave surface.
        
        However, for a shaker both inner convex and outer concave surfaces act as attractors. In other words, a shaker has almost equal affinity towards the strong convex and weak concave surfaces (Fig.~\ref{fig:vex_zero_rep_shaker}--\ref{fig:cave_zero_rep_shaker}). Therefore, shaker exhibits an initial condition dependent trajectory. If the shaker is located close to the inner convex surface, it gets trapped on the inner convex surface. However, the attraction by the confining concave surface results in periodic drifts from the convex surface.  \kvsc{These periodic drifts result in an asymmetric trajectory around a small convex post. As the size of the convex post increases, the trajectory becomes symmetric but oscillatory. This trapping of the shaker on the inner convex post irrespective of $\alpha_i$ is in contrast with the behavior depicted by the proximity parameter (Fig.~\ref{fig:prox_0_rep}). This apparent contrast can be understood as follows. In the absence of the outer confinement, the shaker escapes from the weak convex post ($\alpha = -1$) after a hydrodynamic collision. However, the presence of the outer concave confinement slows down and reverses the direction of the shaker which results in trapping on the inner convex post. }         If the shaker is initialised away from the inner surface, the affinity towards the inner convex surface decreases while affinity towards the outer concave surface increases. Beyond a certain distance, the effect of outer concave curvature dominates and shaker gets trapped by the outer concave surface.  
        
        \section{Conclusion and outlook}
        
        This work analysed the dynamics of a microswimmer (modelled using squirmer model) in  the neighbourhood of a curved wall in two dimensions.  The squirmer trajectories are constructed based on the results by \citet{dario_gareth}, and the corresponding flow fields are obtained from simulations using lattice Boltzmann method (LBM). An implicit scheme to update the squirmer dynamics is proposed in conjunction with LBM, and a close match between analytical and numerical solutions is found. \kvsc{Both instantaneous picture (in terms of translational and angular velocities of the squirmer) and long time effects (in terms of trajectories, fixed points on dynamical space, proximity parameter, retention time, scattering angle) are discussed in a phase plane spanning the activity ($\beta$) and non-dimensional curvature ($\alpha$) thus covering all types of swimmers (pushers, neutral swimmers, and pullers) present in the vicinity of all types of curvatures (convex, flat plate and concave) with the intention of providing a unified understanding of the effect of curvature on swimmer dynamics.}  Finally, we analyzed the squirmer dynamics in an annular confinement to understand the combined effect of convex and concave curvatures.
        
        \kvsc{The analysis of the velocity field has shown that the presence of a nearby boundary results in the formation of eddies in the flow field. In other words the streamlines near the squirmer are closed. The number and extend of circulatory flows depend upon the type and strength of the squirmer as well as the curvature of the neighbouring boundary. For a shaker, it has been found that the asymmetry in the velocity field varies non-monotonically with the curvature, $i.e.,$ asymmetry is minimum in the two limiting cases, \textit{i.e.,} in the limit $\alpha \to -\infty$ and  in the limit $\alpha \to 1$.}

       \par \kvsc{The microswimmer trajectories near a curved surface are analyzed in both the physical space and the dynamical space. The latter was used to understand the sensitivity of the microswimmer trajectories to initial conditions.} It was found that a microswimmer exhibits \kvsc{mainly} three kinds of trajectories near a curved boundary: (i) bouncing, (ii) oscillatory, and (iii) crawling trajectories. This is consistent with the observations near a convex post reported in the experiments of \citet{convex_expts}, and in the simulations of M. Kuron \textit{et al.} (refer Fig.~4 in \cite{convex_simulations}). \kvsc{The trajectories were first analyzed in terms of fixed points. We found two types of fixed points in the phase space that respectively characterize the crawling and hovering trajectories. We analyzed the dynamics of these fixed points in the $\alpha$--$\beta$ phase plane, and found that the number of crawling fixed points increases with increase in either wall curvature or activity. Owing to the inability of this approach to distinguish bouncing and oscillatory trajectories in the vicinity of the curved surface, we also introduced new measures with experimental relevance to characterize the trajectories,} namely average proximity parameter ($\langle\phi\rangle$) and average retention time ($\langle t_r\rangle$) which are obtained by averaging $\phi$ and $t_r$ respectively over all possible initial orientations. The proximity parameter and the retention time respectively measure  the average distance and time that the microswimmer spends in the neighbourhood of a wall.

        \par \kvsc{In agreement with the observations from dynamical portraits,} we found that the proximity parameter $\langle\phi\rangle$ increases with the wall curvature $\alpha$. Therefore microswimmers, irrespective of their type and strength (characterised by activity $\beta$), have a greater affinity towards a concave boundary compared to a convex boundary. Our findings are consistent with earlier observations - (i)  \citet{concave_expts} reported that the probability of finding a \textit{C. reinhardtii} in an elliptical confinement increases with increase in the local curvature of the boundary, and (ii) \citet{convex_simulations}  reported that the affinity of a microswimmer towards a convex boundary increases with increase in the curvature $\alpha$. It has been found that the presence of repulsive forces on the curved walls enhances the affinity of the microswimmer towards them. It happens irrespective of $\alpha$ and $\beta$, and in such cases, pullers have a slightly more affinity towards the convex and weak concave walls compared to pushers. In stronger concave confinements, squirmer dynamics is found to be independent of activity $\beta$.  The higher affinity of pullers towards a convex surface is also consistent with the observations of \citet{convex_simulations} made using their 3D simulations. As the velocity of the squirmer is not constant along its path, the affinity of the squirmer towards a curved surface is also quantified in terms of retention time. We observed that the dependence of average retention time  on $\alpha$ and $\beta$ is similar to that of the proximity parameter.
        \par The near wall behavior of  a squirmer is characterized in terms of its orientation with the surface tangent and its tangential velocity. The squirmer was found to maintain an angle close to $53^{\circ}$ with the neighbouring surface, irrespective of $\alpha$ and $\beta$. On the other hand, the tangential velocity in proximity of a curved surface is observed to decrease with increase in $\alpha$, \textit{i.e.,} it is larger near a convex curvature compared to a concave curvature. Larger tangential velocity near a convex surface indicates an easy escape of the squirmer while that near a concave surface can lead to further trapping of the squirmer. It has been found that irrespective of the curvature, the average tangential velocity of pushers is found to be slightly larger than pullers near convex and weak concave walls. This last observation is consistent with the study by \citet{arbitrary_curvature}, which included the role of thermal fluctuations as well. 
        \par The non-trapping dynamics of a squirmer is characterized in terms of scattering angle while bouncing off from the curved surface. The scattering angle was observed to \kvsc{increase with increase in the initial orientation ($\theta_s^i$) $i.e.,$ angle with respect to the separation vector. As $\theta_s^{i}$ increases, the closest distance that the squirmer can approach the nearby boundary decreases, therefore it experiences a smaller angular velocity and it deflects with a larger scattering angle.} The scattering angle also decreases with increase in $\alpha$, $i.e.,$ as the curvature changes from convex to concave. The larger values of scattering angle in the presence of a convex curvature is due to the undeflected trajectories of the microswimmer, that resulted from the weaker hydrodynamic interaction between the microswimmer and the convex boundary.
        \par The combined effects of convex and concave curvatures on microswimmer dynamics was analysed by placing the squirmer in a concentric annular confinement. Compared to that of an unconfined squirmer, the velocity field due to a confined squirmer is different which includes loss of symmetry of the flow field around the squirmer and the generation of secondary eddies. \kvsc{With increase in eccentricity, \textit{i.e.}, as squirmer gets closer to the outer concave confinement, such deviations increase indicating a large viscous dissipation. For neutral swimmers, larger dissipation corresponds to smaller swimming velocity while for shakers larger dissipation translates to larger swimming velocities (see Fig.~\ref{fig:ann_inst_vel}).} This difference in the flow fields and the squirmer dynamics due to disturbance generated by a convex and a concave boundary are also evident when a microswimmer in the neighbourhood of a single curved wall is analyzed. The exact expressions that govern the squirmer dynamics in an annular confinement are not available, hence we constructed approximate equations by superposition of induced velocities due to convex and concave curvatures. Extending the approximate solution approach to complex geometries like porous media \cite{Porous_media} have to be the subject of future investigations.
        The instantaneous velocities obtained from the approximate expressions are in good agreement with those obtained from lattice Boltzmann simulations for a neutral swimmer. The agreement is weaker for a shaker, possibly due to the slower decay of its velocity field. On analyzing the trajectories, it was found that a neutral swimmer exhibits a bouncing trajectory along the outer confinement. It occurs even if the squirmer is initially oriented towards the inner convex post because it collides with the post and scatters  towards the outer concave boundary. However, the trajectory of a shaker is found to depend on the initial location. If shaker is initialized near the inner convex surface, it gets trapped on it, which happens only beyond a critical radius of the inner convex surface. If squirmer is initialized close to the outer concave surface, it gets trapped on it, irrespective of its radius of curvature.
        \par Finally we note that the analysis presented here may provide guidelines in designing geometrical confinements that can be used to drive the motion of a microswimmer and thus to control its trajectory. Moreover, different measures introduced in the work can be used in a variety of contexts: (i) if it is desirable to trap a microswimmer for a particular application, results from the proximity parameter can be used since it provides information about the curvature of the boundary required to trap the microswimmer of particular type (given $\beta$). (ii) The retention time can be correlated with the time required to clean a curved surface to prevent the bio-fouling since retention time provides the average residence time of the microswimmer in proximity of the surfaces. (iii) The observations of the scattering angle may be used to guide the trajectory of a microswimmer by varying the local curvature of the boundary with which it interacts. However, caution must be exercised in quantitative comparisons as the calculations presented here are strictly in two dimensions. Our results are qualitatively consistent with several experiments and numerical simulations reported in the literature, but future investigations must be carried out in three dimensions to approach experimental conditions, and to make quantitative predictions. Similarly relaxing the assumption of circular or spherical shape of the squirmer may also be important. Such investigations will also help us to determine the relative importance of hydrodynamics in the collision process of a microswimmer with a curved wall more accurately.
        
        \begin{appendices}
        \section*{Appendix A}
        \refstepcounter{section} 
        \label{appendix:appendixA}
        \noindent\textbf{Implicit scheme for updating linear and angular velocities of the squirmer:}
        In this appendix, we extend the implicit scheme proposed for a passive particle by \citet{Frenkel}, to an active particle, squirmer with first two tangential modes to update the translational and angular velocities in the numerical scheme based on lattice Boltzmann method. It is straight forward to extend present scheme to higher order squirmer modes.
        \par The idea of implicit scheme is to use the particle velocity at the new time step to calculate the boundary velocity of the particle given by Eq.~\ref{eqn:boundarynodevel}, rather than using the velocity in the previous time step. We summarise the steps involved in this process. First, the net momentum exchange between the particle and the fluid is calculated for the implemented boundary conditions (bounce back scheme \cite{Timms_LBM_book}). This result is then used to derive the expressions for force and torque acting on a squirmer of mass, $M$, and moment of inertia, $I$. Both force and torque will be functions of two unknown velocity components (in the $x$ and $y$ directions) namely $V_x$ and $V_y$, and the unknown angular velocity of the particle in the z-direction $\Omega_z$. 
        Newton's second law is used here to relate the forces and torques acting on the squirmer to the rate of change of linear and angular momentum. Thus, finally, a set of three linear equations are obtained, 
        $ \left( \begin{array}{ccc}
        a_{11} & a_{12} & a_{13} \\
        a_{21} & a_{22} & a_{23} \\
        a_{31} & a_{32} & a_{33} \end{array} \right)$ $ \left( \begin{array}{c}
        V_x \\
        V_y  \\
        \Omega_z  \end{array} \right)$  = $ \left( \begin{array}{c}
        b_1 \\
        b_2 \\
        b_3  \end{array} \right)$+$ \left( \begin{array}{c}
        c_1 \\
        c_2  \\
        c_3  \end{array} \right)$,
        which has to be solved simultaneously to determine the three unknowns namely $V_x$, $V_y$, and $\Omega_z$.	
        The elements in the matrix are calculated as follows:
        \begin{align*}
        a_{11} &= 1+\frac{6}{M}\sum_{\mathbf{x}_b,i} \rho w_{i}e_{ix}e_{ix} \\
        a_{21} &= a_{12} =  \frac{6}{M}\sum_{\mathbf{x}_b,i} \rho w_{i}e_{ix}e_{iy}\\
        a_{13} &= -\frac{6}{M}\sum_{\mathbf{x}_b,i} (\rho w_{i}e_{ix}(r_{y}e_{ix}- r_{x}e_{iy}))\\
        a_{22} &= 1+\frac{6}{M}\sum_{\mathbf{x}_b,i} \rho w_{i}e_{iy}e_{iy}\\
        a_{23} &= -\frac{6}{M}\sum_{\mathbf{x}_b,i} (\rho w_{i}e_{iy}(r_{y}e_{ix}- r_{x}e_{iy}))
        \end{align*} 
        \begin{align*}
        a_{31} &= -\frac{6}{I}\sum_{\mathbf{x}_b,i} (\rho w_{i}e_{ix}(r_{y}e_{ix}- r_{x}e_{iy}))\\
        a_{32} &= -\frac{6}{I}\sum_{\mathbf{x}_b,i} (\rho w_{i}e_{iy}(r_{y}e_{ix}- r_{x}e_{iy}))\\
        a_{33} &= 1+\frac{6}{I}\sum_{\mathbf{x}_b,i} (\rho w_{i}e_{ix}(r_{y}e_{ix}- r_{x}e_{iy})^{2})\\
        b_1 &= V_x(t)+\frac{2}{M}\sum_{\mathbf{x}_b,i} f_{i}^{*}e_{ix}\\
        b_2 &= V_y(t)+\frac{2}{M}\sum_{\mathbf{x}_b,i} f_{i}^{*}e_{iy}\\
        b_3 &= \Omega_z(t)+\frac{2}{I}\sum_{\mathbf{x}_b,i} (r_{x}f_{i}^{*}e_{iy}-r_{y}f_{i}^{*}e_{ix})\\
        c_1 &= \frac{6}{M}\sum_{\mathbf{x}_b,i} w_{i} \rho (R_x e_{ix}+R_y e_{iy}) e_{ix}\\
        c_2 &= \frac{6}{M}\sum_{\mathbf{x}_b,i} w_{i} \rho (R_x e_{ix}+R_y e_{iy}) e_{iy}\\
        c_3 &= \frac{6}{M}\sum_{\mathbf{x}_b,i} w_{i} \rho (R_x e_{ix}+R_y e_{iy}) (r_{x} e_{iy}-r_{y} e_{ix})\\
        R_x &= \Big(B_{1} \sin(\theta_c)+B_{2} \sin(2\theta_c)\Big) \sin\theta\\
        R_y &= \Big(B_{1} \sin(\theta_c)+B_{2} \sin(2\theta_c)\Big) \cos\theta\\
        \end{align*} 
        
        \vspace{-0.5cm}
        
        where $\theta$ is the polar angle.
        In the above expressions, $\mathbf{r}$ is the position vector of a point relative to the squirmer centre, and $V_x(t)$, $V_y(t)$, and $\Omega_z(t)$ are the known values at the present time $t$. The summation is to be carried out over each boundary node ($\mathbf{x}_b$) and over all relevant directions ($\mathbf{e}_{i}$).  The proposed scheme can be extended easily to any number of squirmers as well.
        
        \section*{Appendix B}
        \refstepcounter{section} 
        \label{appendix:appendixB}
        
        In this appendix, the instantaneous velocities of a squirmer in annular confinement obtained from the approximate expressions, Eq.~\ref{eqn:ann_V},~\ref{eqn:ann_omega} are compared with the values obtained from full numerical, lattice Boltzmann simulations. 
        \begin{figure}[h!]
        \centering
          \subfigure[]{
          \includegraphics[height=5.0cm]{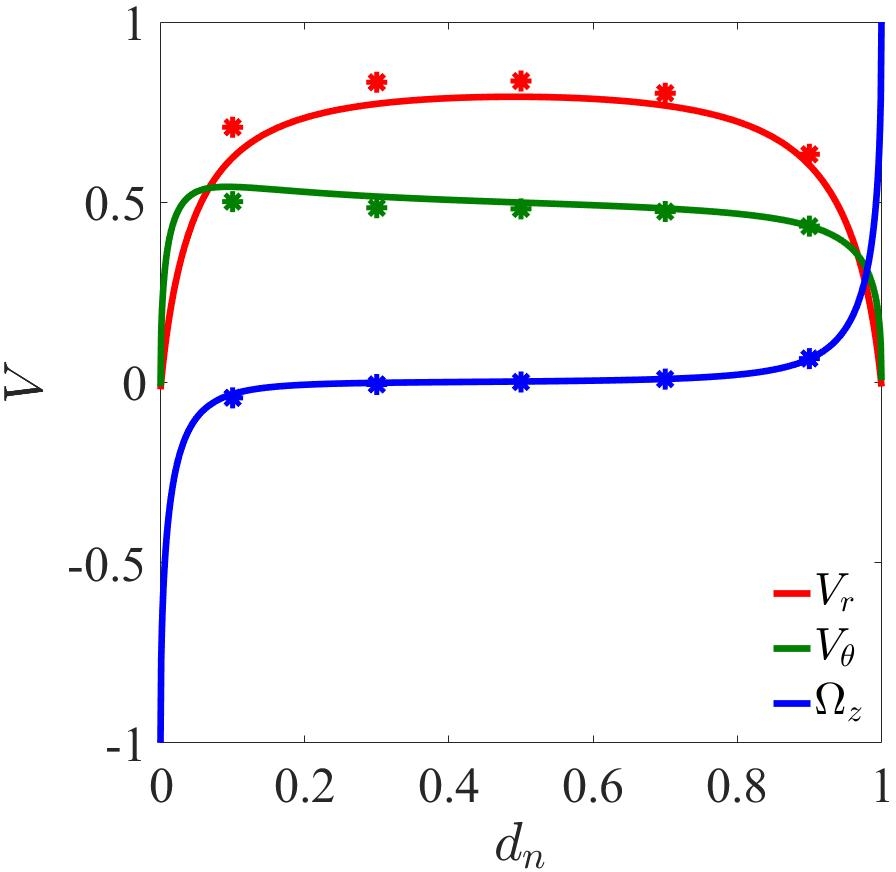}
              \label{fig:B1_ann_inst_vel}}\quad
        \subfigure[]{
          \includegraphics[height=5.0cm]{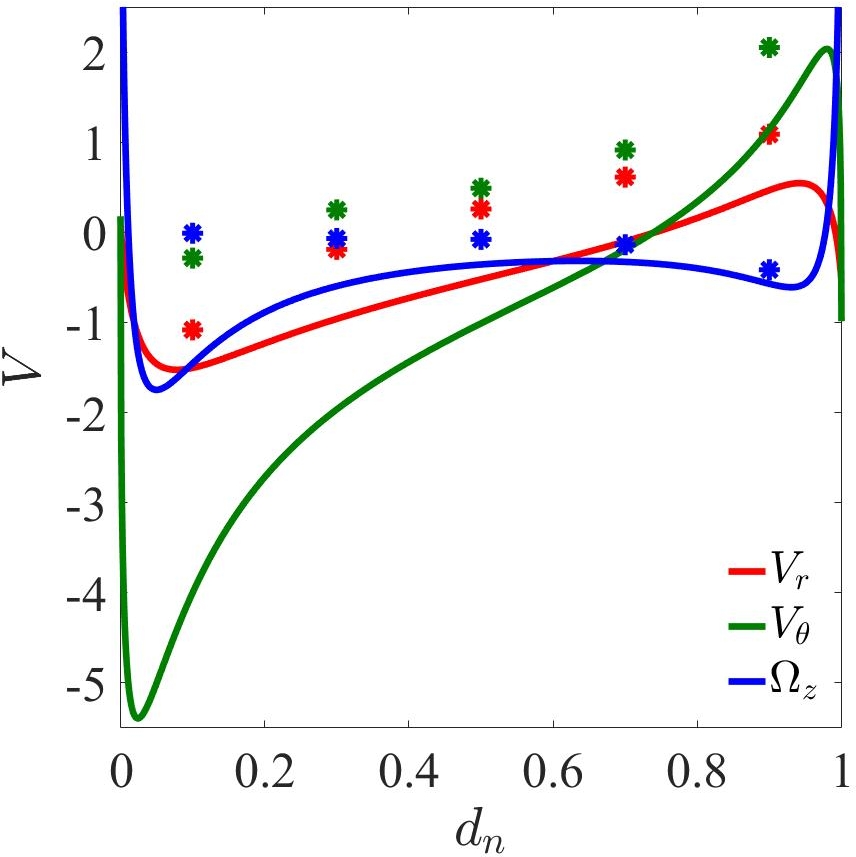}
          \label{fig:B2_ann_inst_vel}}\quad 
           \caption{Instantaneous velocities of a squirmer in an annular confinement as a function of eccentricity for an orientation of $\theta_s = 30^{\circ}$. (a) neutral swimmer ($B_1>0$, $B_2 = 0$) and (b) shaker ($B_1=0$, $B_2 \neq 0$). Continuous lines are expressions based on simple superposition (Eq.~\ref{eqn:ann_V}--\ref{eqn:ann_omega}), and markers indicate the results obtained from the numerical simulations.} 
             \label{fig:ann_inst_vel}
        \end{figure}
        
        The results are illustrated in Fig.~\ref{fig:ann_inst_vel}.
        The match between the simulations and the approximate expressions is better for the case of a neutral swimmer ($B_1>0$, $B_2 = 0$) compared to a shaker ($B_1=0$, $B_2 \neq 0$). This better match for a neutral swimmer is due to the rapid decay of the velocity field generated by it. In an unbounded domain, the velocity field generated by $B_1$ mode decays as $1/r^2$. Moreover, in the annular confinement, the strength of the Stokeslet velocity field generated by the convex post in response to the disturbance flow generated by the neutral swimmer is weaker, and is further suppressed by the outer concave confinement. 
        Thus the approximate expressions obtained by superposition of velocities due to concave and convex surfaces agree well with the results from simulations in the case of neutral swimmer.
        
        However, in the case of a $B_2$ mode, the  velocity field generated by the squirmer decays slower as $1/r$ in an unbounded fluid domain. Moreover, in the annular confinement, the strength of the disturbance velocity field generated by the convex post (Stokeslet) in response to the flow generated by the shaker is also stronger. Of course, the concave confinement will suppress this disturbance velocity field. However, these hydrodynamic interactions are not taken into account in the approximate expressions and thus they deviate from the full numerical results for shaker more compared to a neutral swimmer. \kvsc{Hence, it can be concluded that the error in the approximate solutions increases with
        increase in $\beta$ and in the limit of large $\beta$ numerical approach may give more accurate results compared to the
        approximate solution by superposition.}
        
        \end{appendices}
        
        \setcounter{secnumdepth}{0}
        \section{Acknowledgements}
        SPT acknowledges the support by Department of Science and Technology, India via the research grant CRG/2018/000644. 
        \newpage

        \bibliography{bib-pre}
\end{document}